\documentclass[american,english,pra, reprint, superscriptaddress, showkeys, eqsecnum, showpacs, nofootinbib, raggedbottom]{revtex4-1}
\usepackage[T1]{fontenc}
\usepackage[latin9]{inputenc}
\usepackage{color}
\usepackage{array}
\usepackage{booktabs}
\usepackage{units}
\usepackage{mathrsfs}
\usepackage{amsmath}
\usepackage{amssymb}
\usepackage{graphicx}

\makeatletter

\providecommand{\tabularnewline}{\\}

\usepackage{graphicx,url,layouts,xcolor}
\usepackage{cases}     % For numbering in cases format
\definecolor{blueviolet}{rgb}{0.2, 0.2, 0.6}
\usepackage[pdftex,          % FOR PDFLATEX ONLY
    bookmarks=false,          % Bookmark bar
    colorlinks=true,
    allcolors=blueviolet,
    pdfstartview={FitH},  % FitBH
    ]{hyperref}
\usepackage[preset=reset]{phfnote}

\makeatother

\usepackage{babel}
\begin{document}
\phfnoteSaveDefs{origcmds}{H,c,k,l,b,d,r,aa,u,v,t,o}

\global\long\def\half{\frac{1}{2}}
\global\long\def\bra{\langle}
\global\long\def\ket{\rangle}
\global\long\def\a{\alpha}
\global\long\def\b{\beta}
\global\long\def\g{\gamma}
\global\long\def\c{\chi}
\global\long\def\d{\delta}
\global\long\def\o{\omega}
\global\long\def\m{\mu}
\global\long\def\n{\nu}
\global\long\def\z{\zeta}
\global\long\def\l{\lambda}
\global\long\def\e{\epsilon}
\global\long\def\x{\chi}
\global\long\def\r{\rho}
\global\long\def\t{\theta}
\global\long\def\G{\Gamma}
\global\long\def\S{\Sigma}
\global\long\def\D{\Delta}
\global\long\def\di{\hat{\Delta}}
\global\long\def\parr{\hat{\Pi}}
\global\long\def\O{\Omega}
\global\long\def\L{\Lambda}
\global\long\def\P{\Phi}
\global\long\def\T{\Theta}
\global\long\def\dg{\dagger}
\global\long\def\s{\sigma}
\global\long\def\dag{\dagger}
\global\long\def\lket#1{\left|#1\right\rangle }
\global\long\def\k{\kappa}
\global\long\def\tr{\text{Tr}}
\global\long\def\H{H}

\global\long\def\aa{a}
\global\long\def\bb{b}
\global\long\def\ph{\hat{n}}
\global\long\def\phm{\hat{m}}
\global\long\def\F{F}
\global\long\def\DD{\mathcal{D}}
\global\long\def\HH{\mathcal{H}}

\global\long\def\ko{\kappa_{\text{I}}}
\global\long\def\lo{\mathcal{D}_{\text{I}}}
\global\long\def\fs{F_{\text{I}}}
\global\long\def\op{\mathbf{P}}
\global\long\def\os{\boldsymbol{\mathrm{Q}}}

\global\long\def\kt{\kappa_{\text{II}}}
\global\long\def\lt{\mathcal{D}_{\text{II}}}
\global\long\def\ft{F_{\text{II}}}
\global\long\def\tp{\boldsymbol{P}}
\global\long\def\ts{\boldsymbol{Q}}

\global\long\def\p{P}
\global\long\def\pr{\text{prob}}
\global\long\def\nb{\bar{n}}
\global\long\def\E{E}
\global\long\def\ds{\mathscr{D}}
%%% NUMBER IN CIRCLE %%%
\DeclareRobustCommand{\one}{\hyperref[l:one]{\raisebox{.5pt}{\textcircled{\raisebox{-.9pt}{1}}}}}
\DeclareRobustCommand{\two}{\hyperref[l:two]{\raisebox{.5pt}{\textcircled{\raisebox{-.9pt}{2}}}}}
\DeclareRobustCommand{\three}{\hyperref[l:thr]{\raisebox{.5pt}{\textcircled{\raisebox{-.9pt}{3}}}}}
\DeclareRobustCommand{\four}{\hyperref[l:fou]{\raisebox{.5pt}{\textcircled{\raisebox{-.9pt}{4}}}}}
\DeclareRobustCommand{\five}{\hyperref[l:fiv]{\raisebox{.5pt}{\textcircled{\raisebox{-.9pt}{5}}}}}
\DeclareRobustCommand{\six}{\hyperref[l:six]{\raisebox{.5pt}{\textcircled{\raisebox{-.9pt}{6}}}}}\global\long\def\cc{\textsf{concat}}
\global\long\def\pc{\textsf{pair-cat}}

\newcommand\crule[1][black]{\textcolor{#1}{\rule{4.5pt}{4.5pt}}}
\definecolor{lightgray}{rgb}{0.85, 0.85, 0.85}
\definecolor{green}{RGB}{0, 128, 0}

\title{Pair-cat codes: autonomous error-correction with low-order nonlinearity}

\author{Victor~V.~Albert}

\thanks{Equal contribution.}

\affiliation{Yale Quantum Institute, Departments of Applied Physics and Physics,
Yale University, New Haven, Connecticut 06520, USA}

\affiliation{Institute for Quantum Information and Matter and Walter Burke Institute
for Theoretical Physics, California Institute of Technology, Pasadena,
CA 91125, USA}

\author{Shantanu~O.~Mundhada}

\thanks{Equal contribution.}

\affiliation{Yale Quantum Institute, Departments of Applied Physics and Physics,
Yale University, New Haven, Connecticut 06520, USA}

\author{Alexander~Grimm}

\affiliation{Yale Quantum Institute, Departments of Applied Physics and Physics,
Yale University, New Haven, Connecticut 06520, USA}

\author{Steven~Touzard}

\affiliation{Yale Quantum Institute, Departments of Applied Physics and Physics,
Yale University, New Haven, Connecticut 06520, USA}

\author{Michel~H.~Devoret}

\affiliation{Yale Quantum Institute, Departments of Applied Physics and Physics,
Yale University, New Haven, Connecticut 06520, USA}

\author{Liang~Jiang}

\affiliation{Yale Quantum Institute, Departments of Applied Physics and Physics,
Yale University, New Haven, Connecticut 06520, USA}

\date{\today}
\begin{abstract}
We introduce a driven-dissipative two-mode bosonic system whose reservoir
causes simultaneous loss of two photons in each mode and whose steady
states are superpositions of pair-coherent/Barut-Girardello coherent
states. We show how quantum information encoded in a steady-state
subspace of this system is exponentially immune to phase drifts (cavity
dephasing) in both modes. Additionally, it is possible to protect
information from arbitrary photon loss in either (but not simultaneously
both) of the modes by \textit{continuously} monitoring the difference
between the expected photon numbers of the logical states. Despite
employing more resources, the two-mode scheme enjoys two advantages
over its one-mode cat-qubit counterpart with regards to implementation
using current circuit QED technology. First, monitoring the photon
number difference can be done without turning off the currently implementable
dissipative stabilizing process. Second, a lower average photon number
per mode is required to enjoy a level of protection at least as good
as that of the cat-codes. We discuss circuit QED proposals to stabilize
the code states, perform gates, and protect against photon loss via
either active syndrome measurement or an autonomous procedure. We
introduce quasiprobability distributions allowing us to represent
two-mode states of fixed photon number difference in a two-dimensional
complex plane, instead of the full four-dimensional two-mode phase
space. The two-mode codes are generalized to multiple modes in an
extension of the stabilizer formalism to non-diagonalizable stabilizers.
The $M$-mode codes can protect against either arbitrary photon losses
in up to $M-1$ modes or arbitrary losses and gains in any one mode.
\end{abstract}

\maketitle

\section{Introduction\label{sec:Introduction}}

\subsection{Motivation \& outline}

The search for how to realize the first fault-tolerant quantum computer
is currently underway. Due to the fragility of quantum information,
one has to encode said information redundantly into physical degrees
of freedom in order to be able to protect it from noise. In the field
of continuous-variable (CV) quantum information processing \cite{Braunstein2005,Weedbrook2012,cvbook,serafinibook},
one encodes information in the space corresponding to the occupation
(photon) number of a harmonic oscillator. A CV \textit{quantum code}
is then a subspace of the oscillator Hilbert space that is used to
protect quantum information against errors.

Beginning with the two-mode ``dual-rail'' encoding in 1995 \cite{Chuang1995},
there are currently several CV codes on the market. One can characterize
them by the oscillator basis states that most conveniently expresses
the code: Fock/number states $\{|n\ket\}_{n=0}^{\infty}$ \cite{Chuang1997,Knill2001,Ralph2005,Wasilewski2007,Bergmann2016a,bin,Niu2017},
position and momentum eigenstates $\{|x\ket\}_{x\in\mathbb{R}}$ and
$\{|p\ket\}_{p\in\mathbb{R}}$ \cite{Lloyd1998,Braunstein1998,Gottesman2001,Menicucci2014,Hayden2016,Ketterer2016},
or a few coherent states $\{|\a\ket\}_{\a\in S}$ (for some finite
set $S$) \cite{Cochrane1999,Niset2008,Leghtas2013b,Lacerda2016}.
There also exist hybrid schemes which couple an oscillator to other
systems \cite{Lee2013,Kapit2016}. In addition to the continuing focus
on optical cavity implementations, a few of the recent efforts \cite{Leghtas2013b,bin,Niu2017}
are tailoring codes for use in microwave cavities (\textit{modes})
coupled to Josephson junctions \cite{girvinbook,Gu2017}. In particular,
a class of single-mode codes known as the \textit{cat codes} \cite{Cochrane1999,Leghtas2013b}
(see also \cite{cats,Albert2015,Bergmann2016,Li2016}) has enjoyed
rapid experimental progress in the microwave paradigm \cite{Leghtas2014,Ofek2016,Heeres2016}
and may be applicable to protect against dephasing in phononic systems
\cite{Arrangoiz-Arriola2018,Chu2017}. It is thus natural to consider
similarly-tailored generalizations of this class to multiple modes.

In this manuscript, we present both a new code family --- the pair-cat
codes --- and a proposal for its realization using \textit{reservoir-engineered}
(a term coined in Ref.~\cite{Poyatos1996}) microwave cavities. We
show that the pair-cat code offers a promising balance between protection
from errors and near-term realizability. Namely, it is tailored to
protect from the largest incoherent source of error of microwave cavities
--- photon loss --- and its implementation provides several advantages
over previous designs.

Let $\DD[F]$ be a dissipator \cite{Belavin1969,Lindblad1976,Gorini1976a},
\begin{equation}
\DD[F](\r)=\F\r\F^{\dg}-\half\{\F^{\dg}\F,\r\},\label{eq:lme}
\end{equation}
where $F$ is a jump operator and $\r$ a density matrix. We consider
two schemes with respective jump operators \begin{subequations}
\begin{eqnarray}
\fs & = & \aa^{4}-\a^{4}\label{eq:jumps}\\
\ft & = & \aa^{2}\bb^{2}-\g^{4}\,.
\end{eqnarray}
\end{subequations}Above, $\{\aa,\bb\}$ are the two oscillator mode
operators, but we also use them to label the modes, and $\{\a,\g\}$
are complex parameters. The modes obey the standard commutation relations
$[\aa,\aa^{\dg}]=[\bb,\bb^{\dg}]=1$ and $[\aa,\bb^{\dg}]=0$ and
we denote photon number operators $\ph=\aa^{\dg}\aa$ and $\phm=\bb^{\dg}\bb$.
Storage of at least a qubit worth of information as well as suppression
of error processes requires a certain degree of symmetry, which is
main reason for why $\F_{\text{I,II}}$ are high-order (quartic) processes.
Scheme I has already been thoroughly studied \cite{Leghtas2013b,cats}
and we only review it here in a context that allows for a direct analogy
with the new scheme II.

Time evolution of a one- or two-mode density matrix is then governed
by the Lindbladian
\begin{equation}
\dot{\r}=\kappa_{\#}\DD_{\#}(\r)+\cdots\,,\label{eq:maindiss}
\end{equation}
where $\DD_{\#}=\DD[F_{\#}]$ and $\#\in\{\text{I},\text{II}\}$ corresponds
to jump the respective operators in Eq.~(\ref{eq:jumps}-b), $\kappa_{\#}$
is a non-negative rate and ``$\cdots$'' represent competing error
processes. The competing error processes include \textit{loss errors},
caused by dissipators of the form $\k_{\aa}\DD[\aa]$ and $\k_{b}\DD[\bb]$,
and \textit{dephasing errors}, caused by dissipators of the form $\k_{n}\DD[\ph]$
and $\k_{m}\DD[\phm]$. Quantum information is encoded in certain
steady states of $\DD_{\#}$, i.e., states $\r$ such that $\DD_{\#}(\r)=0$,
which form a decoherence-free subspace of $\DD_{\#}$ \cite{Duan1997,Zanardi1997,Lidar1998},
represented by its projection $P_{\text{\#}}$. In the cases considered
here, the code subspace satisfies $F_{\#}P_{\text{\#}}=0$, meaning
that $F_{\#}$ annihilates all states that are in the subspace (i.e.,
all states $\r$ for which $\r=P_{\text{\#}}\r P_{\text{\#}}$).

We continue this section by discussing the advantages of scheme II
and describing how to analyze errors and gates for both schemes. In
Sec.~\ref{sec:Background:-single-mode-cat-code}, we review code
properties and gates for scheme I. In Sec.~\ref{sec:Two-mode-pair-cat-code},
we do the same for scheme II. In Sec.~\ref{sec:Visualization-techniques},
we introduce techniques to visualize two-mode states in a two-dimensional
plane. In Sec.~\ref{sec:Multi-mode-generalizations-and}, we comment
on multimode generalizations and make contact with the stabilizer
formalism. In Sec.~\ref{sec:Continuous-vs-loss}, we develop the
experimental realization for scheme II. We conclude in Sec.~\ref{sec:Conclusion}.

\subsection{Advantages of pair-cat codes}

In this work, we introduce a complete error-correction method for
the two-mode scheme II. A side-by-side comparison to scheme I is in
Table \ref{tab:1}. The leading uncorrectable errors for both schemes
are of the same order, $\aa^{2}$ for scheme I and $\aa\bb$ for scheme
II, so the code subspaces in both schemes are of comparable quality.
However, while retaining all of the benefits of the cat codes, scheme
II enjoys several advantages, including most importantly a drastic
reduction of the order of the nonlinearity required for realization.
Three- and higher-mode extensions of scheme II further increase the
error-correcting properties of the codes, e.g., an $M$-mode code
for $M\geq2$ enjoys a leading-order uncorrectable loss error of $a_{1}a_{2}\cdots a_{M}$.
We summarize these advantages below.

\subsubsection{Discrete QEC against photon loss}

One can show that a dominant dissipative term $\k_{\#}\mathcal{D}_{\#}$
(\ref{eq:maindiss}) is able to continuously suppress (or, in the
sense of Ref.~\cite{Terhal2015}, passively protect from) any dephasing
error processes without the need for error syndrome measurement and
recovery operations. In this work, we refer to an error-correction
process that is continuous in time and that does not require active
measurement and feedback operations as \textit{continuous quantum
error correction (QEC)} \cite{Paz1998,Barnes2000,Ahn2002,Sarovar2005,Oreshkov2007,Kerckhoff2010,Kerckhoff2011,Sarma2013,Kapit2016,Jae-MoLihmKyungjooNoh}.\footnote{Continuous means ``continuous in time'' and autonomous means ``without
measurement and feedback'' \cite{cohenthesis}, but we use the terms
interchangeably since all of our continuous QEC is also autonomous.} Both schemes also admit \textit{discrete QEC} (i.e., conventional
protection via non-demolition measurements of error syndromes and
adaptive control) against photon loss, but only scheme II can perform
both QEC processes simultaneously using currently available techniques. 

The scheme I syndrome is the photon number parity,
\begin{equation}
\parr=\left(-1\right)^{\ph}\,,\label{eq:parity}
\end{equation}
and parity measurements \cite{Sun2014} and full-blown discrete QEC
\cite{Ofek2016} for scheme I have been implemented using current
superconducting circuit technologies. Separately, continuous QEC against
dephasing has been achieved for the simplest cat-code with jump operator
$\aa^{2}-\a^{2}$ \cite{Leghtas2014} (such a cat code cannot protect
from photon loss). However, it is impossible to perform both discrete
and continuous QEC for scheme I \textit{simultaneously} with current
technologies. The established measurement technique implements an
entangling gate $e^{iHt}$ generated by the naturally occurring cross-Kerr
interaction $H=\x\ph\s_{z}$ (where $\s_{z}$ acts on an ancillary
junction). The dissipator $\fs$ commutes with $e^{iHt}$ only at
$t=\nicefrac{\pi}{\chi}$ and not at any other intermediate time.
Therefore, the protective dissipation due to $F_{\text{I}}$ has to
be turned off during the measurement.

The scheme II syndrome is the photon difference, 
\begin{equation}
\di=\phm-\ph\,.\label{eq:diff}
\end{equation}
Unlike the photon parity, $\di$ is \textit{quadratic} in the bosonic
ladder operators. This mathematical fact yields a practical advantage:
discrete and continuous QEC can be implemented simultaneously using
the same circuit QED measurement scheme used for scheme I, namely,
reading out of the syndrome using an ancillary transmon. In other
words, if we were to use the now two-mode cross-Kerr interaction $H=(\x_{a}\ph+\chi_{b}\phm)\s_{z}$
to generate an entangling gate, then fine-tuning the two parameters
$\chi_{b}=-\chi_{a}=\chi$ generates an interaction $H=\x\di\s_{z}$
whose exponential $e^{iHt}$ commutes with $F_{\text{II}}$ \textit{for
all} $t$. Thus, the the stabilization process $\DD_{\text{II}}$
can remain on during measurement. Since fine tuning the nonlinearities
can only be done during fabrication, we introduce another scheme avoiding
such fine-tuning. This new scheme implements discrete QEC by substituting
the transmon with a cavity and coupling the syndrome to the amplitude
of the cavity coherent state.

\subsubsection{Continuous QEC against photon loss}

One way to circumvent the problem of scheme I is to correct photon
loss continuously using the Hamiltonian $H\propto\parr$ (\ref{eq:parity}).
Such a Hamiltonian can be synthesized using superinductances formed
by arrays of Josephson junctions \cite{Cohen2016} (see also {[}\citealp{cohenthesis},
Sec.~4.2.2{]}). Besides requiring such technology, this requires
an infinite-order nonlinearity (since $\parr$ is an infinite expansion
in powers of $\ph$) and a significantly higher number of photons
to guarantee that there are no spurious logical operations. On the
other hand, an analogous procedure for scheme II requires the Hamiltonian
$H\propto\di$ (\ref{eq:diff}) that is only bilinear in $\aa,\bb$.
Since such a Hamiltonian is readily available, realization of the
required jump operators is simpler and applicable to technologies
other than circuit QED. We provide a continuous QEC proposal against
loss for scheme II using Superconducting Nonlinear Asymmetric Inductive
eLements (SNAILs) \cite{Frattini2017} which, other than that and
the fact that the syndrome is bilinear, is similar in spirit to the
superinductance-based proposal for scheme I. 

\subsubsection{Realizing jump operators $F_{\#}$}

While the jump operators $\F_{\text{I}},\F_{\text{II}}$ are both
quartic in the lowering operators $\aa,\bb$, the latter is only quadratic
in the lowering operators of each mode. Qualitatively, this allows
us to spread the degree of nonlinearity required to realize the scheme
over two modes instead of ``concentrating'' it in one mode. The
quantitative advantage is that the dissipative part of scheme II requires
less photons per mode to enjoy a comparable protection against dephasing
and a slightly lower probability of the leading uncorrectable loss
error. Moreover, while our proposed experimental design suffers from
an undesirable error-causing dissipator, errors due to this dissipator
can in principle be measured and corrected. This is not the case for
a similar design of scheme I \cite{Mundhada2017}, which introduces
dissipation consisting of uncorrectable two-photon-loss errors.

\subsubsection{Advantages of more modes}

While the two-mode pair-cat code has mostly experimentally relevant
advantages over single-mode cat codes, $M$-mode pair-cat codes correct
even more errors as $M$ increases. In Sec. \ref{sec:Multi-mode-generalizations-and},
we show that our three-mode code has the ability to \textit{either}
correct arbitrary losses in any two modes \textit{or} to correct arbitrary
gains or losses in any one mode. We compare this code to two other
multi-mode bosonic codes, $\chi^{(2)}$ codes \cite{Niu2017} and
noon codes \cite{Bergmann2016a}, showing that it has a larger set
of correctable errors. We also provide a numerical comparison of our
three-mode code to a three-mode code consisting of the simplest single-mode
cat-code concatenated with a repetition code. The latter, whose codes
states are GHz states consisting of coherent state components \cite{Munro2002},
has been proposed as a candidate for a future bosonic qubit {[}\citealp{cohenthesis},
Sec. 4.3{]}, assuming that the aforemenetioned superinductance technology
necessary to reliably measure its syndromes is developed. While not
at all complete due to the difficulty of running numerics on the large
three-mode Hilbert space, our comparison suggests that the pair-cat
code outperforms the concatenated cat code in the regime where $\approx1$
photon per mode is used.

\begin{table*}[t]
\begin{tabular}{>{\raggedright}p{0.27\textwidth}ll}
\toprule 
~ & Single-mode cat code \cite{Leghtas2013b,cats} & Two-mode pair-cat code\tabularnewline\addlinespace[2pt]
\midrule 
Error syndrome \& projections & Photon number parity $\parr=\left(-1\right)^{\ph}$ & Photon number difference $\hat{\D}=\phm-\ph$\tabularnewline\addlinespace[2pt]
 & $\mathbf{P}_{{\color{red}\Pi}}=\frac{1}{2}[1+(-1)^{\ph-{\color{red}\Pi}}]$
(\ref{eq:parity_projections}); ${\color{red}\Pi}\in\{0,1\}$~~~ & $\boldsymbol{P}_{{\color{red}\D}}=\intop_{0}^{2\pi}\frac{d\t}{2\pi}\exp[i(\di-{\color{red}\D})\t]$
(\ref{eq:deltaprojections}); ${\color{red}\D}\in\mathbb{Z}$\tabularnewline\addlinespace[2pt]
\midrule 
Code state components & Cat states $|{\color{green}\a}_{{\color{red}\Pi}}\ket\propto\mathbf{P}_{{\color{red}\Pi}}|{\color{green}\a}\ket$ & Pair-coherent states $|{\color{green}\g}_{{\color{red}\D}}\ket\propto\boldsymbol{P}_{{\color{red}\D}}|{\textstyle {\color{green}\g}},{\color{green}\g}\ket$\tabularnewline\addlinespace[2pt]
 & $\parr|{\color{green}\a}_{{\color{red}\Pi}}\ket=\left(-1\right)^{{\color{red}\Pi}}|{\color{green}\a}_{{\color{red}\Pi}}\ket$ & $\hat{\D}|{\color{green}\g}_{{\color{red}\D}}\ket={\color{red}\D}|{\color{green}\g}_{{\color{red}\D}}\ket$\tabularnewline\addlinespace[2pt]
\midrule 
Code states ${\color{blue}\m}\in\{0,1\}$ & $|{\color{blue}\m}_{{\color{green}\a},{\color{red}\Pi}}\ket\sim\frac{1}{\sqrt{2}}[|{\color{green}\a}_{{\color{red}\Pi}}\ket+\left(-1\right)^{{\color{blue}\m}}|{\color{green}i\a}_{{\color{red}\Pi}}\ket]$ & $|{\color{blue}\m}_{{\color{green}\g},{\color{red}\D}}\ket\sim\frac{1}{\sqrt{2}}[|{\color{green}\g}_{{\color{red}\D}}\ket+\left(-1\right)^{{\color{blue}\m}}\left(-i\right)^{\D}|{\color{green}i\g}_{{\color{red}\D}}\ket]$\tabularnewline\addlinespace[2pt]
\midrule 
Correctable loss errors & $\aa$ & $\{\aa^{k},\,\bb^{\ell}\,|\,k,\ell\geq0\}$\tabularnewline\addlinespace[2pt]
Uncorrectable loss error & $\aa^{2}$ & $\aa\bb$\tabularnewline\addlinespace[2pt]
\vspace{-2cm}
How errors act on codespace & ~~~~~~\includegraphics[height=0.9in]{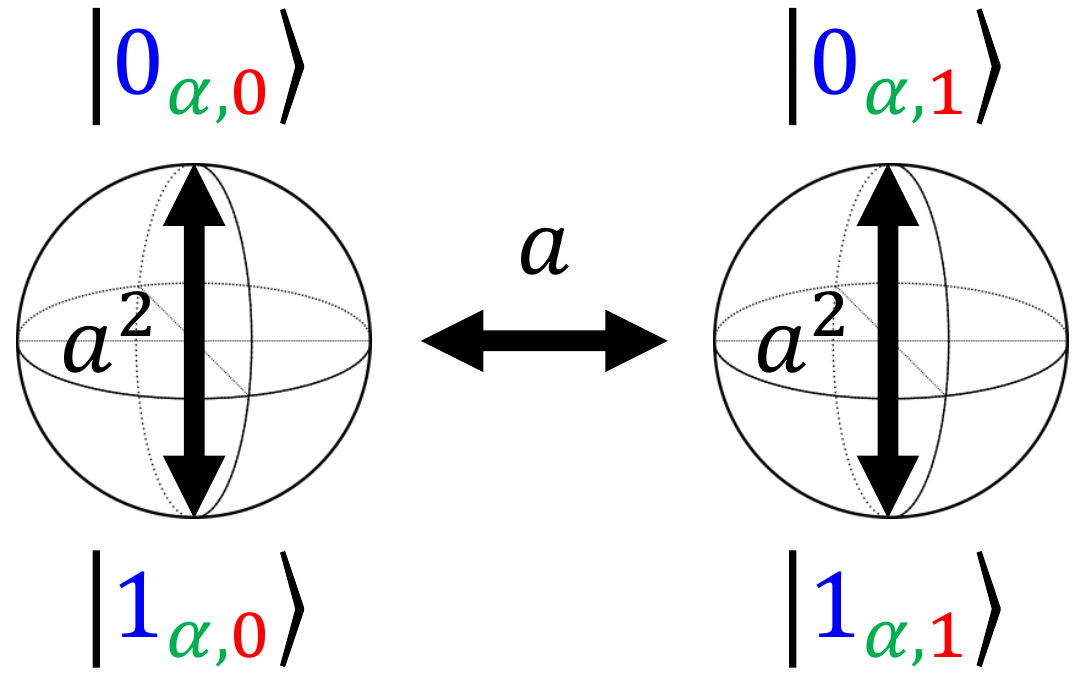} & \includegraphics[height=0.9in]{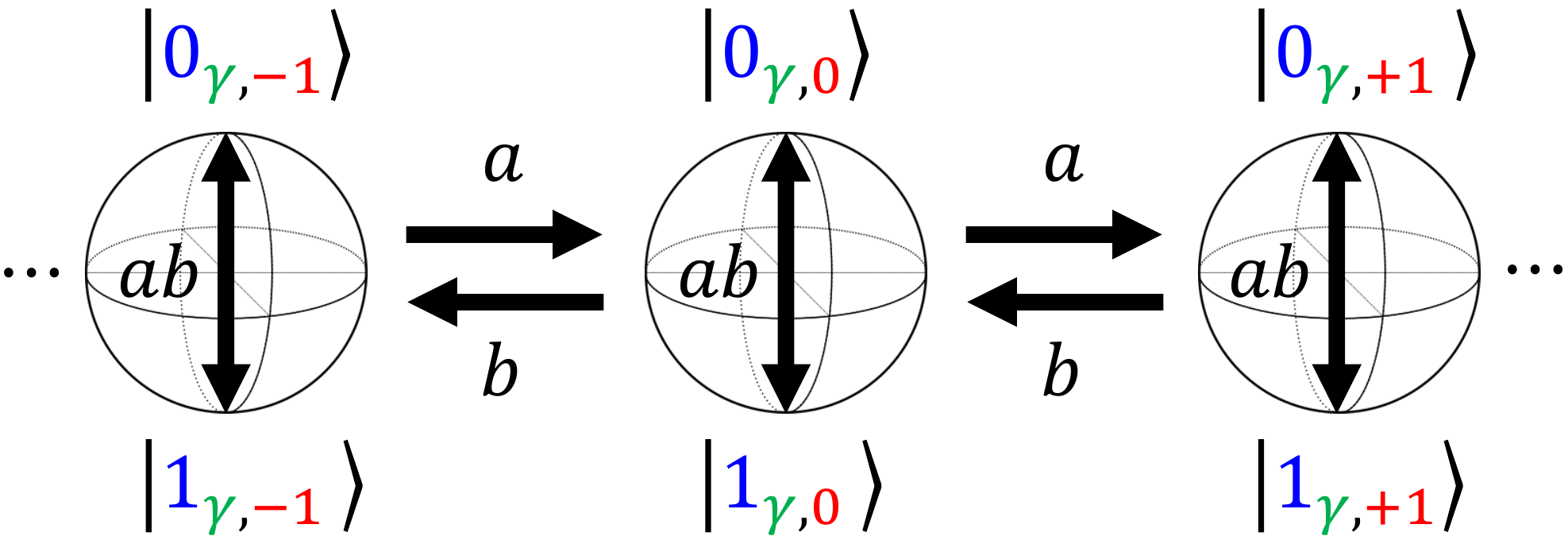}\tabularnewline\addlinespace[2pt]
\midrule 
Stabilizing jump operator & $\fs=\aa^{4}-{\color{green}\a}^{4}$ & $\ft=\aa^{2}\bb^{2}-{\color{green}\g}^{4}$\tabularnewline\addlinespace[2pt]
Dephasing errors suppressed as & ${\color{green}\a}\rightarrow\infty$ & ${\color{green}\g}\rightarrow\infty$\tabularnewline\addlinespace[2pt]
\midrule 
Realizing jump operator & Refs.~\cite{cats,Mundhada2017}; realized for $\aa^{2}-{\color{green}\a}^{2}$
\cite{Leghtas2014} & Sec.~\ref{sec:Continuous-vs-loss}\tabularnewline\addlinespace[2pt]
Realizing discrete QEC vs. loss & Ref.~\cite{cats}; realized \cite{Ofek2016} & Sec.~\ref{sec:Discrete-QEC-against}\tabularnewline\addlinespace[2pt]
Realizing cont. QEC vs. loss & Ref.~\cite{Cohen2016} & Sec.~\ref{sec:Continuous-QEC-against}\tabularnewline\addlinespace[2pt]
\midrule 
Hamiltonian $X$-gate & $\H_{\text{I}}^{X}=g_{X}(\aa^{2}+\mathrm{h.c.})$ \cite{cats} & $\H_{\text{II}}^{X}=g_{X}(\aa\bb+\mathrm{h.c.})$\tabularnewline\addlinespace[2pt]
Hamiltonian $XX$-gate & $\H_{\text{I}}^{XX}=g_{XX}[(\aa_{1}\aa_{2})^{2}+\mathrm{h.c.}]$ \cite{cats} & $\H_{\text{II}}^{XX}=g_{XX}(\aa_{1}\bb_{1}\aa_{2}\bb_{2}+\mathrm{h.c.})$\tabularnewline\addlinespace[2pt]
Hamiltonian $Z$-gate (in RWA) & $\H_{\text{I}}^{\text{jnct}}=E_{J}\cos\left(\b\aa e^{i\o t}+\mathrm{h.c.}\right)$
\cite{Cohen2016} & $\H_{\text{II}}^{\text{jnct}}=E_{J}\cos\left(\a\aa e^{i\o_{a}t}+\b be^{i\o_{b}t}+\mathrm{h.c.}\right)$\tabularnewline\addlinespace[2pt]
Holonomic $Z$-gate & $U_{\text{I}}^{\text{hol}}:\a\rightarrow0\rightarrow\a e^{i\phi}\rightarrow\a$
\cite{Albert2015} & $U_{\text{II}}^{\text{hol}}:\g\rightarrow0\rightarrow\g e^{i\phi}\rightarrow\g$\tabularnewline\addlinespace[2pt]
\midrule 
Kerr $\nicefrac{\pi}{2}$ $Z$-rotation & $U_{\text{I}}^{Z}=\exp[i\frac{\pi}{8}(\ph-\Pi)^{2}]$ \cite{cats} & $U_{\text{II}}^{Z}=\exp[i\frac{\pi}{8}(\ph+\phm-\D)^{2}]$\tabularnewline\addlinespace[2pt]
Kerr control-phase gate & $U_{\text{I}}^{CZ}=\exp[i\frac{\pi}{4}(\ph_{1}-\Pi_{1})(\ph_{2}-\Pi_{2})]$
\cite{zhang2017} & $U_{\text{II}}^{CZ}=\exp[i\frac{\pi}{4}(\ph_{1}+\phm_{1}-\D_{1})(\ph_{2}+\phm_{2}-\D_{2})]$\tabularnewline\addlinespace[2pt]
Control engineering & Ref.~\cite{Heeres2016} (experiment) & Ref.~\cite{bin}, Appx.~G\tabularnewline\addlinespace[2pt]
\bottomrule
\end{tabular}\caption{\label{tab:1}Comparison between the single-mode cat code \citep{cats}
and the two-mode pair-cat code. The last three entries represent gates
which have to be implemented with the stabilizing jump operator $F_{\#}$
turned off.}
\end{table*}

\subsection{Error analysis and recipe for logical gates\label{sec:Introduction-1}}

This paper is structured such that both schemes I and II are analyzed
in the framework of quantum error-correcting codes \cite{Bennett1996,Knill1997}
(see also \cite{nielsen_chuang}, Thm.~10.1). Namely, we analyze
the error-correcting properties of the codes from both schemes in
terms of the quantum error-correction conditions, extending notions
of weight and distance from traditional multi-qubit quantum error
correction. A \textit{quantum error-correcting code} is a subspace
of the full (one- or two-mode) Hilbert space that is used to store
a quantum state in order to prevent its quantum information from changing
without notice. The subspace corresponding to code \# (with $\#\in\{\text{I},\text{II}\}$)
is determined uniquely by its corresponding projection 
\begin{equation}
\p_{\#}=|0_{\#}\ket\bra0_{\#}|+|1_{\#}\ket\bra1_{\#}|\,,
\end{equation}
where $|\m_{\#}\ket$ ($\m\in\{0,1\}$) are the logical states of
the code. (One can easily check that $\p_{\#}$ is invariant under
changes of basis.) All errors in a set $\{E_{\ell}\}$ are correctable
if and only if, for all $\ell,\ell^{\prime}$,
\begin{equation}
P_{\#}E_{\ell}^{\dg}E_{\ell^{\prime}}P_{\#}=c_{\ell\ell^{\prime}}P_{\#}\,,\label{eq:qec}
\end{equation}
where $c_{\ell\ell^{\prime}}\in\mathbb{R}$ (and can be zero). In
other words, products of errors $E_{\ell}^{\dg}E_{\ell^{\prime}}$
must act trivially \textit{within the code space} (i.e., must act
independently of the code words when projected onto the code space).
For generic errors not satisfying the error-correction conditions,
Eq.~(\ref{eq:qec}) becomes
\begin{equation}
P_{\#}E_{\ell}^{\dg}E_{\ell^{\prime}}P_{\#}=c_{\ell\ell^{\prime}}P_{\#}+x_{\ell\ell^{\prime}}X_{\#}+y_{\ell\ell^{\prime}}Y_{\#}+z_{\ell\ell^{\prime}}Z_{\#}\,,\label{eq:qec-generic}
\end{equation}
with the latter three matrix basis elements defined in terms of outer
products of the code states:\begin{subequations}
\begin{align}
Z_{\#} & =|0_{\#}\ket\bra0_{\#}|-|1_{\#}\ket\bra1_{\#}|\label{eq:logz}\\
X_{\#} & =|0_{\#}\ket\bra1_{\#}|+|1_{\#}\ket\bra0_{\#}|\label{eq:logx}\\
Y_{\#} & =|1_{\#}\ket\bra0_{\#}|-|0_{\#}\ket\bra1_{\#}|\,.\label{eq:logy}
\end{align}
\end{subequations}Since the codes we consider consist of real vectors
and the error Kraus operators (\ref{eq:loss}-\ref{eq:deph}) are
real when written in the Fock-state basis, the matrices are defined
as such in order to avoid complex numbers.

We analyze the effect of various dephasing and loss errors by checking
whether $x_{\ell\ell^{\prime}}=y_{\ell\ell^{\prime}}=z_{\ell\ell^{\prime}}=0$
in Eq.~(\ref{eq:qec-generic}), i.e., the quantum error-correction
conditions hold. The errors we consider can be expressed in terms
of the Kraus operators of the respective processes, which we define
only for the first mode since they are the same for the second mode.
The error channel for an error $\text{err}\in\{\aa,\ph\}$ and acting
for a time $t$ can be written as
\begin{align}
e^{\k_{\text{err}}t\DD[\text{err}]}(\r) & =\sum_{\ell=0}^{\infty}E_{\text{err}}^{\ell}\r E_{\text{err}}^{\ell\dg}\,,
\end{align}
where $\r$ is a state and the Kraus operators for loss \cite{Ueda1989,Lee1994,Chuang1997,klimov_book}
and dephasing\footnote{One can use the same techniques as from, e.g., Ref.~\cite{klimov_book};
the calculations dramatically simplify since all terms in $\DD[\ph]$
commute.} are
\begin{align}
E_{\aa}^{\ell} & =\sqrt{\frac{(1-e^{-\k_{\aa}t})^{\ell}}{\ell!}}e^{-\half\k_{\aa}t\ph}\aa^{\ell}\label{eq:loss}\\
E_{\ph}^{\ell} & =\sqrt{\frac{(\k_{n}t)^{\ell}}{\ell!}}e^{-\half\k_{n}t\ph^{2}}\ph^{\ell}\,,\label{eq:deph}
\end{align}
respectively. The operators $E_{\aa}^{\ell=0}$ and $\{E_{\ph}^{\ell}\}_{\ell=0}^{\infty}$
induce exclusively \textit{dephasing errors} $\{\ph^{k}\}_{k=0}^{\infty}$
because they do not contain a power of the loss operator $\aa$ that
is not compensated by the same power of $\aa^{\dg}$. The remaining
operators $\{E_{\aa}^{\ell}\}_{\ell>0}^{\infty}$ are called \textit{loss
errors} since they each contain a decrease of the occupation number
by $\ell$. All errors are written as a superposition of a power of
$\aa$ multiplied by a function which can be expanded in a series
consisting of powers of $\ph$. Therefore, we only have to consider
whether the constituents $\aa^{k}$ and $\ph^{k}$ violate Eq.~(\ref{eq:qec})
when projected onto the code subspace.\footnote{\label{fn:Interested-readers-are}Interested readers are welcome to
browse Ref.~\cite{codecomp}, which performs in-depth calculations
for general cat codes.} Moreover, since an expression consisting of $\{\ph^{k}\}_{k=0}^{\infty}$
can be normal ordered into that consisting of $\{\aa^{\dg k}\aa^{k}\}_{k=0}^{\infty}$,
we instead consider the constituents $\aa^{k}$ and $\aa^{\dg k}\aa^{k}$.

Analysis of gates for our codes is also performed using the above
framework. Namely, given a perturbation Hamiltonian $\e H$ with small
parameter $\e$,\footnote{We are dealing with perturbations of unbounded operators, so an average
photon number constraint or truncation of Fock space need to be imposed
for perturbation theory to be meaningful.} we can determine whether it achieves a rotation within the subspace
$P_{\#}$ by checking its effect within the code space ($P_{\#}HP_{\#}$).
In this case, it is beneficial to violate Eq.~(\ref{eq:qec}) since
otherwise $H$ acts trivially on the code. In other words, say that
$E_{\ell^{\prime}}=I$ (identity) and the remaining $E_{\ell}^{\dg}$
in the product $E_{\ell}^{\dg}E_{\ell^{\prime}}$ fails to satisfy
Eq.~(\ref{eq:qec}). Then, one can interpret $E_{\ell}^{\dg}$ not
only as an uncorrectable error, but as a quantum gate generated by
the corresponding Hamiltonian $H=E_{\ell}+E_{\ell}^{\dg}$. This Hamiltonian,
and more generally any Hamiltonian, can be used to generate rotations
within the codespace $P_{\#}$ in the following way. Let $\HH(\r)=-i[\e H,\r]$
and $\e\ll1$ and consider the Lindbladian
\begin{equation}
{\cal L}=\DD_{\#}+\e\HH\,.
\end{equation}
Then, to the lowest order in $\e$, the effect of $\HH$ within the
code subspace is exactly \cite{ABFJ} (see also \cite{Zanardi2014,Azouit2016})
\begin{equation}
H_{\#}=P_{\#}HP_{\#}\,.\label{eq:zeno}
\end{equation}
This should not come as a surprise since this is exactly the energy
correction term to the subspace $P_{\#}$ in ordinary Hamiltonian-based
perturbation theory, but its extension to steady-state subspaces of
open systems nevertheless required a more careful derivation. We note
that first-order perturbation theory also allows for leakage to occur
outside of the code space, but that effect can be suppressed by a
proper rescaling of the perturbation that can be interpreted as quantum
Zeno dynamics \cite{Facchi2002,Zanardi2014,Arenz2016}. In other words,
if we let $\e=\nicefrac{1}{T}$ with $T$ being the total time that
${\cal H}$ is applied to our system, then at time $T$, the leading-order
term governing leakage out of the code space of order $O(\nicefrac{1}{T})$
while $TP_{\#}HP_{\#}=O(1)$ \cite{Zanardi2014}. As $T\rightarrow\infty$,
the state of the system continues to evolve in the code space under
$P_{\#}HP_{\#}$ and any leakage is suppressed. Unless otherwise specified,
any Hamiltonian-based gates we consider below can be implemented in
this manner.

\section{Background: single-mode cat-code\label{sec:Background:-single-mode-cat-code}}

We first review the cat-code scheme I \cite{cats} using notation
that allows us to generalize to scheme II in a straightforward manner.

\subsection{Primer on cat states}

In order to define the code subspaces for scheme I, we perform a symmetry
analysis \cite{prozen,pub011} of the corresponding jump operator
$\F_{\text{I}}$. Recall that all steady states are annihilated by
$\F_{\text{I}}$ and notice that $\fs$ commutes with the photon number
parity $\parr=\op_{0}-\op_{1}$ (\ref{eq:parity}), where we denote
parity eigenspace projectors
\begin{equation}
\op_{\Pi}=\frac{1+(-1)^{\ph+\Pi}}{2}=\sum_{n=0}^{\infty}|2n+\Pi\ket\bra2n+\Pi|\,,\label{eq:parity_projections}
\end{equation}
$\Pi\in\{0,1\}$. Parity is therefore a ``good quantum number''
and can be used to label the steady states of $\DD_{\text{I}}$ in
each parity sector (similar to angular momentum variables $\mathfrak{l,m}$
labeling eigenstates of the Hydrogen atom). In other words, there
exists a basis for the steady states which consists of elements of
``fixed'' parity $\Pi\in\{0,1\}$. We can construct such a basis
by applying the above projections to the coherent state $|\a\ket$,
which is a steady state ($\F_{\text{I}}|\a\ket=0$) but which does
not have fixed parity. Projecting the coherent state $|\a\ket$ obtains
the single mode cat states \cite{Dodonov1974}:
\begin{equation}
|\a_{\Pi}\ket=\frac{\op_{\Pi}|\a\ket}{\sqrt{N_{\Pi}}}\,\,\,\,\,\,\,\,\,\text{where}\,\,\,\,\,\,\,\,\,N_{\Pi}=\bra\a|\op_{\Pi}|\a\ket\label{eq:cats}
\end{equation}
and $\Pi\in\{0,1\}$ labels the parity of the state. Taking limits
of small and large $\a$ yields\begin{subnumcases}{|\a_\Pi\ket \sim}
|\Pi\ket & $\a\rightarrow0 \label{eq:catlim2}$ \\
\displaystyle{\frac{|\a\ket+(-1)^\Pi\lket{-\a}}{\sqrt{2}}} & 
$\a\rightarrow\infty~.$ \label{eq:catlim3}
\end{subnumcases}For $\a\ll1$, the cat states approach Fock states $|\Pi\ket\in\{|0\ket,|1\ket\}$,
which are the steady states of $\fs$ for $\a=0$. For large $\a$,
they are simply superpositions of the aforementioned coherent states.
Notice that we can also project the coherent state $|i\a\ket$ onto
subspaces of fixed parity to yield the states $|i\a_{\Pi=0}\ket$
and $|i\a_{\Pi=1}\ket$, which are also annihilated by $\F_{\text{I}}$.

\subsection{Cat code states}

For $\a\rightarrow\infty$, it is clear that $\bra\a|i\a\ket=O(e^{-\a^{2}})$
so, in that limit, we can think of the two even parity states $|\a_{\Pi=0}\ket,|i\a_{\Pi=0}\ket$
as being a basis for a two-dimensional subspace (and same for the
odd-parity states $|\a_{\Pi=1}\ket,|i\a_{\Pi=1}\ket$). Therefore,
each pair of fixed-parity states forms a code subspace with projection
$P_{\text{I}}^{(\Pi)}\sim|\a_{\Pi}\ket\bra\a_{\Pi}|+|i\a_{\Pi}\ket\bra i\a_{\Pi}|$
(where we use the mathematician's definition of ``$\sim$'' \cite{vaughn_book}
and with the limit being $\a\rightarrow\infty$). We only need to
consider one of the code subspaces in order to store a qubit, but
we will see later that loss errors transport the quantum information
between these subspaces. Also, which subspace best protects from loss
errors is dependent on $\a$ \cite{Li2016,codecomp}, so we analyze
both in order to not lose generality.

In order to provide a basis for all values of $\a$ (instead of just
large $\a$), we can take $\pm$ linear superpositions of the respective
pair of fixed-parity states. This turns out to be equivalent to applying
the following projections
\begin{align}
\os_{2\m+\Pi} & =\frac{1}{4}\sum_{k=0}^{3}\exp[i\frac{\pi}{2}(\ph-2\m-\Pi)k]\label{eq:smproj}\\
 & =\sum_{n=0}^{\infty}|4n+2\m+\Pi\ket\bra4n+2\m+\Pi|\nonumber 
\end{align}
onto only $\op_{\Pi}|\a\ket$. In other words, the code states $\m\in\{0,1\}$
for each subspace $\Pi$ and for any $\a$ are\footnote{The presence of $\op_{\Pi}$ in the definition of $|\m_{\a,\Pi}\ket$
is redundant for this single-mode case, but makes a nice analogy with
the two-mode case, which does require two projections to define this
way.}
\begin{align}
|\m_{\a,\Pi}\ket & =\frac{\os_{2\m+\Pi}\op_{\Pi}|\a\ket}{\sqrt{N_{\m,\Pi}}}=\frac{|\a_{\Pi}\ket+\left(-1\right)^{\m}|i\a_{\Pi}\ket}{4\sqrt{N_{\m,\Pi}/N_{\Pi}}}\,,\label{eq:smcat}
\end{align}
where the normalization factor is
\begin{equation}
N_{\m,\Pi}=\bra\a|\os_{2\m+\Pi}\op_{\Pi}|\a\ket\,.\label{eq:norm unsimplified}
\end{equation}
For example, for odd parity $\Pi=1$, $|0_{\a,1}\ket$ lies in the
span of Fock states $|1\ket,|5\ket,|9\ket,\cdots$ while $|1_{\a,1}\ket$
lies in the span of $|3\ket,|7\ket,|11\ket,\cdots$. In the limit
of large $\a$, these become superpositions of the even- and odd-parity
cat states, respectively:\begin{subnumcases}{|\m_{\a,\Pi}\ket\sim}
|2\m+\Pi\ket & $\a\rightarrow 0 \label{eq:smlima}$ \\
\displaystyle{\frac{|\a_\Pi\ket+(-1)^\m|i\a_\Pi\ket}{\sqrt{2}}} & $\a\rightarrow\infty~.$ \label{eq:smlimb}
\end{subnumcases}In the small $\a$ limit, the code states become even- and odd-parity
Fock states, thereby preserving the parity for all $\a$. We have
thus constructed the basis of code states for each of the single-mode
cat codes $\Pi\in\{0,1\}$, whose projections can now be exactly expressed
as
\begin{equation}
\p_{\text{I}}^{(\Pi)}=|0_{\a,\Pi}\ket\bra0_{\a,\Pi}|+|1_{\a,\Pi}\ket\bra1_{\a,\Pi}|\,.
\end{equation}
We will see that a large-enough $\a$ suppresses certain errors, so
we consider the large $\a$ limit. In this limit, each cat state $|\a_{\Pi}\ket$
and $|i\a_{\Pi}\ket$ becomes an equal superposition of well-separated
coherent states, so the code states $|\m_{\a,\Pi}\ket$ become equal
superpositions of the four well-separated coherent states $\{|i^{k}\a\ket\}_{k=0}^{3}$.
We now proceed to project various errors onto the code spaces using
the above projections to determine which errors are protected by the
codes.

\subsection{Cat code error analysis\label{subsec:Cat-code-error}}

To set up the error-correction calculations, let us first calculate
the effect of $\aa$ on a code state $|\m_{\a,\Pi}\ket$ to show the
utility of the representation (\ref{eq:smcat}) in terms of a projected
coherent state. We know that $|\a\ket$ is an eigenstate of $\aa$,
so all that is left is to permute $\aa$ through the two projections.
A simple calculation using the representation (\ref{eq:smproj}) of
$\os_{2\m+\Pi}$ in terms of $e^{i\frac{\pi}{2}\ph}$ shows that\begin{subequations}
\begin{align}
\aa\op_{\Pi} & =\op_{\Pi+1}\aa\label{eq:smproj1}\\
\aa\os_{2\m+\Pi} & =\os_{2\m+\Pi-1}\aa=\os_{2(\m+\Pi+1)+\Pi+1}\aa\,,\label{eq:smproj2}
\end{align}
\end{subequations}where $\Pi+1$ is evaluated modulo 2 and $2\m+\Pi-1$
modulo 4. On the right-hand side of Eq.~(\ref{eq:smproj2}), the
parity $\Pi+1$ and code state index $\m+\Pi+1$ are both evaluated
modulo 2. This can be verified by explicitly plugging in $\m,\Pi\in\{0,1\}$,
showing that losing a single photon corresponds to binary subtraction
with carry. The reason for this manipulation is to separate out the
effect of the error on the parity $\Pi$ from that on the code index
$\m$. For example, if $\m=0$ and $\Pi=0$, then $\aa$ takes the
even-parity subspace to the odd parity subspace ($\Pi=0\rightarrow1$)
while at the same time performing a logical bit flip on the logical
qubit ($\m=0\rightarrow1$). If $\Pi=1$, then $\aa$ causes one to
go back to the even-parity subspace, but this time without the logical
bit flip. In summary, starting with the representation (\ref{eq:smcat})
of the code states, permuting $\aa$ through the projections using
Eqs.~(\ref{eq:smproj1}-b), recalling that $|\a\ket$ is an eigenstate
of $\aa$, and renormalizing yields
\begin{equation}
\aa|\m_{\a,\Pi}\ket=\a\sqrt{\frac{N_{\m+\Pi+1,\Pi+1}}{N_{\m,\Pi}}}|\m+\Pi+1_{\a,\Pi+1}\ket\,,\label{eq:smerrora}
\end{equation}
where both $\m+\Pi+1$ and $\Pi+1$ are evaluated modulo 2. We thus
see that, up to the extra parity-dependent bit flip, the effect of
$\aa$ is take the state from the even- to the odd-parity subspace
(and visa versa).

Now let us examine the square root factor above in the large $\a$
limit. The explicit formula for the normalizations (\ref{eq:norm unsimplified})
is easily calculated to be
\begin{equation}
N_{\m,\Pi}={\textstyle \half N_{\Pi}+\half\left(-1\right)^{\m}e^{-\a^{2}}\cos\left(\a^{2}-\frac{\pi}{2}\Pi\right)\,,}\label{eq:pimup}
\end{equation}
where $N_{\Pi}=\half[1+\left(-1\right)^{\Pi}e^{-2\a^{2}}]$ is the
normalization factor of the cat states from Eq.~(\ref{eq:cats}).
Recall that we do not want the quantum information stored in a superposition
of $|0_{\a,\Pi}\ket$ and $|1_{\a,\Pi}\ket$ to become distorted,
so we would prefer that the effect of the error $\aa$ is independent
of $\m$. Luckily, we find that the $\m$-dependent piece of $N_{\m,\Pi}$
is suppressed exponentially with $\a^{2}$. Similarly, the $\Pi$-dependent
part of $N_{\Pi}$ also disappears at the same rate, yielding (as
$\a\rightarrow\infty$) 
\begin{equation}
N_{\m,\Pi}={\textstyle \frac{1}{4}}+\left(-1\right)^{\m}O(e^{-\a^{2}})\,.
\end{equation}
Therefore, the square-root factor in Eq.~(\ref{eq:smerrora}) quickly
approaches 1 in the large $\a$ limit. We now apply what we have learned
to the error-correction conditions (\ref{eq:qec}).

\subsubsection{Dephasing errors\label{subsec:Dephasing-errors}}

Let us first analyze those errors from Eqs.~(\ref{eq:loss}-\ref{eq:deph})
which only cause dephasing. Recall from the text below those equations
that the Kraus operators for such errors can be expressed as a sum
of products of elements from $\{\aa^{\dg k}\aa^{k}\}_{k=0}^{\infty}$.
Therefore, we need only project $\aa^{\dg k}\aa^{k}$ onto the code
subspaces to see whether the error-correction conditions (\ref{eq:qec})
hold. We are interested in the effect of small error rates $\k_{n}\ll\k_{\text{I}}$,
so only the first few values of $k$ in the expansion of the Kraus
operators are necessary.$^{\ref{fn:Interested-readers-are}}$ Calculating
$\p_{\text{I}}^{(\Pi)}\aa^{\dg k}\aa^{k}\p_{\text{I}}^{(\Pi)}$ requires
determining the matrix elements $\bra\m_{\a,\Pi}|\aa^{\dg k}\aa^{k}|\n_{\a,\Pi}\ket$
for $\m,\n\in\{0,1\}$, which we can depict using a diagonal $2\times2$
matrix. Generalizing the calculation above and taking the $\a\rightarrow\infty$
limit yields\begin{subequations}
\begin{align}
\p_{\text{I}}^{(\Pi)}\aa^{\dg k}\aa^{k}\p_{\text{I}}^{(\Pi)} & =\a^{2k}\p_{\text{I}}^{(\Pi)}+O(\a^{2k}e^{-\a^{2}})Z_{\text{I}}^{(\Pi)}\,,\label{eq:deph-1}
\end{align}
\end{subequations}where $Z_{\text{I}}^{(\Pi)}$ is the logical $Z$-operator
(\ref{eq:logz}) for the code $\p_{\text{I}}^{(\Pi)}$. The $Z$-operator
comes from the $\m$-dependence of the ratios of normalizations. As
we saw, this $\m$-dependence is suppressed exponentially with $\a^{2}$,
so the cat codes can \textit{approximately} correct all dephasing
errors $k$ such that $2k\ll\a^{2}$ since such errors satisfy Eq.~(\ref{eq:qec})
up to exponential corrections. 

We note that the exponential suppression is not quite the whole story
and that cat codes can gain extra protection if $\a$ is fine tuned
to certain values. Namely, given a power $k$, the \textit{exact}
coefficient in front of $Z_{\text{I}}^{(\Pi)}$ contains an order
$O(1)$ trigonometric function of $\a^{2}$ which can be exactly zero
at certain values of the argument \cite{Li2016,codecomp}. In an experimental
setting, where low values of $\a$ are achievable more easily, such
values can make a significant difference in helping suppress dephasing
errors.

We have just shown that cat codes, \textit{in principle}, protect
well against dephasing errors in the large $\a$ limit. But how does
the autonomous error correction of scheme I against such errors work
\textit{in practice}? It turns out that the dissipator $\k_{\text{I}}\DD_{\text{I}}$
does the job of protecting against dephasing errors. To show this,
assume that we have turned on $\k_{\text{I}}\DD_{\text{I}}$, stabilized
the initial state into one of the subspaces of fixed parity $\Pi$,
and then turned on a small perturbation in the form of a dephasing
error $\k_{n}\DD[\ph]$ (for $\k_{n}\ll\k_{\text{I}}$). We can then
calculate the effect of dephasing perturbation within the code space
by projecting the dephasing superoperator onto the code. Since $\ph$
respects the parity symmetry of $\DD_{\text{I}}$, the perturbation
only causes decay of the $X,Y$-components of the cat-qubit Bloch
sphere and leaves the (diagonal) $Z$-component unaffected. However,
the calculation reveals that this leading-order $X,Y$ decay rate
is suppressed exponentially (with a power of $\a$) for large $\a$
\{\cite{cats}, Fig.~A1(b)\}. Moreover, increasing the rate $\k_{n}$
such that it is no longer a perturbative process still reveals an
exponential suppression of the effect within the code space, as long
as $\a$ is sufficiently large (\cite{thesis}, Ch.~8). Intuitively,
the dephasing process $\DD[\ph]$ merely diffuses the four coherent
states making up the cat code around the perimeter of the circle of
radius $\a$ in phase space. Since the coherent states are well-separated,
one has to perform significant diffusion in order to make them overlap
with each other. Moreover, for any given dephasing parameter $\k_{n}t$,
there exists a sufficiently large $\a$ such that the diffusion is
insufficient to make the coherent states overlap. Analytical perturbative
calculations for the open system with jump $\F=\aa^{2}-\a^{2}$ \cite{cats}
and closed system with Hamiltonian $F^{\dg}F$ \cite{Puri2017} corroborate
this reasoning, providing strong evidence that dephasing processes
are not a concern in the large $\a$ limit for any cat-code.

\subsubsection{Loss errors\label{subsec:Loss-errors}}

Let us return to the effect of $\aa$ on the code states from Eq.~(\ref{eq:smerrora}).
Recall that $\aa$ changes the parity of the states, mapping the subspace
of fixed $\Pi$ onto $\Pi+1$ modulo 2. Therefore, projecting back
onto the $\Pi$ subspace produces
\begin{equation}
\p_{\text{I}}^{(\Pi)}\aa\p_{\text{I}}^{(\Pi)}=0\,.
\end{equation}
Combined with the above protection from dephasing, the cat codes can
protect from a single loss error $E_{\aa}^{\ell=0}$ (\ref{eq:loss}).
However, the application of two loss errors is uncorrectable due to
the extra bit flip ($\m\rightarrow\m+1$ modulo 2) described in the
beginning of this Subsection. Performing the calculation and taking
the large $\a$ limit yields
\begin{equation}
\p_{\text{I}}^{(\Pi)}\aa^{2}\p_{\text{I}}^{(\Pi)}=\a^{2}X_{\text{I}}^{(\Pi)}+O(\a^{2}e^{-\a^{2}})Y_{\text{I}}^{(\Pi)}\,,\label{eq:smtp}
\end{equation}
where $X_{\text{I}}$ (\ref{eq:logx}) and $Y_{\text{I}}$ (\ref{eq:logy})
are logical operators. Therefore, $\aa^{2}$ acts nontrivially on
the code space and is thus the first uncorrectable error of the code.

In the limit $\k_{a}t\rightarrow0$ in Eq.~(\ref{eq:loss}), continuous
(i.e., Lindbladian-based\footnote{For example, a three-qubit repetition code with projection $P$ admits
a Lindbladian with error-correcting jumps $F_{i}=PX_{i}$, where $X_{i}$
is a bit-flip on qubit $i\in\{1,2,3\}$ \cite{Ippoliti2014}. Proposals
exist to implement such a code for discrete-variable (i.e., multi-qubit)
systems in, e.g., trapped ion \cite{Reiter2017} and superconducting
qubit \cite{Cohen2014} setups.}) protection from loss can in principle be done by initializing the
system in $\p_{\text{I}}^{(\Pi=0)}$ and implementing the jump operator
\begin{equation}
\F_{\text{I}}^{\text{loss}}=|0_{\a,0}\ket\bra1_{\a,1}|+|1_{\a,0}\ket\bra0_{\a,1}|\label{eq:loss1}
\end{equation}
alongside $\F_{\text{I}}$ (\ref{eq:jumps}). This jump operator acts
only on the $\Pi=1$ parity subspace and maps the state back to the
$\Pi=0$ subspace while reversing the bit flip caused by a loss event
$\aa$. In an alternative scenario, the recovery channel in Ref.~\cite{Li2016}
can be implemented after the state has evolved under photon loss for
finite $\k_{a}t$. Such a channel can be implemented continuously
via the procedure in Sec.~III.D of Ref.~\cite{ABFJ}. However, the
jumps are more difficult to implement in both continuous QEC scenarios,
so current cat-code error-correction procedures rely on discrete QEC
by measuring and tracking the photon number parity $\parr$ (\ref{eq:parity}).
The extension of the superinductance-based proposal \cite{Cohen2016}
to continuous QEC {[}\citealp{cohenthesis}, Sec.~4.2.2{]} does however
realize the related jump $\aa^{\dg}\op_{\Pi=1}$. Note that tracking
$\parr$ allows one to avoid having to move a $\Pi=1$ state back
to the $\Pi=0$ codespace, akin utilizing Pauli frames in the conventional
stabilizer formalism \cite{Knill2005}. The same holds for scheme
II.

\subsection{Cat code gates}

We now provide an overview of some of the ways to perform gates on
the code spaces for scheme I. For Subsecs. \ref{subsec:Hamitonian--and}
and \ref{subsec:Hamiltonian--gate}, we utilize Zeno dynamics caused
by the perturbation within the codespace stabilized by $\ko\lo$:
recall from Subsec.~\ref{sec:Introduction-1} that the first-order
(in $\e\ll\k_{\text{I}}$) effect of a Hamiltonian $\e H$ within
the code spaces $\p_{\text{I}}^{(\Pi)}$ is simply (\ref{eq:zeno})
\begin{equation}
H_{\text{I}}^{(\Pi)}=\p_{\text{I}}^{(\Pi)}H\p_{\text{I}}^{(\Pi)}\,.
\end{equation}
For Subsecs. \ref{subsec:Self-Kerr---rotation}-\ref{subsec:Cross-Kerr-control-phase-gate},
we turn off $\ko\lo$ and evolve directly. Since such evolution does
not cause leakage outside of the codespace for the times $t$ that
we consider, the above formula remains valid.

\subsubsection{Hamiltonian $X$ and $XX$ gates\label{subsec:Hamitonian--and}}

Here we review how to perform $X$ and $XX$ rotations of arbitrary
angle on the cat codes \cite{cats}. In Eq.~(\ref{eq:smtp}), we
find that $\aa^{2}$ is an uncorrectable error on our code since it
acts nontrivially within the code. However, we can turn ``trash into
treasure'' by utilizing this feature to perform a gate on the code.
According to Eq.~(\ref{eq:smtp}), applying a squeezing Hamiltonian
$\H_{\text{I}}^{X}=g_{X}(\aa^{2}+\mathrm{h.c.})$ yields exactly the
generator of $X_{\text{I}}^{(\Pi)}$-rotations when projected onto
the code space.

We can straightforwardly scale up this idea into a two-qubit $XX$-gate.
Let $\p_{\text{I},1}^{(\Pi_{1})}$ and $\p_{\text{I},2}^{(\Pi_{2})}$
be projections on codes of fixed parities $\Pi_{1},\Pi_{2}$ and code
parameters $\a_{1},\a_{2}$ in modes 1 and 2, respectively. Let the
Hamiltonian now be $\H_{\text{I}}^{XX}=g_{XX}(\aa_{1}^{2}\aa_{2}^{2}+\mathrm{h.c.})$.
We can perform the same projection calculation, noting that $\Pi_{1}$
does not have to be equal to $\Pi_{2}$ and $\a_{1}$ does not have
to be identical to $\a_{2}$ as long as both are sufficiently large
to protect from dephasing noise:
\begin{align}
\p_{\text{I},1}^{(\Pi_{1})}\p_{\text{I},2}^{(\Pi_{2})}\H_{\text{I}}^{XX}\p_{\text{I},1}^{(\Pi_{1})}\p_{\text{I},2}^{(\Pi_{2})} & \sim2g_{XX}\a_{1}^{2}\a_{2}^{2}X_{\text{I},1}^{(\Pi_{1})}X_{\text{I},2}^{(\Pi_{2})}\,,
\end{align}
with corrections exponentially suppressed in $\a_{1,2}^{2}$.

\subsubsection{Hamiltonian $Z$-gate\label{subsec:Hamiltonian--gate}}

Usually in superconducting circuits, expansion of the Josephson junction
Hamiltonian 
\begin{equation}
\H_{\text{I}}^{\text{jnct}}=E_{J}\cos\left(\b\aa e^{i\o t}+\mathrm{h.c.}\right)
\end{equation}
and the rotating-wave approximation are used to produce the anharmonic
terms of a desired Hamiltonian. Above, $\o$ is the drive frequency
of the mode, $E_{J}$ is the Josephson energy, and $\b$ is the drive's
amplitude and phase. However, Ref.~\cite{Cohen2016} proposed a way
of using the entire Hamiltonian (i.e., without expansion but still
in the RWA) to generate a $Z$-rotation. Recall that the above cosine
can be thought of as a sum of two displacement operators, $\H_{\text{I}}^{\text{jnct}}=\half E_{J}(D_{\b\exp(i\o t)}+D_{\b\exp(i\o t)}^{\dg})$,
where $D_{\a}|0\ket=|\a\ket$. If we now write the displacement operators
as matrices in Fock space, we will see that, for $\o\neq0$, the only
time-independent terms will be those which are diagonal in Fock space.
This means that the diagonal terms will be the dominant contributions
in the RWA and we can ignore the rest, yielding $\H_{\text{I}}^{\text{jnct}}\approx E_{J}\overline{D_{\b}}$,
where
\begin{equation}
\overline{D_{\b}}=e^{-\half\left|\b\right|^{2}}\sum_{n=0}^{\infty}L_{n}\left(|\b|^{2}\right)|n\ket\bra n|\label{eq:dispRWA}
\end{equation}
is the displacement operator after the RWA and $L_{n}$ is the Laguerre
polynomial. Projecting $\overline{D_{\b}}$ on the code is simpler
if we instead use the Fock state representation of the states,
\begin{equation}
|\m_{\a,\Pi}\ket=\frac{e^{-\half\a^{2}}}{\sqrt{N_{\m,\Pi}}}\sum_{n=0}^{\infty}\frac{\a^{4n+2\m+\Pi}}{\sqrt{(4n+2\m+\Pi)!}}|4n+2\m+\Pi\ket\,.\label{eq:catfock}
\end{equation}
Since $\overline{D_{\b}}$ is diagonal and the above code states are
superpositions of two different sets of Fock states, projecting $\overline{D_{\b}}$
onto the codespace can only yield terms which are diagonal w.r.t.
the code basis,
\begin{equation}
\p_{\text{I}}^{(\Pi)}\overline{D_{\b}}\p_{\text{I}}^{(\Pi)}=C_{+}^{(\Pi)}\p_{\text{I}}^{(\Pi)}+C_{-}^{(\Pi)}Z_{\text{I}}^{(\Pi)}\,,
\end{equation}
where $C_{\pm}^{(\Pi)}=\bra0_{\a,\Pi}|\overline{D_{\b}}|0_{\a,\Pi}\ket\pm\bra1_{\a,\Pi}|\overline{D_{\b}}|1_{\a,\Pi}\ket$.
Since generically $C_{-}^{(0)}\neq C_{-}^{(1)}$, this gate is parity-dependent,
meaning that any loss events occurring during the gate will change
the gate's effect.\footnote{More precisely \cite{Cohen2016}, $C_{-}^{(0)}\approx C_{-}^{(1)}$
at a region around $\b=2\a\gtrsim8$, but at that value of $\a$ there
are about $\a^{2}\approx16$ photons in the cavity. This means that
error correction has to be performed extremely quickly because there
is a large probability of losing two of more photons \cite{codecomp}.} However, one can introduce additional junctions with respective Hamiltonians
of the same form as $\H_{\text{I}}^{\text{jnct}}$, but with independent
tunable parameters. Clever calibration then allows one to make sure
that the projection on the codespace generates a parity-independent
$Z$-gate.

\subsubsection{Holonomic $Z$-gate\label{subsec:Holonomic--gate}}

Here we review an additional gate \cite{Albert2015} which allows
for the active parity measurements to occur, thereby protecting from
loss errors. However, while the dissipation $\DD_{\text{I}}$ remains
on throughout this gate, this gate utilizes the small $\a$ limit
of the code spaces and thus does not allow protection from dephasing.

This gate involves an adiabatic variation of the code parameter $\a$
in the following sequence: $\a\rightarrow0\rightarrow\a e^{i\phi}\rightarrow\a$
(for $\a\gg1$ and some angle $\phi$). In the superoperator adiabatic
limit (see Ref.~\cite{ABFJ} and refs. therein), the effective holonomy
due to variation of $\k_{\text{I}}\DD_{\text{I}}$ is determined by
the non-Abelian \cite{Wilczek1984} Berry connection $\vec{A}_{\m\n}=\bra\m_{\a,\Pi}|\vec{\nabla}|\n_{\a,\Pi}\ket$
(akin to $\p_{\text{I}}^{(\Pi)}H\p_{\text{I}}^{(\Pi)}$ for Hamiltonian
perturbations $H$), where $\vec{\nabla}=(\partial_{\left|\a\right|},\partial_{\arg\a})$.
However, instead of calculating the Berry connections (done in the
supplement of Ref.~\cite{Albert2015}), here we offer a heuristic
account of the effective operation. The only nontrivial part of the
gate occurs during the step $0\rightarrow\a e^{i\phi}$ of the sequence.
In this step, the new steady states of $\DD_{\text{I}}$ are $|\m_{\a\exp(i\phi),\Pi}\ket$,
whose $\a=0$ limit is $\exp[i(2\m+\Pi)\phi]|2\m+\Pi\ket$. However,
the initial states for this step consist of just the Fock states $|2\m+\Pi\ket$
without the extra phase. Thus, to compensate for including the extra
phase during the step $0\rightarrow\a e^{i\phi}$, one will have $|2\m+\Pi\ket\rightarrow\exp[-i(2\m+\Pi)\phi]|\m_{\a\exp(i\phi),\Pi}\ket$.
The entire sequence thus performs an effective $Z$-rotation 
\begin{equation}
\p_{\text{I}}^{(\Pi)}U_{\text{I}}^{\text{hol}}\p_{\text{I}}^{(\Pi)}=e^{-i\phi\Pi}\begin{pmatrix}1 & 0\\
0 & e^{-2i\phi}
\end{pmatrix}\,.
\end{equation}
The $\Pi$-dependent phase is an overall phase since the qubit is
entirely in a code space of fixed parity, so the effect of the gate
is independent of $\Pi$.

\subsubsection{Self-Kerr $\nicefrac{\pi}{2}$ $Z$-rotation\label{subsec:Self-Kerr---rotation}}

Another gate from Ref.~\cite{cats} utilizes a strong self-Kerr nonlinearity
$H_{K}=K\left(\ph-\Pi\right)^{2}$ (with $K\in\mathbb{R}$) to perform
a $Z$-rotation for an exact angle of $\nicefrac{\pi}{2}$. Note that
this gate is parity ($\Pi$) \textit{dependent}, meaning that either
(1) it has to be performed quickly enough ($K\gg\k_{\aa}$) so that
loss errors do not occur or (2) it has to be followed by a rotation
$e^{i\t\ph}$ where $\t$ is chosen to compensate any rotations induced
by the nonlinearity {[}\citealp{cats}, Sec.~3.4{]}; the second option
requires continuous monitoring of the parity). Moreover, since equation
(\ref{eq:deph-1}) tells us that $\p_{\text{I}}^{(\Pi)}H_{K}\p_{\text{I}}^{(\Pi)}$
acts trivially on the codespace at large $\a$, the leading-order
effect of perturbing the dissipator $\lo$ with $H_{K}$ is not sufficient
to implement a gate. Therefore, this gate can only be performed if
we turn off $\DD_{\text{I}}$ and freely evolve under $H_{K}$ to
a time $t=\nicefrac{\pi}{8K}$, yielding 
\begin{equation}
U_{\text{I}}^{Z}=\exp(i{\textstyle \frac{\pi}{8K}}H_{K})=\exp[i{\textstyle \frac{\pi}{8}}(\ph-\Pi)^{2}]\,.
\end{equation}
Applying $U_{\text{I}}^{Z}$ onto each Fock state in the representation
(\ref{eq:catfock}) allows us to substitute $4n+2\m+\Pi$ for $\ph$.
Performing some algebra then yields the phase $\exp(i\frac{\pi}{8}\m^{2})$
for each Fock state. Projecting onto the code space, this $\m$-dependent
phase translates to a $\nicefrac{\pi}{2}$ rotation around the $Z$
axis:
\begin{equation}
\p_{\text{I}}^{(\Pi)}U_{\text{I}}^{Z}\p_{\text{I}}^{(\Pi)}=\begin{pmatrix}1 & 0\\
0 & i
\end{pmatrix}\,.
\end{equation}

Since $\lo$ is off, one may worry about errors caused by the dephasing
process (\ref{eq:deph}) during the gate. However, there is no need
to be concerned because, at sufficiently large $\a$, the dephasing
process will not have enough time to induce tunneling between the
well-separated coherent states $\{|i^{k}\a\ket\}_{k=0}^{3}$ making
up the code. Recall that dephasing induces diffusion of the phase
of each coherent state, and this diffusion would need to occur for
a time $\sim\nicefrac{\a\pi}{4}$ in order to cause overlap between
neighboring coherent states. Therefore, such errors are still suppressed
once $\DD_{\text{I}}$ is used to stabilize back to the codespace
after the gate.

\subsubsection{Cross-Kerr control-phase gate\label{subsec:Cross-Kerr-control-phase-gate}}

Along similar lines as the above self-Kerr rotation, Ref.~\cite{zhang2017}
has proposed a two-qubit control-phase gate for a simpler version
of the cat code. (Recall that such a gate should produce $|1_{\a_{1},\Pi_{1}},1_{\a_{2},\Pi_{2}}\ket\rightarrow-|1_{\a_{1},\Pi_{1}},1_{\a_{2},\Pi_{2}}\ket$
while leaving the remaining two-qubit components unchanged.) Here,
we extend this gate to the cat code described here. We turn off dissipation
and evolve under the unitary $U_{\text{I}}^{CZ}=\exp[i\frac{\pi}{4}(\ph_{1}-\Pi_{1})(\ph_{2}-\Pi_{2})]$,
which is generated by a cross-Kerr nonlinearity. Just like the self-Kerr
rotation, this gate is also parity dependent, so we assume that the
parities of the two cat-qubits $\p_{\text{I},1}^{(\Pi_{1})}$ and
$\p_{\text{I},2}^{(\Pi_{2})}$ are $\Pi_{1}$ and $\Pi_{2}$, respectively.
Projecting this unitary onto the two-qubit codespace yields
\begin{equation}
\p_{\text{I},1}^{(\Pi_{1})}\p_{\text{I},2}^{(\Pi_{2})}U_{\text{I}}^{CZ}\p_{\text{I},1}^{(\Pi_{1})}\p_{\text{I},2}^{(\Pi_{2})}=\begin{pmatrix}1 & 0 & 0 & 0\\
0 & 1 & 0 & 0\\
0 & 0 & 1 & 0\\
0 & 0 & 0 & -1
\end{pmatrix}\,.
\end{equation}
This is proven by noting that a two-qubit state $|\m_{\a_{1},\Pi_{1}},\n_{\a_{2},\Pi_{2}}\ket$
is expressed using Fock states $|4n_{1}+2\m+\Pi_{1},4n_{2}+2\n+\Pi_{2}\ket$
(with $\n\in\{0,1\}$ and $n_{1},n_{2}\geq0$), substituting the Fock
state numbers into $\ph_{1}$ and $\ph_{2}$ in $U_{\text{I}}^{CZ}$,
and noting that $U_{\text{I}}^{CZ}$ reduces to $(-1)^{\m\n}$. As
with the self-Kerr gate, the reasoning regarding protection from dephasing
also holds here.

\subsubsection{Control engineering}

Another way to engineer gates for the single-mode cat code is to utilize
a time-dependent drive $\e_{C}(t)\aa+\mathrm{h.c.}$ on the oscillator
and $\e_{T}(t)\s_{+}+\mathrm{h.c.}$ on an ancilla transmon qubit
in combination with the dispersive nonlinearity $\ph\s_{z}$ coupling
the two \cite{Krastanov2015,Heeres2015,Heeres2016}. This is sufficient
for universal control, and an optimization routine can be used to
determine which values of the drives to pick at each increment of
time. This particular scheme was realized experimentally in Ref.~\cite{Heeres2016}.
It is likely that such control needs to be performed with $\lo$ turned
off. More generally, arbitrary quantum processes can be achieved using
only an ancilla qubit with non-demolition readout and adaptive control
\cite{Shen2016} (see also \cite{Lloyd2001,Andersson2008,Iten2016}).

\section{Two-mode pair-cat code\label{sec:Two-mode-pair-cat-code}}

In the previous Section, we have reviewed the single-mode cat code
\cite{Leghtas2013b} and its associated reservoir engineering scheme
I \cite{cats}. In this Section, we introduce the two-mode pair-cat
code and its associated scheme II in completely analogous fashion.
The respective code states, gates, and protected errors of both schemes
are listed side-by-side in Table \ref{tab:1}.

\subsection{Primer on pair-coherent states\label{subsec:Primer-on-pair-coherent}}

We now perform a symmetry analysis of the jump operator $\F_{\text{II}}=\aa^{2}\bb^{2}-\g^{4}$
for scheme II in order to determine the components which will be used
to construct this scheme's code states. A more gentle exposition is
presented in Ch.~8 of Ref.~\cite{thesis}.

Observe that $\F_{\text{II}}|\a,\frac{\g^{2}}{\a}\ket=0$ for any
two-mode coherent state $|\a,\frac{\g^{2}}{\a}\ket$ and $\a\neq0$.
Such coherent states and their counterparts $|\frac{\g^{2}}{\b},\b\ket$
for $\b\neq0$ can be used to determine a continuous basis for the
subspace annihilated by $\ft$. However, such a basis is not terribly
illuminating. Instead, one can construct a basis which has one discrete
and continuous index, just like the basis of states $|\a_{\Pi}\ket$
(with discrete parity index $\Pi\in\{0,1\}$ and continuous index
$\a$) for the subspaces of fixed parity $\left(-1\right)^{\ph}$.
Instead of the single-mode parity, the ``good quantum number'' used
to define the discrete index is the photon number difference $\D$,
determined by the operator $\di=\phm-\ph$ (\ref{eq:diff}). This
operator commutes with $\aa\bb$ and therefore commutes with $\F_{\text{II}}$.
Thus, the space of states annihilated by the jump can be spanned by
a basis of states with fixed eigenvalues $\D\in\mathbb{Z}$. To determine
such states, first let us define projections onto sectors of fixed
$\D$,\begin{subequations}
\begin{align}
\tp_{\D} & =\intop_{0}^{2\pi}\frac{d\t}{2\pi}\exp[i(\di-\D)\t]\label{eq:deltaprojections}\\
 & =\begin{cases}
{\displaystyle \sum_{n=0}^{\infty}}|n,n+\D\ket\bra n,n+\D| & \D\geq0\\
\text{SWAP}\,\tp_{\left|\D\right|}\,\text{SWAP} & \D<0
\end{cases}\,,
\end{align}
\end{subequations}where the SWAP operator ($\text{SWAP}|n,m\ket=|m,n\ket$)
is 
\begin{equation}
\text{SWAP}=\exp\left[i\frac{\pi}{2}(\aa^{\dg}-\bb^{\dg})\left(\aa-\bb\right)\right]\,.\label{eq:exch}
\end{equation}
\foreignlanguage{american}{From now on, we assume that $\D\geq0$,
remembering that an application of $\text{SWAP}$ yields the corresponding
results for $\D<0$.} Notice that the two-mode coherent state $|\g,\g\ket$
is annihilated by $\F_{\text{II}}$. We now apply the above projections
to this state with the goal of determining our basis for the code
space. Projection yields the pair-coherent/Barut-Girardello \foreignlanguage{american}{\cite{Barut1971,Agarwal1986,Agarwal1988}}
state (defined here for complex $\g$)\footnote{\label{fn:In-the-general}In contrast to Ref.~\cite{Agarwal1988},
we include the extra phase $\exp(i\frac{\D}{2}\arg\g^{2})$ in order
to express pair-coherent states as projected coherent states. We also
set the eigenvalue of $\aa\bb$ to $\g^{2}$ instead of $\g$ because
that leads to more visual similarity of $|\g_{\D}\ket$ to $|\a_{\Pi}\ket$
in Sec.~\ref{sec:Visualization-techniques}.}\begin{subequations}
\begin{align}
|\g_{\D}\ket & =\frac{\tp_{\D}|\g,\g\ket}{\sqrt{N_{\D}}}\\
 & =\frac{1}{\sqrt{I_{\D}(2|\g|^{2})}}\sum_{n=0}^{\infty}\frac{\g^{2n+\D}}{\sqrt{n!\left(n+\D\right)!}}|n,n+\D\ket\,,\label{eq:paircoh}
\end{align}
\end{subequations}with $I_{\D}$ being a modified Bessel function
of the first kind and normalization
\begin{equation}
N_{\D}={\textstyle \bra\g,\g|\tp_{\D}|\g,\g\ket}=e^{-2|\g|^{2}}I_{\D}(2|\g|^{2})\,.\label{eq:pidelta}
\end{equation}
Since $\aa\bb$ commutes with $\tp_{\D}$, it is simple to show that
\begin{equation}
\aa\bb|\g_{\D}\ket=\g^{2}|\g_{\D}\ket\,.\label{eq:eigen}
\end{equation}

Pair-coherent states resolve the identity for a given $\D$:
\begin{equation}
\tp_{\D}=\int d^{2}\g\s(\g)|\g_{\D}\ket\bra\g_{\D}|\,,\label{eq:overcomp}
\end{equation}
where the measure is 
\begin{equation}
\s(\g)=\frac{4}{\pi}\left|\g\right|^{2}I_{\D}(2\left|\g\right|^{2})K_{\D}(2\left|\g\right|^{2})\label{eq:measure}
\end{equation}
and $K_{\D}$ is the modified Bessel function of the second kind.
The $|\g_{\D}\ket$ are an overcomplete basis for the blocks in the
block-diagonal form of $\aa\bb=\sum_{\D\in\mathbb{Z}}\tp_{\D}\aa\bb\tp_{\D}$,
and similarly for $\aa^{\dg}\bb^{\dg}$ and $\ph+\phm$. From the
point of view of group theory, $\{\aa\bb,\aa^{\dg}\bb^{\dg},\ph+\phm\}$
form a reducible two-mode representation of the Lie algebra $\mathfrak{su}(1,1)$,
and $\D$ labels all of the irreducible two-mode representations.
Similarly, cat states $|\a_{\Pi}\ket$ with $\Pi\in\{0,1\}$ span
the two ($\Pi\in\{0,1\}$) irreducible representation spaces for $\{\aa^{2},\aa^{\dg2},\ph\}$,
a single-mode reducible $\mathfrak{su}(1,1)$ representation. As a
result of this group-theoretical connection, $|\g_{\D}\ket$ share
several features with $|\a_{\Pi}\ket$ (summarized in Table \ref{tab:2}):
both are eigenstates of lowering operators ($\aa^{2}$ and $\aa\bb$,
respectively), behave similarly under rotations, and have exponentially
suppressed overlap. We will see in Sec.~\ref{sec:Visualization-techniques}
that $|\g_{\D}\ket$ are also visually similar to $|\a_{\Pi}\ket$
if the former's $Q$-function is plotted vs. $\{\g^{2},\g^{2\star}\}$.
For the remainder of this Section, we go back to assuming $\g$ is
real and consider only the states $\lket{\g_{\D}}$ and $\lket{i\g_{\D}}$. 

The pair-coherent states are not to be confused with two-mode squeezed
states (also called Perelomov \cite{perelomov_book} coherent states)
\begin{align}
|\xi^{(\D)}\ket & =\exp[\xi(\aa^{\dg}\bb^{\dg}-\aa\bb)]|0,\D\ket\label{eq:squeezed}\\
 & \propto\sum_{n=0}^{\infty}\sqrt{{n+\D \choose n}}\tanh^{m}\xi|n,n+\D\ket\,,\nonumber 
\end{align}
which is another extension of ordinary coherent states to two-mode
systems. We define them for real $\xi$ for simplicity, and extension
to complex values can be done by applying a two-mode rotation. These
states are not eigenvectors of $\aa\bb$, but (as seen above) are
generated by the exponential of $\aa\bb$ and its conjugate. Of course,
ordinary coherent states $|\a\ket$ are both eigenstates of $\aa$
and satisfy $|\a\ket=D_{\a}|0\ket$ for a displacement $D_{\a}$.

\begin{table}
\begin{tabular}{cc}
\toprule 
Cat states & Pair-coherent states\tabularnewline
\midrule
$\aa^{2}|\a_{\Pi}\ket=\a^{2}|\a_{\Pi}\ket$ & $\aa\bb|\g_{\D}\ket=\g^{2}|\g_{\D}\ket$\tabularnewline
$e^{i\t\ph}|\a_{\Pi}\ket=|(\a e^{i\t})_{\Pi}\ket$ & $e^{i\t(\ph+\phm)}|\g_{\D}\ket=|(\g e^{i\t})_{\D}\ket$\tabularnewline
$|\bra\a_{\Pi}|\b_{\Pi^{\prime}}\ket|^{2}\sim\d_{\Pi\Pi^{\prime}}e^{-|\a-\b|^{2}}$~~ & ~~$|\bra\g_{\D}|\d_{\D^{\prime}}\ket|^{2}\sim\d_{\D\D^{\prime}}e^{-2|\g-\d|^{2}}$\tabularnewline
\bottomrule
\end{tabular}\caption{\label{tab:2}Similarities between cat (\ref{eq:cats}) and pair-coherent
(\ref{eq:paircoh}) states. The ``$\sim$'' means asymptotically
equal in the limit $|\protect\a|,|\protect\b|,|\protect\g|,|\protect\d|\rightarrow\infty$
and, for the left column, the additional limit $|\protect\a-\protect\b|\ll|\protect\a+\protect\b|$.}
\end{table}

\subsection{Pair-cat code states}

We see from Table \ref{tab:2} that, for each sector of fixed $\D$
and for large $\g$, there exists a two-dimensional subspace spanned
by $\lket{\g_{\D}}$ and $\lket{i\g_{\D}}$. The projections on these
subspaces, $\p_{\text{II}}^{(\D)}\sim|\g_{\D}\ket\bra\g_{\D}|+\lket{i\g_{\D}}\bra i\g_{\D}|$,
are thus our code spaces (for $\g\rightarrow\infty$). However, unlike
the parity $\parr$, which only had two distinct eigenvalues, now
the number of values of $\D$ (and thus the numbers of code spaces)
is infinite! We proceed to determine a basis for the code spaces which
is valid for all values of $\g$. 

Recall that the subspace we are in consists of Fock states $\{|n,n+\D\ket\}_{n=0}^{\infty}$.
As with the single mode space, we can develop a notion of parity for
these states by dividing them into those with even and odd $n$, i.e.,
two sets of states $\{|2n+\m,2n+\m+\D\ket\}_{n=0}^{\infty}$ for $\m\in\{0,1\}$.
The parity index $\m$ is then exactly the logical index for the pair-cat
states and corresponds to the parity of the states of the first mode
for $\D\geq0$. (Recall that $\D<0$ is handled by the SWAP operator
(\ref{eq:exch}), so $\m$ becomes the parity of the states of the
second mode in that case.) The extra projection we need to apply onto
$\tp_{\D}|\g,\g\ket$ to project onto these two sets is\begin{subequations}
\begin{align}
\ts_{2\m+\D} & =\frac{1}{4}\sum_{k=0}^{3}\exp[i\frac{\pi}{2}(\ph+\phm-2\m-\D)k]\\
 & =\sum_{n,m=0}^{\infty}\d_{n+m,2\m+\D}^{\text{mod\,}4}|n,m\ket\bra n,m|\,.
\end{align}
\end{subequations}Above, $\d_{n_{1},n_{2}}^{\text{mod\,}4}=1$ if
and only if $n_{1}=n_{2}$ modulo 4. Letting $\m\in\{0,1\}$, the
code states are
\begin{equation}
|\m_{\g,\D}\ket=\frac{\ts_{2\m+\D}\tp_{\D}|\g,\g\ket}{\sqrt{N_{\m,\D}}}=\frac{|\g_{\D}\ket+(-1)^{\m}\left(-i\right)^{\D}\lket{i\g_{\D}}}{2\sqrt{N_{\m,\D}/N_{\D}}}\label{eq:paircatdef}
\end{equation}
with normalization\begin{subequations}
\begin{align}
N_{\m,\D} & ={\textstyle \bra\g,\g|\ts_{2\m+\D}\tp_{\D}|\g,\g\ket}\label{eq:renorm}\\
 & =e^{-2\g^{2}}\frac{I_{\D}(2\g^{2})+\left(-1\right)^{\m}J_{\D}(2\g^{2})}{2}\label{eq:renorm2}
\end{align}
\end{subequations}and with $J_{\D}$ being the Bessel function of
the first kind, the logical state index $\m$ defined modulo 2, and
integer subspace index $\D$. When applied to $\tp_{\D}|\g,\g\ket$,
$\ts_{2\m+\D}$ is designed to map the index $n$ in the sum (\ref{eq:paircoh})
to $2n+\m$. The Fock space representation of the code states is thus\begin{widetext}
\begin{equation}
|\m_{\g,\D}\ket=\frac{\sqrt{2}}{\sqrt{I_{\D}(2\g^{2})+\left(-1\right)^{\m}J_{\D}(2\g^{2})}}\sum_{n=0}^{\infty}\frac{\g^{4n+2\m+\D}}{\sqrt{\left(2n+\m\right)!\left(2n+\m+\D\right)!}}|2n+\m,2n+\m+\D\ket\,.\label{eq:paircat}
\end{equation}
\end{widetext}Once again, we have only two distinct parameter regimes:
small and large $\g$. The behavior of the code states is thus reminiscent
of the single-mode code states,\begin{subnumcases}{\!\!\!\!\!\!\!\!\!\!\!\!\!\!\!\!\!|\m_{\g,\D}\ket\sim}
|\m,\m+\D\ket & $\g\rightarrow 0 \label{eq:smlima}$ \\
\displaystyle{\frac{|\g_{\D}\ket+(-1)^{\m}\left(-i\right)^{\D}\lket{i\g_{\D}}}{\sqrt{2}}} & $\g\rightarrow\infty~.$ \label{eq:tmlim}
\end{subnumcases}As a result, one should consider the code states as cat-state-like
superpositions of pair-coherent states, so we refer to them as ``pair-cat''
states (noting that they have previously been studied in quantum optics
\cite{Gerry1995,Gou1996,Liu2001,Choi2008}). Note also the connection
to NOON states in the $\g\ll1$ limit. One slight complication in
our definition is the $\D$-dependent phase between the superpositions
of $\lket{\g_{\D}}$ and $\lket{i\g_{\D}}$, but this is a mere bookkeeping
issue due to the unavoidable presence of the phase in the states'
definition.$^{\ref{fn:In-the-general}}$ We once again will focus
on the large $\g$ limit since that is when $\lket{\g_{\D}}$ and
$\lket{i\g_{\D}}$ become approximately orthogonal and when pair-cat
codes allow for protection against dephasing errors. The code projections
defined for all $\g$ are then
\begin{equation}
\p_{\text{II}}^{(\D)}=|0_{\g,\D}\ket\bra0_{\g,\D}|+|1_{\g,\D}\ket\bra1_{\g,\D}|\,.\label{eq:proj2}
\end{equation}
We will not fix the value of $\D$ in order to maintain generality.
In an experimental setting however, the most natural value of $\D$
is likely zero, and such a state will also enjoy symmetry under exchange
of the modes.

\subsection{Pair-cat code error analysis\label{subsec:Pair-cat-code-error}}

Analysis of errors on pair-cat codes follows closely that of the cat
codes, but the workload is ``doubled'' since we have to account
for two modes. We first determine the action of $\aa,\bb$ on our
codes. Here is where a key difference develops, namely, $\aa$ and
$\bb$ compensate each other by shifting $\D$ in opposite directions:\begin{subequations}
\begin{align}
\aa\tp_{\D} & =\tp_{\D+1}\aa\label{eq:smproj1-1}\\
\bb\tp_{\D} & =\tp_{\D-1}\bb\,.\label{eq:smproj2-1}
\end{align}
\end{subequations}In this way, losses in both modes counteract each
other and help keep $\D$ centered at zero. For the other projection,\begin{subequations}
\begin{align}
\aa\ts_{2\m+\D} & =\ts_{2\m+\D-1}\aa=\ts_{2(\m+1)+\D+1}\aa\\
\bb\ts_{2\m+\D} & =\ts_{2\m+\D-1}\bb\,,
\end{align}
\end{subequations}where we have added 4 in the subscript of $\ts$
in the first line (since the entire subscript is defined modulo 4)
in order to match the positive shift in $\D$ with that of Eq.~(\ref{eq:smproj1})
and in order to have a positive shift in $\m$ (for convention). Note
that the $\m+1$ part of the subscript $2(\m+1)+\D+1$ is defined
modulo 2, denoting a bit flip on the qubit. We thus see that application
of $\aa$ shifts $\Delta$ up by one while at the same time applying
a logical bit flip $\m+1$, while application of $\bb$ shifts $\Delta$
down by one without the extra bit flip.

Armed with the above equations, we can now apply the techniques from
Sec.~\ref{subsec:Cat-code-error} to these codes. Let us now determine
the effects of losses $\aa$ and $\bb$ exactly. Permuting $\aa,\bb$
through the projections in the definition (\ref{eq:paircatdef}),
applying them to the two-mode coherent state, and renormalizing yields\begin{subequations}
\begin{align}
\aa|\m_{\g,\D}\ket & =\g\sqrt{\frac{N_{\m+1,\D+1}}{N_{\m,\D}}}|\m+1_{\g,\D+1}\ket\label{eq:oneloss}\\
\bb|\m_{\g,\D}\ket & =\g\sqrt{\frac{N_{\m,\D-1}}{N_{\m,\D}}}|\m_{\g,\D-1}\ket\,.
\end{align}
\end{subequations}Therefore, unlike single-mode cat codes, here losses
on either mode take one to completely orthogonal subspaces. We will
see later that this is what allows one to correct arbitrary losses
in either mode.

Let us now examine the ratios of the normalizations $N$ in the above
equation in the large $\g$ limit. As with the cat codes, we would
like the $\m$-dependent factors to be suppressed. It turns out they
in fact are suppressed due to the differing asymptotic behaviors of
the two Bessel functions $I_{\D},J_{\D}$ making up $N_{\m,\D}$ (\ref{eq:renorm2}).
As $\g\rightarrow\infty$, $I_{\D}$ grows as order $O(e^{2\g^{2}}/\g)$
while $J_{\D}$ falls off as $O(1/\g)$. Therefore, just like the
cat codes, the $\m$-dependence (and, consequently, dephasing errors
within the code space) falls off exponentially with $\g^{2}$:
\begin{equation}
N_{\m,\D}={\textstyle \half}N_{\D}+\left(-1\right)^{\m}O(\g^{-1}e^{-2\g^{2}})\,.
\end{equation}
However, unlike the $\Pi$-dependence of $N_{\Pi}$ (\ref{eq:pimup})
falling off exponentially in the case of the cat codes, the $\D$-dependence
of $N_{\D}$ (\ref{eq:pidelta}) falls off only algebraically as $O(1/\g)$
(due to the $e^{-2\g^{2}}$ canceling the exponential growth of $I_{\D}$),
so $N_{\m,\D}$ does not become asymptotically constant very quickly.
Nevertheless, this will not present a problem since $N_{\D}$ is independent
of the qubit index and so does not violate the error-correction conditions
(\ref{eq:qec}). We now proceed to determine the matrix elements $\bra\m_{\g,\D}|O|\n_{\g,\D}\ket$
of the $2\times2$ matrix $\p_{\text{II}}^{(\D)}O\p_{\text{II}}^{(\D)}$
for various components $O$ of loss and dephasing errors for both
modes.

\subsubsection{Dephasing errors\label{subsec:Dephasing-errors-1}}

As with cat codes, we will see that dephasing is suppressed as $\g\rightarrow\infty$.
Projecting $\aa^{\dg k}\aa^{k}$ and $\bb^{\dg k}\bb^{k}$ onto our
code spaces using $\p_{\text{II}}^{(\D)}$ (\ref{eq:proj2}) yields\begin{subequations}
\begin{align}
\p_{\text{II}}^{(\D)}\aa^{\dg k}\aa^{k}\p_{\text{II}}^{(\D)} & =\g^{2k}\begin{pmatrix}\frac{N_{k,\D+k}}{N_{0,\D}} & 0\\
0 & \frac{N_{k+1,\D+k}}{N_{1,\D}}
\end{pmatrix}\label{eq:depht-exact-1}\\
\p_{\text{II}}^{(\D)}\bb^{\dg k}\bb^{k}\p_{\text{II}}^{(\D)} & =\g^{2k}\begin{pmatrix}\frac{N_{0,\D-k}}{N_{0,\D}} & 0\\
0 & \frac{N_{1,\D-k}}{N_{1,\D}}
\end{pmatrix}\,.\label{eq:depht-exact-2}
\end{align}
\end{subequations}In the large $\g$ limit, we expand to obtain\begin{subequations}
\begin{align}
\p_{\text{II}}^{(\D)}\aa^{\dg k}\aa^{k}\p_{\text{II}}^{(\D)} & =\g^{2k}\frac{N_{\D+k}}{N_{\D}}\p_{\text{II}}^{(\D)}+O\left(\g^{2k}e^{-2\g^{2}}\right)Z_{\text{II}}^{(\D)}\label{eq:dephtm}\\
\p_{\text{II}}^{(\D)}\bb^{\dg k}\bb^{k}\p_{\text{II}}^{(\D)} & =\g^{2k}\frac{N_{\D-k}}{N_{\D}}\p_{\text{II}}^{(\D)}+O\left(\g^{2k}e^{-2\g^{2}}\right)Z_{\text{II}}^{(\D)}\,,
\end{align}
\end{subequations}where $Z_{\text{II}}$ (\ref{eq:logz}) is the
logical $Z$-operator of the code. As with the analogous Eq.~(\ref{eq:deph-1})
for cat codes, logical $Z$-errors on the code spaces due to dephasing
are suppressed exponentially with $\g^{2}$. The main difference is
the presence of the ratio of normalization factors $N_{\D}$. However,
these only affect the coefficient in front of $\p_{\text{II}}^{(\D)}$
and thus do not violate the error-correction conditions.

We note here that, as with cat codes \cite{bin,Li2016,codecomp},
fine tuning of $\g$ can also help suppress logical errors due to
dephasing even at small $\g$. We consider the $\D=0$ code space,
which is special because it is invariant under the exchange operator
$\E$ from Eq.~(\ref{eq:exch}). This means that the effects of errors
for both modes, Eqs.~(\ref{eq:depht-exact-1}) and (\ref{eq:depht-exact-2}),
should be identical for this code space. The identity required to
show this, $N_{\m,-\D}=N_{\m+\D,\D}$ (where $\m+\D$ is evaluated
modulo two), comes from the properties of the two Bessel functions
under changes of sign of $\D$: $I_{-\D}(x)=I_{\D}(x)$ and $J_{-\D}(x)=\left(-1\right)^{\D}J_{\D}(x)$.
Using this identity and letting $c\in\{\aa,\bb\}$, we thus have
\begin{equation}
\p_{\text{II}}^{(0)}c^{\dg k}c^{k}\p_{\text{II}}^{(0)}=\g^{2k}\left(C_{k}^{+}\p_{\text{II}}^{(0)}+C_{k}^{-}Z_{\text{II}}^{(0)}\right)\,,
\end{equation}
where $C_{k}^{\pm}=\frac{N_{k,k}}{N_{0,0}}\pm\frac{N_{k+1,k}}{N_{1,0}}$.
For simplicity, let us consider $k=1$ and numerically minimize the
undesired effect $C_{1}^{-}$. It turns out that $C_{1}^{-}\approx0$
at an optimal value of $\g\approx1.3$, so lowest-order dephasing
errors in both modes are suppressed at that value! This $\g$ yields
an average occupation number of 
\begin{equation}
\half\tr\{\p_{\text{II}}^{(0)}c^{\dg}c\}\approx1.3
\end{equation}
for both modes one and two. By comparison, the occupation number for
the cat code which minimizes the $k=1$ error is $\half\tr\{\p_{\text{I}}^{(\Pi=0)}\ph\}\approx2.3$,
corresponding to $\a\approx1.5$ \cite{bin,Li2016,codecomp}.

\begin{figure}
\centering \includegraphics[width=1\columnwidth]{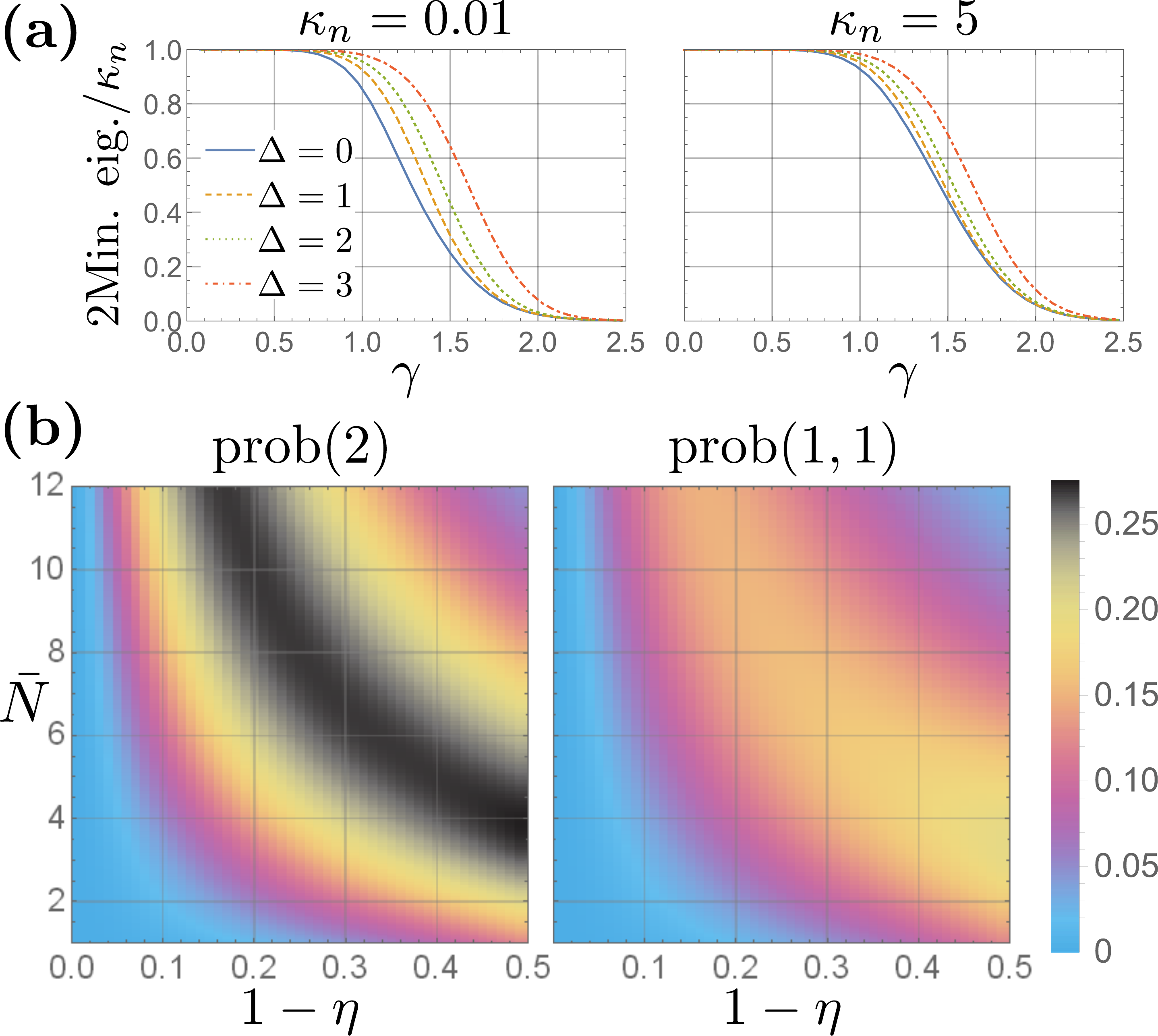} \caption{\label{fig:errors} \textbf{(a) }Plot of the logical dephasing rate
(scaled by $\protect\k_{n}/2$) of subspaces of $\protect\D\in\{0,1,2,3\}$
vs. $\protect\g$ at $\protect\k_{n}=0.01$ (left) and $\protect\k_{n}=5$
(right). This shows the exponential suppression of scheme II against
dephasing (cf. Fig.~A1 in Ref.~\cite{cats} and Fig.~7.4 in Ref.~\cite{thesis})
that persists for non-perturbative values of $\protect\k_{n}$. \textbf{(b)}
Probability of the leading uncorrectable loss errors, $\protect\pr(2)$
(\ref{eq:p2}) for scheme I and $\protect\pr(1,1)$ (\ref{eq:p11})
for scheme II, versus total occupation number $\bar{N}$ (\ref{eq:nbar-1})
and dimensionless cavity loss rate $1-\eta$ (\ref{eq:transmissivity})
(assuming equally lossy cavities, $\protect\k_{a}=\protect\k_{b}=\protect\k$).
For all values of the parameters, the probability (\ref{eq:p2}) of
a maximally-mixed cat-code state to lose two photons is greater than
the probability (\ref{eq:p11}) of a maximally-mixed paircat-code
state to lose one in each mode. Both probabilities follow the Poisson
distribution in the $(1-\eta)\bar{N}\rightarrow\infty$ limit.}
\end{figure}

As with the cat codes, it turns out that the dissipator $\k_{\text{II}}\DD_{\text{II}}$
does the job of protecting against dephasing errors for both modes.
Since $\DD[\ph]$ and $\DD[\phm]$ both commute with $\di$, dephasing
does not connect code spaces for different values of $\D$. (We could
have also inferred this much from the above error-correction conditions.)
Therefore, we only need to see how dephasing acts within each subspace
of fixed $\D$. The rates of dephasing-induced logical errors are
determined by the eigenvalues of the superoperator $\k_{\text{II}}\DD_{\text{II}}+\k_{n}\DD[\ph]+\k_{m}\DD[\phm]$,
and one can numerically plot those eigenvalues and observe that they
are suppressed exponentially as $O(e^{-c\g^{2}})$, where $c$ is
a constant. It is easy to show that the effect of $\DD[\ph]$ and
$\DD[\phm]$ within a fixed-$\D$ subspace is identical, so we consider
only $\k_{n}\DD[\ph]$ and plot the dephasing error rates vs. $\g$
for two values of $\k_{n}$ and four values of $\D$ in Fig.~\ref{fig:errors}(a).

\subsubsection{Loss errors}

Now let us turn to loss errors and show that arbitrary instances of
$a^{k}$ and $\bb^{\ell}$ (for $\ell\geq0$) are correctable. Equations
(\ref{eq:oneloss}-b) readily tell us that the value of $\D$ is shifted
in different directions upon the respective loss events. Therefore,
projecting back onto the original code space yields
\begin{equation}
\p_{\text{II}}^{(\D)}\aa^{k}\p_{\text{II}}^{(\D)}=\p_{\text{II}}^{(\D)}\bb^{\ell}\p_{\text{II}}^{(\D)}=0\,.
\end{equation}
The code thus corrects all \textit{individual} loss errors in both
modes. However, the leading uncorrectable error is a \textit{simultaneous}
loss in both modes. Due to the extra bit flip induced by a loss in
mode one, $\aa\bb$ induces a bit flip within the code:\footnote{\label{fn:If-we-change}If we change the error channel in a way that
allows us to know how many total photons were lost (``channel monitoring''
\cite{Niu2017}), we can track even this error: if the measured $\D=0$
but two photons were lost in total, then $\aa\bb$ had to have occurred.}
\begin{equation}
\p_{\text{II}}^{(\D)}\aa\bb\p_{\text{II}}^{(\D)}\sim\g^{2}X_{\text{II}}^{(\D)}\,.
\end{equation}
Note that there is no ratio of normalizations this time because the
value of $\D$ is unchanged.

Strictly speaking, the leading uncorrectable cat code error $\aa^{2}$
is of the same order as the leading uncorrectable pair-cat code error
$\aa\bb$. So what is the advantage of scheme II over scheme I? While
there is no qualitative information-theoretic advantage, the probability
of the leading uncorrectable error is slightly lower for scheme II
than for scheme I when evaluated for the maximally mixed state for
both codes. (We use the maximally mixed state so as to not give preference
to any particular superposition of code states. The probabilities
below should thus be interpreted as averaged over the code space.)
For code I, the probability of losing $\ell$ photons is
\begin{equation}
\pr(\ell)=\half\tr\{\p_{\text{I}}^{(\Pi)}E_{\aa}^{\ell\dg}E_{\aa}^{\ell}\}\,,\label{eq:prop_l}
\end{equation}
where $E_{\aa}^{\ell}$ (\ref{eq:loss}) are the Kraus operators for
the loss channel. The distribution $\{\pr(\ell)\}_{\ell=0}^{\infty}$
becomes approximately Poissonian in the large $\a$ limit \cite{codecomp},
but we keep things exact to consider experimentally relevant $\a$.
Similarly, the probability of $\ell$ loss events in mode $\aa$ and
$\ell^{\prime}$ in mode $\bb$ for the maximally mixed code II state
is
\begin{equation}
\pr(\ell,\ell^{\prime})=\half\tr\{\p_{\text{II}}^{(\D)}E_{\aa}^{\ell\dg}E_{\aa}^{\ell}E_{b}^{\ell^{\prime}\dg}E_{b}^{\ell^{\prime}}\}\,.\label{eq:prop_llp}
\end{equation}
The leading uncorrectable error probabilities for schemes I and II
are thus $\pr(2)$ to $\pr(1,1)$, respectively. We plot them in Fig.~\ref{fig:errors}(b)
versus the dimensionless\textit{ loss rate} $1-\eta$, with transmissivity
\begin{equation}
\eta\equiv e^{-\k t}\label{eq:transmissivity}
\end{equation}
and assuming equally lossy cavities ($\k_{a}=\k_{b}=\k$), and the
total occupation number,
\begin{equation}
\bar{N}=\begin{cases}
\half\tr\{\p_{\text{I}}^{(\Pi)}\ph\} & \text{Scheme I}\\
\half\tr\{\p_{\text{II}}^{(\D)}(\ph+\phm)\} & \text{Scheme II}
\end{cases}\,,\label{eq:nbar}
\end{equation}
for scheme I code $\Pi=0$ and scheme II code $\D=0$. One can calculate
these analytically, yielding\begin{subequations}
\begin{align}
\pr(2) & =\frac{(1-\eta)^{2}\a^{4}}{4}e^{-(1-\eta)\a^{2}}\left(\frac{N_{1,\Pi=0}^{\eta\a}}{N_{0,\Pi=0}}+\frac{N_{0,\Pi=0}^{\eta\a}}{N_{1,\Pi=0}}\right)\label{eq:p2}\\
\pr(1,1) & =\frac{(1-\eta)^{2}\g^{4}}{2}e^{-2(1-\eta)\g^{2}}\left(\frac{N_{1,\D=0}^{\eta\g}}{N_{0,\D=0}}+\frac{N_{0,\D=0}^{\eta\g}}{N_{1,\D=0}}\right),\label{eq:p11}
\end{align}
\end{subequations}where $N_{\m,\Pi}^{\eta\a}$ is the cat normalization
factor $N_{\m,\Pi}$ (\ref{eq:pimup}) with $\a\rightarrow\eta\a$
and $N_{\m,\D}^{\eta\g}$ is the pair-cat normalization factor $N_{\m,\D}$
(\ref{eq:renorm2}) with $\g\rightarrow\eta\g$. The respective total
photon numbers are (for $\p_{\text{I}}^{\Pi=0}$ and $\p_{\text{II}}^{\D=0}$
codes)
\begin{equation}
\bar{N}=\begin{cases}
\half\a^{2}\left(\frac{N_{1,\Pi=1}}{N_{0,\Pi=0}}+\frac{N_{0,\Pi=1}}{N_{1,\Pi=0}}\right) & \text{Scheme I}\\
\g^{2}\left(\frac{N_{1,\D=1}}{N_{0,\D=0}}+\frac{N_{0,\D=1}}{N_{1,\D=0}}\right) & \text{Scheme II}
\end{cases}\,.\label{eq:nbar-1}
\end{equation}

Let us compare the uncorrectable error probabilities. At a loss rate
$1-\eta=0.03$ and at the optimal (for dephasing) values of the two
codes introduced in the previous subsection, we have $\pr(2)\approx2.4\%$
at the optimal $\nb\approx2.3$ for scheme I and $\pr(1,1)\approx2.1\%$
at the optimal $\nb\approx2.6$ for scheme II. While the difference
is not large, it shows that the protection from loss of scheme II
is no worse than that of scheme I. At $1-\eta=0.20$ and $\nb=10$,
the difference is more pronounced: $\pr(2)\approx27\%$ for scheme
I and $\pr(1,1)\approx15\%$ for scheme II. A low loss probability
allows one to take more time during syndrome measurement, resulting
in improved measurement accuracy {[}\citealp{Ofek2016}, Supplementary
Information Sec.~4.1{]}.

Protection from loss events can be implemented in a continuous manner
using additional jump operators
\begin{equation}
\F_{\text{II},\D}^{\text{loss}}=\begin{cases}
|0_{\g,0}\ket\bra1_{\g,\D}|+|1_{\g,0}\ket\bra0_{\g,\D}| & \D<0,\,\D\text{ odd}\\
|0_{\g,0}\ket\bra0_{\g,\D}|+|1_{\g,0}\ket\bra1_{\g,\D}| & \text{otherwise}
\end{cases}\label{eq:loss2}
\end{equation}
for $\D\neq0$, provided that $\k t\ll1$ and the state is initialized
in the $\D=0$ codespace {[}cf. Eq.~(\ref{eq:loss1}){]}. The positive
odd $\D$ case corrects the bit-flip induced by $\bb^{\D}$. The alternative
cat-code correction scenario from Sec.~\ref{subsec:Loss-errors}
can also be extended to scheme II by $\g\rightarrow\g\exp(-\half\k t)$
in the bras of Eq.~(\ref{eq:loss2}). We propose to realize related
jumps of the form $F(1)=\aa^{\dg}\tp_{\D=1}$ (\ref{eq:autojumpplus1})
and $F(-1)=\bb^{\dg}\tp_{\D=-1}$, adding photons conditional on a
nonzero $\D$. These jumps are analogous to those stemming from the
continuous QEC proposal of scheme I {[}\citealp{cohenthesis}, Sec.~4.2.2{]}.

\subsection{Pair-cat code gates}

Let us introduce the gates for the setup of scheme II, which are all
in complete analogy to those of scheme I.

\subsubsection{Hamiltonian $X$ and $XX$ gates}

We can once again leverage the fact that $\aa\bb$ is an uncorrectable
error and create a gate. Via the same techniques described for the
cat codes, the Hamiltonian $\H_{\text{II}}^{X}=g_{X}(\aa\bb+\mathrm{h.c.})$
generates rotations around the $X$-axis:
\begin{equation}
\p_{\text{II}}^{(\D)}\H_{\text{II}}^{X}\p_{\text{II}}^{(\D)}\sim2g_{X}\g^{2}X_{\text{II}}^{(\D)}\,,
\end{equation}
with corrections exponential in $\g^{2}$. A two-qubit gate can similarly
be created using the Hamiltonian $\H_{\text{II}}^{XX}=g_{XX}(\aa_{1}\bb_{1}\aa_{2}\bb_{2}+\mathrm{h.c.})$
for qubit systems $1$ and $2$ with respective parameters $\{\g_{1},\D_{1}\}$
and $\{\g_{2},\D_{2}\}$:
\begin{equation}
\p_{\text{II},1}^{(\D_{1})}\p_{\text{II},2}^{(\D_{2})}\H_{\text{II}}^{XX}\p_{\text{II},1}^{(\D_{1})}\p_{\text{II},2}^{(\D_{2})}\sim2g_{XX}\g_{1}^{2}\g_{2}^{2}X_{\text{II},1}^{(\D_{1})}X_{\text{II},2}^{(\D_{2})}\,.
\end{equation}

\subsubsection{Hamiltonian $Z$-gate}

We sketch an extension of the scheme \cite{Cohen2016} from cat to
pair-cat codes. The Josephson junction Hamiltonian now couples to
both modes,
\begin{equation}
H_{\text{II}}^{\text{jnct}}=E_{J}\cos\left(\a\aa e^{i\o_{a}t}+\b be^{i\o_{b}t}+\mathrm{h.c.}\right)\,,\label{eq:junct2}
\end{equation}
where $E_{J}$ is the Josephson energy, $\o_{a},\o_{b}$ are drive
frequencies, and $\a$ ($\b$) is the amplitude and phase of the drive
of mode one (two). Following Sec.~\ref{subsec:Hamiltonian--gate},
we apply the RWA to the above Hamiltonian, yielding an operator that
is diagonal in Fock space, $\overline{D_{\a}}\otimes\overline{D_{\b}}$,
where $\overline{D_{\b}}$ is defined in Eq.~(\ref{eq:dispRWA}).
Once again, since the two basis states (\ref{eq:paircat}) we use
to represent the code are superpositions of different sets of Fock
states, there will be no $X_{\text{II}}^{(\D)}$ or $Y_{\text{II}}^{(\D)}$
terms when $H_{\text{II}}^{\text{jnct}}$ under the RWA is projected
into the codespace in this basis,
\begin{equation}
\p_{\text{II}}^{(\D)}(\overline{D_{\a}}\otimes\overline{D_{\b}})\p_{\text{II}}^{(\D)}=C_{+}^{(\D)}\p_{\text{II}}^{(\D)}+C_{-}^{(\D)}Z_{\text{II}}^{(\D)}\,,
\end{equation}
where $C_{\pm}^{(\D)}=\bra0_{\g,\D}|\overline{D_{\a}}\otimes\overline{D_{\b}}|0_{\g,\D}\ket\pm\bra1_{\g,\D}|\overline{D_{\a}}\otimes\overline{D_{\b}}|1_{\g,\D}\ket$.
Just like the analogous single mode gate is $\Pi$-dependent, this
gate is $\D$-dependent. However, following Ref.~\cite{Cohen2016},
we can in principle combine multiple junctions, each with a Hamiltonian
like $H_{\text{II}}^{\text{jnct}}$, but with their own tunable parameters.
While making a completely $\D$-independent gate is outside the scope
of this work, we anticipate that there are sufficient degrees of freedom
in such a combination to allow for $C_{-}^{(\D)}$ to be $\D$-independent
for at least $\D\in\{0,\pm1\}$.

\subsubsection{Holonomic $Z$-gate}

Alternatively to the above $Z$-gate, we can maintain protection from
loss events but suppress protection from dephasing events by adiabatically
changing $\g$ in the sequence $\g\rightarrow0\rightarrow\g e^{i\phi}\rightarrow\g$.
Using the decoherence Hamiltonian $\ft^{\dg}\ft$ and following {[}\citealp{ABFJ},
Sec.~IV C{]}, one can verify that the Lindbladian remains gapped
throughout the entire adiabatic path. Thus, the leading-order effect
is the holonomy induced on the states after the path, which comes
from the part $0\rightarrow\g e^{i\phi}$. In this step, the new steady
states of $\DD_{\text{II}}$ are $|\m_{\g\exp(i\phi),\D}\ket$, whose
$\g=0$ limit is $\exp[i(2\m+\D)\phi]|\m,\m+\D\ket$. However, the
initial states for this step consist of just the Fock states $|\m,\m+\D\ket$
without the extra phase. Thus, to compensate for including the extra
phase during the step $0\rightarrow\g e^{i\phi}$, one will have $|\m,\m+\D\ket\rightarrow\exp[-i(2\m+\D)\phi]|\m_{\g\exp(i\phi),\D}\ket$.
The entire sequence thus performs an effective $Z$-rotation
\begin{equation}
\p_{\text{II}}^{(\D)}U_{\text{II}}^{\text{hol}}\p_{\text{II}}^{(\D)}=e^{-i\phi\D}\begin{pmatrix}1 & 0\\
0 & e^{-2i\phi}
\end{pmatrix}\,.
\end{equation}
The $\D$-dependent phase is an overall phase since the qubit is entirely
in a code space of fixed occupation number difference, so the effect
of the gate is independent of $\D$. 

\subsubsection{Kerr $\nicefrac{\pi}{2}$ $Z$-rotation}

As with cat codes, we can utilize Kerr nonlinearities to form a Hamiltonian
$K(\ph+\phm-\D)^{2}$ and create a $\frac{\pi}{2}$-rotation around
the $Z$-axis of the pair-cat qubit. However, this is less practical
than the cat-code gate since it requires coupling several fine-tuned
junctions to each mode. Running this evolution for a time $t=\frac{\pi}{8K}$
yields the operation
\begin{equation}
\p_{\text{II}}^{(\D)}U_{\text{II}}^{Z}\p_{\text{II}}^{(\D)}=\begin{pmatrix}1 & 0\\
0 & i
\end{pmatrix}\,.
\end{equation}
This can be proven by substituting the labels for sets of Fock states
$2n+\m$ and $2n+\m+\D$ for $\ph$ and $\phm$, respectively, in
$U_{\text{II}}^{Z}$. Unfortunately, as with cat codes, this does
require a relatively large $K\gg\k_{\aa},\k_{\bb}$ so that no loss
events occur during the running of this gate. Alternatively, one can
track loss events during the gate by measuring $\di$ and compensate
by applying rotations $\exp[i(\t\ph+\phi\phm)]$ afterwards.

\subsubsection{Kerr control-phase gate}

Rounding out Table \ref{tab:1}, we can evolve under another combination
of Kerr nonlinearities for four modes,
\begin{equation}
U_{\text{II}}^{CZ}=\exp[i\frac{\pi}{4}(\ph_{1}+\phm_{1}-\D_{1})(\ph_{2}+\phm_{2}-\D_{2})]\,,
\end{equation}
where $\D_{1}$ ($\D_{2}$) is the occupation number difference and
$\ph_{1},\phm_{1}$ ($\ph_{2},\phm_{2}$) are the occupation number
operators for pair-cat qubit 1 (2). Substituting the Fock state components
of the two-qubit basis elements $|\m_{\D_{1},\g_{1}},\n_{\D_{2},\g_{2}}\ket$
into the four occupation number operators yields an effect gate
\begin{equation}
\p_{\text{II},1}^{(\D_{1})}\p_{\text{II},2}^{(\D_{2})}U_{\text{II}}^{CZ}\p_{\text{II},1}^{(\D_{1})}\p_{\text{II},2}^{(\D_{2})}=\begin{pmatrix}1 & 0 & 0 & 0\\
0 & 1 & 0 & 0\\
0 & 0 & 1 & 0\\
0 & 0 & 0 & -1
\end{pmatrix}\,.
\end{equation}

\subsubsection{Control engineering}

It turns out that one can use drives $\e_{C}(t)\aa+\mathrm{h.c.}$
and $\e_{C}(t)\bb+\mathrm{h.c.}$ on cavity one and two respectively,
an ancilla transmon drive $\e_{T}(t)\s_{+}+\mathrm{h.c.}$, and the
two-cavity dispersive Hamiltonian $(\chi_{1}\ph+\chi_{2}\phm)\s_{z}$
to implement a universal set of gates for both cavities (\cite{bin},
Appx.~G). Similar schemes have already been implemented experimentally
\cite{Wang2016}, and one could consider using such schemes for pair-cat
manipulation. However, as with cat-codes, it is likely that these
procedures will have to be performed without the engineered dissipation
$\lt$.

\begin{figure*}[t]
\centering \includegraphics[width=1\textwidth]{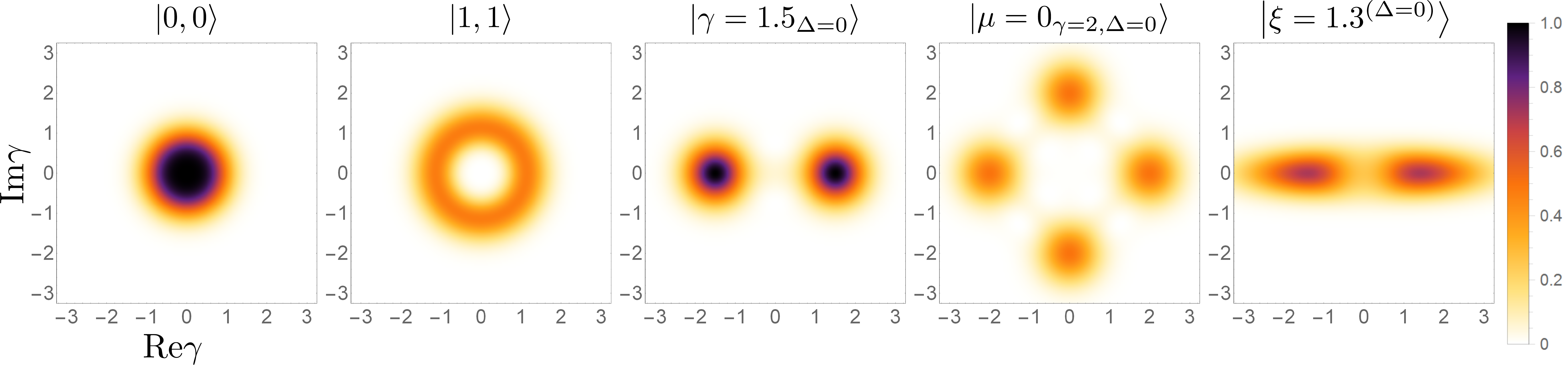} \caption{\label{fig:husimi} From left to right, $Q$-functions $|\protect\bra\protect\g_{\protect\D}|\psi\protect\ket|^{2}$
(\ref{eq:qfunc}) vs. $\protect\g$ for states $|\psi\protect\ket$
being the Fock state $|0,0\protect\ket$, Fock state $|1,1\protect\ket$,
a pair-coherent state $|\protect\g_{\protect\D}\protect\ket$ (\ref{eq:paircoh}),
a pair-cat state $|\protect\m_{\protect\g,\protect\D}\protect\ket$
(\ref{eq:paircat}), and a two-mode squeezed state $|\xi^{(\protect\D)}\protect\ket$
(\ref{eq:squeezed}). Note that the usual fringes between peaks are
not present here because these are not Wigner distributions. All states
are contained in the subspace of $\protect\D=0$, and we find similar
shapes for analogous states at other $\protect\D$. We also find that
these two-mode $Q$-functions look similar to their counterparts in
single-mode phase space: the Fock states $|0\protect\ket$, $|1\protect\ket$,
the cat state $|\protect\a_{\Pi}\protect\ket$ (\ref{eq:cats}), the
``four-cat'' code state $|\protect\m_{\protect\a,\Pi}\protect\ket$
(\ref{eq:smcat}), and a single-mode squeezed state $\exp[\protect\half\xi(\protect\aa^{2}-\protect\aa^{\protect\dg2})]|0\protect\ket$.
This visualization strategy avoids having to deal with the entire
four-dimensional two-mode phase space while also preserving the intuition
of single-mode phase space. Note that all plots will be symmetric
under $\protect\g\rightarrow-\protect\g$ as a result of our convention.}
\end{figure*}

\section{Quasiprobability distributions for fixed-$\protect\D$ subspaces\label{sec:Visualization-techniques}}

A generic two-mode state can be represented using two-mode analogues
of the $P$-, $Q$-, or $W$igner quasiprobability distributions \cite{Hillery1984}.
However, the phase space of the full two-mode system is four-dimensional,
so one has to judiciously pick the right two-dimensional cross-sections
to study the state. We show here that, if one is restricted to a fixed-$\Delta$
sector, a two-dimensional space is sufficient to represent the state.
Given pair-coherent states $\{|\g_{\D}\ket\}_{\g\in\mathbb{C}}$,
this space is the complex plane represented by $\g$. One should think
of this as the fixed-$\D$ two-mode analogue of the $\a$ complex
plane for a single mode. We\textit{ }suggest not to call this a phase
space \cite{Werner2016} since the lowest-order physically motivated
operators --- $\tp_{\D}\aa\bb\tp_{\D}$ and $\tp_{\D}\aa^{\dg}\bb^{\dg}\tp_{\D}$
--- do not commute to a constant; we instead refer to it as the $\g$\textit{-plane}.
The derivations below can be repeated for the two fixed-parity subspaces
of a single mode using the cat states $\{|\a_{\Pi}\ket\}_{\a\in\mathbb{C}}$.

The eigenvalue equation (\ref{eq:eigen}) and overcompleteness of
$|\g_{\D}\ket$ (\ref{eq:overcomp}) are sufficient to define informationally
complete analogues of $P$- and $Q$-distribution functions in the
$\g$-plane \cite{Brif1994,DusanPopov2001}. Along similar lines and
following standard procedures \cite{Hillery1984}, here we also define
a generalized $W$-representation. In order to help simplify these
distributions, we define 
\begin{equation}
\Gamma\equiv\g^{2}
\end{equation}
and employ the more conventional set of pair-coherent states
\begin{equation}
|\widetilde{\G_{\D}}\ket=|\sqrt{\G}_{\D}\ket\,\,\,\,\,\,\,\,\,\,\,\,\,\,\text{with}\,\,\,\,\,\,\,\,\,\,\,\,\,\,\aa\bb|\widetilde{\G_{\D}}\ket=\G|\widetilde{\G_{\D}}\ket\,.\label{eq:newpaircoh}
\end{equation}
This convention allows us to avoid dealing with $\g^{2}$ whenever
we act on these states with $\aa\bb$. These states also resolve the
identity:
\begin{align}
\tp_{\D} & =\intop d^{2}\G\widetilde{\s_{\D}}(\G)|\widetilde{\G_{\D}}\ket\bra\widetilde{\G_{\D}}|\,,\label{eq:overcomp2}
\end{align}
where $\widetilde{\s_{\D}}(\G)=\frac{2}{\pi}K_{\D}(2|\sqrt{\G}|)I_{\D}(2|\sqrt{\G}|)$
{[}this measure differs from $\s_{\D}(\g)$ (\ref{eq:measure}) by
the Jacobian $2\G${]}. Below, we define our distributions $\ds(\G;\r)$
(with $\ds\in\{P,Q,W\}$) using $|\widetilde{\G_{\D}}\ket$, but convert
back to our convention by examining $\ds(\g^{2};\r)$ instead. The
reason we do this is because we have found $\ds(\g^{2};\r)$ more
visually similar to their corresponding single-mode quasiprobability
distributions.

Given a state $\r$ and a fixed occupation number difference $\D$,
the respective distributions are\begin{subequations}
\begin{align}
Q(\G;\r) & =\frac{1}{\widetilde{\s_{\D}}(\G)}\intop\frac{d^{2}\eta}{\pi^{2}}e^{\eta^{\star}\G-\G^{\star}\eta}\tr_{\D}\left\{ \r e^{-\eta^{\star}\aa\bb}e^{\eta\aa^{\dg}\bb^{\dg}}\right\} \label{eq:dists}\\
P(\G;\r) & =\frac{1}{\widetilde{\s_{\D}}(\G)}\intop\frac{d^{2}\eta}{\pi^{2}}e^{\eta^{\star}\G-\G^{\star}\eta}\tr_{\D}\left\{ \r e^{\eta\aa^{\dg}\bb^{\dg}}e^{-\eta^{\star}\aa\bb}\right\} \label{eq:dists2}\\
W(\G;\r) & =\frac{1}{\widetilde{\s_{\D}}(\G)}\intop\frac{d^{2}\eta}{\pi^{2}}e^{\eta^{\star}\G-\G^{\star}\eta}\tr_{\D}\left\{ \r e^{\eta\aa^{\dg}\bb^{\dg}-\eta^{\star}\aa\bb}\right\} \,,\label{eq:dists3}
\end{align}
\end{subequations}where $\tr_{\D}\{\r\}=\tr\{\tp_{\D}\r\}$. The
three traces are called the characteristic functions of the state,
and the distributions are simply their Fourier transforms. These are
normalized, $\intop d^{2}\G\widetilde{\s_{\D}}\left(\G\right)\ds\left(\G\right)=1$,
which is easily seen using the identity
\begin{equation}
\intop\frac{d^{2}\eta}{\pi^{2}}e^{\eta^{\star}\G-\G^{\star}\eta}=\d^{2}(\G)\,.
\end{equation}

Applying Eqs.~(\ref{eq:newpaircoh}-\ref{eq:overcomp2}) and the
above identity to Eq.~(\ref{eq:dists}) yields
\begin{equation}
Q(\G;\r)=\bra\widetilde{\G_{\D}}|\r|\widetilde{\G_{\D}}\ket=\bra\g_{\D}|\r|\g_{\D}\ket\,.\label{eq:qfunc}
\end{equation}
This two-mode $Q$-distribution provides us with plots that are visually
similar to the conventional single-mode $Q$-distribution $\bra\a|\r|\a\ket$
for the various states we have tried (see Fig.~\ref{fig:husimi}).
Note that, due to the squaring of the argument, all phase space plots
are invariant under $\g\rightarrow-\g$. This way, pair-coherent states
look like cat states (as opposed to coherent states). We argue this
is more natural due to the close group-theoretical connection between
$|\g_{\D}\ket$ and $|\a_{\Pi}\ket$.

The $P$-distribution provides a diagonal representation for $\r$
in terms of $|\widetilde{\G_{\D}}\ket$,
\begin{align}
\r & =\intop d^{2}\G\widetilde{\s_{\D}}(\G)P(\G;\r)|\widetilde{\G_{\D}}\ket\bra\widetilde{\G_{\D}}|\,.\label{eq:prep}
\end{align}
We can plug in the above equation into Eq.~(\ref{eq:dists2}) and
simplify using Eqs.~(\ref{eq:newpaircoh}-\ref{eq:overcomp2}) to
show that $P(\G;\r)$ is indeed equal to Eq.~(\ref{eq:dists2}).

\begin{figure}[t]
\centering \includegraphics[width=1\columnwidth]{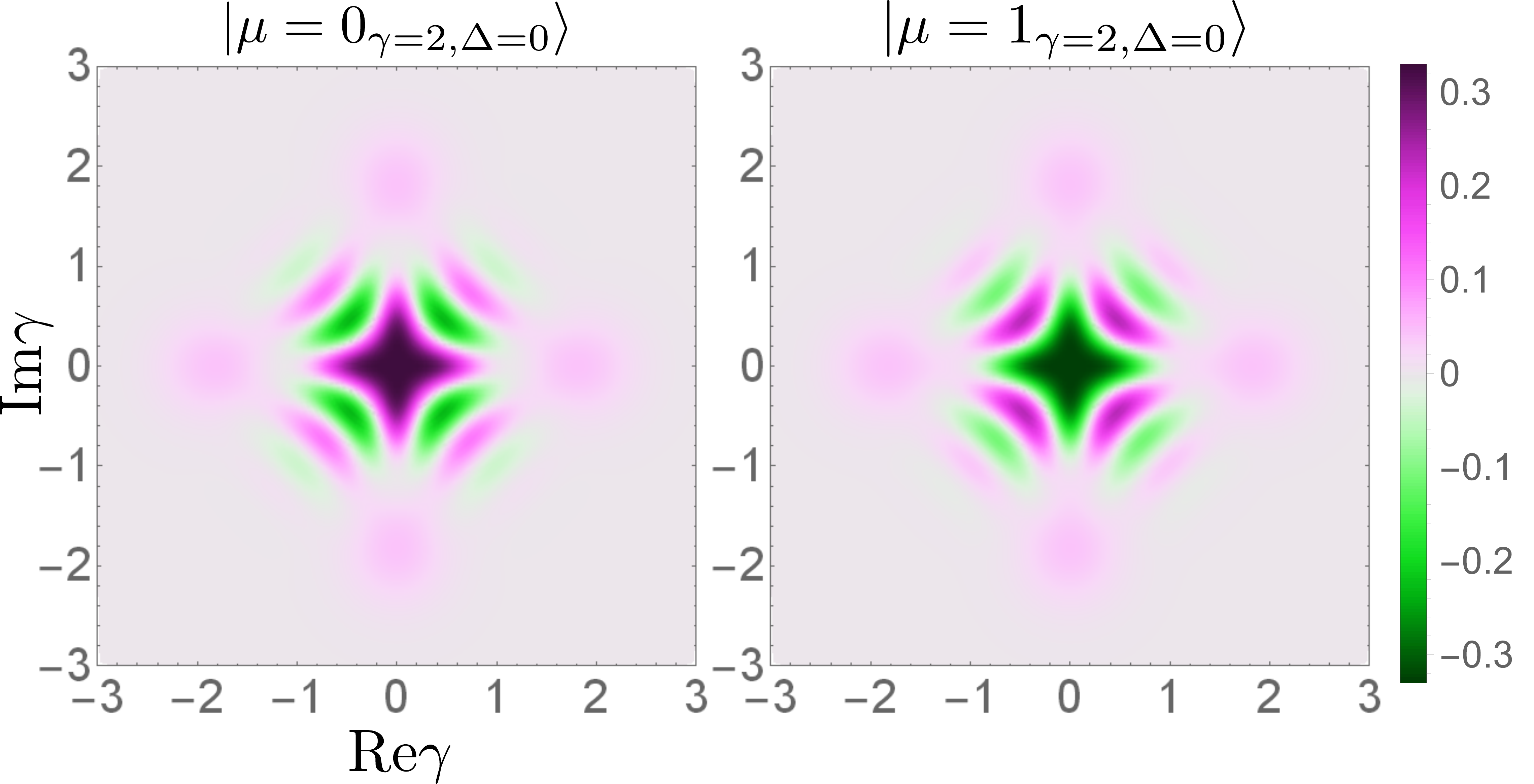} \caption{Unnormalized $W$-distributions $\widetilde{\protect\s_{\protect\D}}(\protect\g^{2})W(\protect\g^{2};\protect\r)$
(\ref{eq:dists3}) for the pair-cat qubit states $|0_{\protect\g,\protect\D}\protect\ket$
and $|1_{\protect\g,\protect\D}\protect\ket$ (\ref{eq:paircat}),
where $\protect\D=0$ and $\protect\g=2$. These plots were obtained
by expressing both $\protect\r$ and the two-mode squeezing operator
in terms of Fock states using Ref.~\cite{Maccone2001}.\label{fig:wigner}}
\end{figure}

The above distributions can also be used for state tomography, in
which the expectation value of an observable $A$ is evaluated using
only the distribution $\ds(\G;\r)$ of the state and the corresponding
dual distribution $\ds^{\star}(\G;A)$ of the observable:
\begin{align}
\tr_{\D}(A\r) & =\intop d^{2}\G\varLambda(\G)\ds^{\star}(\G;A)\ds(\G;\r)\,,\label{eq:dual}
\end{align}
where $\varLambda(\G)$ is a suitable measure. We define $\ds$ to
be informationally complete --- equivalent to the state itself ---
if the above equality is satisfied for some $\ds^{\star}$. Plugging
in Eq.~(\ref{eq:prep}) into the left-hand side of Eq.~(\ref{eq:dual})
easily yields $P^{\star}=Q$ with $\varLambda(\G)=\widetilde{\s_{\D}}(\G)$.
The dual for the characteristic function of the $W$-distribution,
$\tr_{\D}\{\r\exp(\G\aa^{\dg}\bb^{\dg}-\G^{\star}\aa\bb)\}$, was
determined in Ref.~\cite{Maccone2001} (see also \cite{Carmeli2009}).
This means that $W$ itself is also informationally complete, but
it is no longer self-dual as it is in the single-mode case (so it
is technically not a proper Wigner distribution \cite{Ferrie2011}).
There is currently no analogue of the dramatic simplification that
can be done for the conventional Wigner function (see, e.g., \cite{catbook},
Appx.~A.2), resulting in time-consuming numerics. We leave its simplification,
study, and interpretation to future work, but sketch our code states
(\ref{eq:paircat}) in Fig.~\ref{fig:wigner} to reveal fringes characteristic
of the conventional Wigner distribution.

\section{Stabilizers \& generalizations\label{sec:Multi-mode-generalizations-and}}

We comment on higher-mode generalizations of scheme II, making contact
with concepts from stabilizer-based error-correction \cite{gottesman_thesis,nielsen_chuang}
and its extensions \cite{Knill1996,Pollatsek2004,Looi2008,Cross2009,alKruszynska2009,Rossi2013,Ni2015}. 

\subsection{Pair-cat code stabilizers}

Recall that traditional stabilizer codes, denoted by projection $P$,
are defined as unique eigenspaces of eigenvalue one of a set of commuting
operators $\{S\}$ (called \textit{stabilizers}):
\begin{equation}
SP=P\,\,\,\,\,\,\,\,\,\,\,\,\,\,\,\forall S\in\{S\}\,.
\end{equation}
These commuting operators are part of a larger group of operators.
We introduce stabilizers for the two-mode case, but by picking stabilizers
out of the \textit{algebra} of two-mode operators $\{\aa^{\dg k}\aa^{m}b^{\dg l}\bb^{n}\}_{k,m,l,n=0}^{\infty}$
instead of a group. In addition, we relax the usual assumptions that
$\{S\}$ are all Hermitian and involutive (square to the identity).
While our stabilizers commute, a consequence of this algebraic framework
is that some of them are not diagonalizable.

Recall that the logical state set $\{|\m_{\g,\D}\ket\}_{\m=0}^{1}$
is defined by two parameters: real $\g$ and integer $\D$. The logical
subspace is the eigenspace of eigenvalue one of the stabilizers\begin{subequations}
\begin{align}
S_{\g} & =1+\aa^{2}\bb^{2}-\g^{4}\\
S_{\D} & =1+\bb^{\dg}\bb-\aa^{\dg}\aa-\D\,.
\end{align}
\end{subequations}(Similar stabilizers exist for cat codes, $S_{\a}=1+\aa^{4}-\a^{4}$
and $S_{\Pi}=(-1)^{\aa^{\dg}\aa+\Pi}$, with the latter an infinite
sum of elements of the algebra $\{\aa^{\dg n}\aa^{m}\}_{n,m=0}^{\infty}$.)
These stabilizers obviously commute and give 
\begin{equation}
S_{\g}\p_{\text{II}}^{(\D)}=S_{\D}\p_{\text{II}}^{(\D)}=\p_{\text{II}}^{(\D)}
\end{equation}
when applied to the code subspace projection $\p_{\text{II}}^{(\D)}$
(\ref{eq:proj2}). Since $S_{\g}$ is not Hermitian, $\p_{\text{II}}^{(\D)}S_{\g}\neq\p_{\text{II}}^{(\D)}$
(but we do have $\p_{\text{II}}^{(\D)}S_{\g}^{\dg}=\p_{\text{II}}^{(\D)}$)
and we cannot straightforwardly construct Hermitian projections out
of $S_{\g}$. The projection constructed out of exponentials of $S_{\D}$
is of course onto a subspace of fixed $\di$ {[}see Eq.~(\ref{eq:deltaprojections}){]}.
The stabilizer $S_{\D}$ picks the subspace of fixed occupation number
difference $\D$ while $S_{\g}$ selects the two pair-cat states with
the proper value of $\g$ within that subspace. There are only two
pair-coherent states having that value of $\g$ because of the relations
$\aa\bb|\g_{\D}\ket=\g^{2}|\g_{\D}\ket$ and $\lket{(-\g)_{\D}}=\left(-1\right)^{\D}|\g_{\D}\ket$. 

Recall from Sec.~\ref{subsec:Pair-cat-code-error} that the code
$\p_{\text{II}}^{(\D)}$ can detect any number of single-mode losses.
Consider only operators of the form $\aa^{n}\bb^{m}$ and let the
\textit{weight} of an operator $O$ be the number of modes on which
$O$ acts nontrivially. Then, we find that $\p_{\text{II}}^{(\D)}$
detects all weight one errors of this type. However, due to the approximate
satisfaction of the diagonal error-correction conditions, this code
can exactly correct against such errors only in the $\g\rightarrow\infty$
limit; for any finite $\g$, this is an exact error-detecting and
an approximate error-correcting code \cite{Leung1997,Crepeau2005}.
The lowest-weight undetectable error is $\aa\bb$ --- a sort of square-root
of the stabilizer $S_{\g}$.

\subsection{Multimode generalization}

The above framework can easily be generalized to multiple modes and
qudit codes. Given $M$ modes and logical qudit dimension $d$, let
\begin{equation}
S_{\g}=1+\left(\aa^{d}\right)^{\otimes M}-\g^{dM}=1-\g^{dM}+\prod_{m=1}^{M}\aa_{m}^{d}\,,
\end{equation}
where $\aa_{m}$ is the lowering operator for mode $m$. Consider
also the $M-1$ occupation number differences
\begin{equation}
S_{m}=1+\aa_{m+1}^{\dg}\aa_{m+1}-\aa_{m}^{\dg}\aa_{m}-\D_{m}
\end{equation}
for $m\in\{1,2,\cdots,M-1\}$ and a vector of differences 
\begin{equation}
\vec{\D}=(\D_{1},\,\D_{2},\,\cdots\D_{M-1})\in\mathbb{Z}^{\times(M-1)}\label{eq:deltas}
\end{equation}
with the corresponding operators $\hat{\D}_{m}=\aa_{m+1}^{\dg}\aa_{m+1}-\aa_{m}^{\dg}\aa_{m}$.
One can then construct eigenstates of all $\hat{\D}_{m}$,
\begin{equation}
|\g_{\vec{\D}}\ket\propto\tp_{\vec{\D}}\left(|\g\ket^{\otimes M}\right)\,,\label{eq:catm}
\end{equation}
where $\tp_{\vec{\D}}$ is the projection on the multimode subspace
whose nearest-neighbor occupation value differences are fixed by $\vec{\D}$.
The qudit code
\begin{equation}
\{|\g_{\vec{\D}}\ket,|(\g e^{i\frac{2\pi}{dM}})_{\vec{\D}}\ket,\cdots,|(\g e^{i\frac{2\pi}{dM}(d-1)})_{\vec{\D}}\ket\}
\end{equation}
can detect any loss errors of weight $M-1$ or less. Superposition
of such projected coherent states yields the conjugate ``cat'' basis,
\begin{equation}
|\m_{\g,\vec{\D}}\ket\propto\sum_{\n=0}^{d-1}|(\g e^{i\frac{2\pi}{dM}\n})_{\vec{\D}}\ket\,.\label{eq:paircatm}
\end{equation}
The form of such ``cat'' states is especially concise for $\vec{\D}=\vec{0}$,
\begin{equation}
|\m_{\g,\vec{0}}\ket\propto\sum_{n=0}^{\infty}\frac{\g^{M(dn+\m)}}{[(dn+\m)!]^{M/2}}|dn+\m\ket^{\otimes M}\,.\label{eq:3mcode}
\end{equation}
The lowest-weight undetectable error is $\aa^{\otimes M}$. For even
$d$, instead of utilizing the entire $d$-dimensional space for each
$\vec{\D}$ to store information, one can define the two-dimensional
subspace $\mu\in\{0,\nicefrac{d}{2}\}$ as the new logical qubit and
use the complementary subspace to protect said qubit from higher-weight
errors. (In the single mode case, a more judicious choice of qubit
suppresses errors even more \cite{Li2016}; the same is likely true
here, but this is outside the scope of this paper.) For example, the
generalized states for $M=2$, obtained by taking $|\m_{\g,\D}\ket$
(\ref{eq:paircat}) and substituting
\begin{equation}
2n+\m\longrightarrow(S+1)(2n+\m)\,,\label{eq:genpaircat}
\end{equation}
can detect $\aa^{\ell}\bb^{\ell}$ with $\ell\leq S$. In combination
with being able to detect arbitrary single-mode loss events, this
means that generalized pair-cat codes can detect up to $S$ loss errors
in each mode --- $\{\aa^{k}\bb^{\ell}\}_{k+\ell\leq S}$. The \textit{spacing}
$S$ is the same as the spacing discussed in Ref.~\cite{Li2016}
for ``multi-legged'' single-mode cat codes and binomial codes. Details
as to how $S>0$ pair-cat codes detect $(\aa\bb)^{\ell\leq S}$ are
given in Ref.~\cite{thesis}.

We have so far considered only photon losses in our error analysis.
However, we can equivalently consider two-mode gains and losses and
their generalization for multiple modes. In other words, multimode
cat codes can protect either against a set of losses or against a
different set of losses \textit{and} gains. For example, the two-mode
scheme II can protect against either $\{\aa^{k},\bb^{\ell}\}_{k,\ell=0}^{\infty}$
or $\{\aa^{k},\aa^{\dg\ell}\}_{k,\ell=0}^{\infty}$ or $\{\bb^{k},\bb^{\dg\ell}\}_{k,\ell=0}^{\infty}$.
The latter two sets include only one of the two modes, so this analysis
is not particularly useful for $M=2$. However, higher $M$ codes
can in fact protect against all single-mode losses and gains. For
$M\geq3$, a loss (gain) event of $\ell$ photons in mode $1<m<M$
shifts $\D_{m-1}$ down (up) by $\ell$ and $\D_{m}$ up (down) by
$\ell$. The edge cases $m=1$ and $m=M$ are handled by positive
and negative shifts in $\Delta_{1}$ and $\D_{m}$, respectively.
Thus, all single-photon losses and gains correspond to a unique syndrome.

\begin{figure}[t]
\centering \includegraphics[width=0.75\columnwidth]{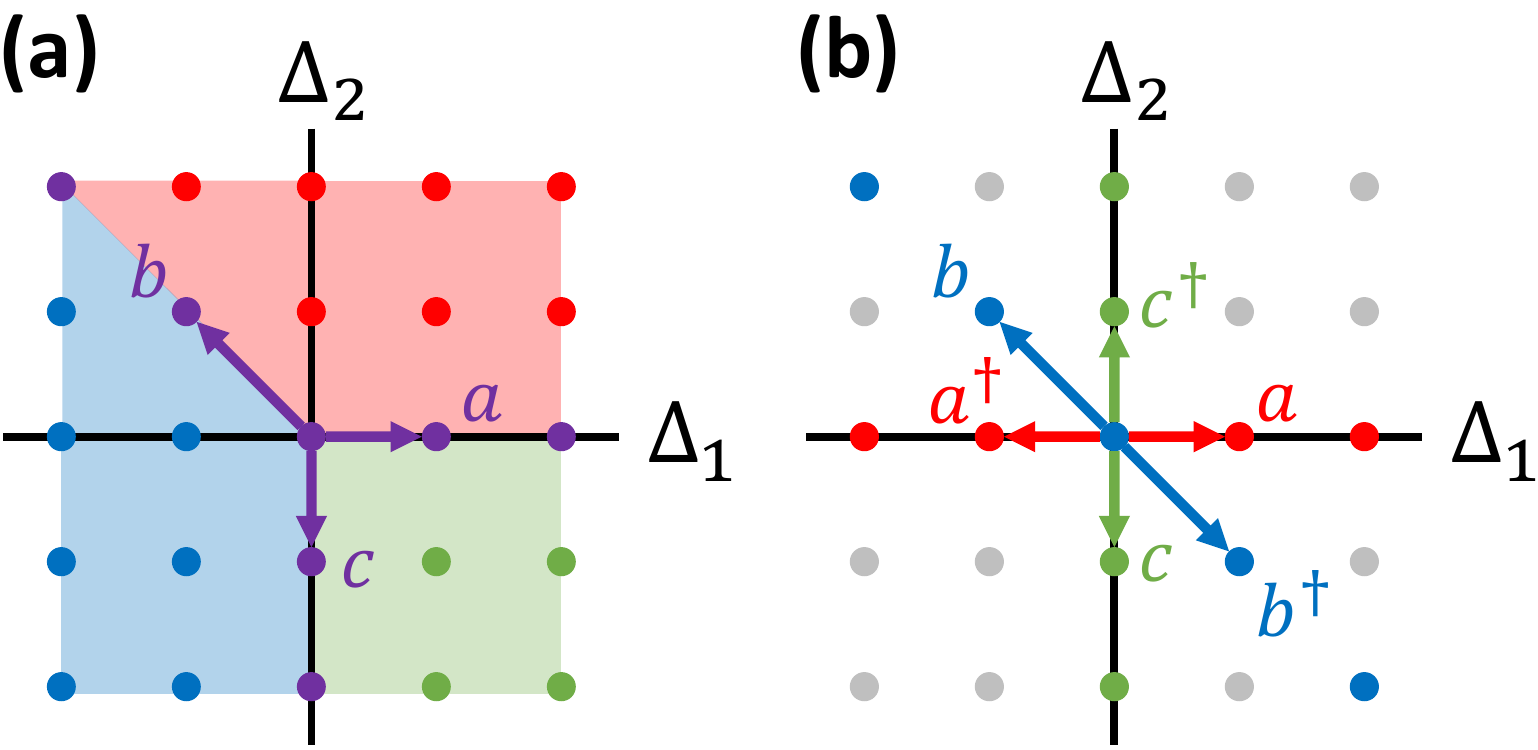} \caption{Lattice of the error subspaces for the three-mode ($M=3$) code (\ref{eq:3mcode}),
characterized by photon number differences $(\protect\D_{1},\protect\D_{2})$
(\ref{eq:deltas}). This code can detect all loss errors up to weight
two $\{\protect\aa^{n}\protect\bb^{m},\protect\bb^{n}c^{m},\protect\aa^{n}c^{m}\}_{n,m=0}^{\infty}$
(\ref{eq:errors1}). Panel\textbf{ (a)} shows the shifts (purple arrows)
that occur after respective loss events $\protect\aa,\protect\bb,c$.
Drawing three lines from the origin to these three points and onward
to infinity divides the lattice into three regions (highlighted in
red, blue, and green), which correspond to the three possible types
of at most weight-two operators $\protect\aa^{n+m}\protect\bb^{m}$,
$\protect\bb^{-n}c^{-(n+m)}$, or $\protect\aa^{n}c^{-m}$. Alternatively,
the same code can detect all single-mode loss and gain errors (\ref{eq:errors2}).
Panel \textbf{(b)} shows the shifts caused by single instances of
such events. The lines formed by the three pairs of antiparallel arrows
form the error subspaces necessary for detection of all single-mode
losses and gains. For both scenarios, the code becomes an error-correcting
code against the respective sets of errors in the limit of large $\protect\g$.\label{fig:lattice}}
\end{figure}

\subsection{Three-mode example}

Consider $d=2$, $M=3$, $S=0$. The range of $\tp_{\D_{1},\D_{2}}$
depends on the values of $\vec{\D}=(\D_{1},\D_{2})$,\[
\begin{array}{ll}
\!\!\{|n,n+\D_{1},n+\D_{1}+\D_{2}\ket\}_{n=0}^{\infty}&\D_{1},\D_{2}\geq0\\
\!\!\{|n+|\D_{1}|,n,n+\D_{2}\ket\}_{n=0}^{\infty}&\D_{1}<0,\,\D_{2}\geq0\\
\!\!\{|n+|\D_{2}|,n+\D_{1}+|\D_{2}|,n+\D_{1}\ket\}_{n=0}^{\infty}&\D_{1}\geq0,\,\D_{2}<0\\
\!\!\{|n+|\D_{1}|+|\D_{2}|,n+|\D_{2}|,n\ket\}_{n=0}^{\infty}&\D_{1},\D_{2}<0\,. \label{eq:3m}
\end{array}
\]The three-mode generalized ``cat'' states $\{|\m_{\D_{1},\D_{2}}\ket\}_{\m=0}^{1}$
(\ref{eq:paircatm}) --- superpositions of the three-mode projected
coherent states $|\g_{\D_{1},\D_{2}}\ket$ (\ref{eq:catm}) --- have
been studied before for this case \foreignlanguage{american}{\cite{an2003}}.

The integer differences $\vec{\D}=(\D_{1},\,\D_{2})\in\mathbb{Z}^{2}$
form the two-dimensional lattice shown in Fig.~\ref{fig:lattice}
and each weight-two loss operator shifts to a unique point on the
lattice. To prove this, observe that losing one photon in mode 1 {[}2,
3{]} shifts you from the origin to the point $A=(1,0)$ {[}$B=(-1,1)$,
$C=(0,-1)${]} on the lattice. Drawing three lines from the origin
to these three points and onward to infinity divides the lattice into
three regions {[}Fig.~\ref{fig:lattice}(a){]}, which correspond
to the three possible types of at most weight-two operators, 
\begin{equation}
\{\aa^{n}\bb^{m},\bb^{n}c^{m},\aa^{n}c^{m}\}_{n,m=0}^{\infty}\,.\label{eq:errors1}
\end{equation}
Given a syndrome $(n,m)\in\mathbb{Z}^{2}$, one first determines which
region it belongs to. Depending on region, the syndrome then corresponds
to an error of $\aa^{n+m}\bb^{m}$, $\bb^{-n}c^{-(n+m)}$, or $\aa^{n}c^{-m}$.
The leading undetectable error is $\aa\bb c$.

Alternatively, let us consider protecting against one-photon losses
and gains for all three modes, 
\begin{equation}
\{\aa^{n},\aa^{\dg n},\bb^{m},\bb^{\dg m},c^{p},c^{\dg p}\}_{n,m,p=0}^{\infty}\,.\label{eq:errors2}
\end{equation}
Considering once more the lattice formed by $(\D_{1},\D_{2})$, $n$
events in mode $\aa$, $\bb$, or $c$ bring about the shifts $(\pm n,0)$,
$(\mp n,\pm n)$, and $(0,\mp n)$, respectively, with the sign signaling
whether the events were losses or gains. Such errors cover three non-parallel
lines in the lattice {[}Fig.~\ref{fig:lattice}(b){]}, so each error
in the above set corresponds to a unique syndrome. Note that in this
case, the full lattice of possible syndromes is not utilized; the
unused error spaces are colored gray in Fig.~\ref{fig:lattice}(b).

\subsection{Comparison to other codes\label{subsec:Comparison-to-other}}

\subsubsection{Noon and $\chi^{(2)}$ codes}

It is useful to compare this family to the $\chi^{(2)}$ codes \cite{Niu2017}
and noon codes \cite{Bergmann2016a} --- two-mode binomial codes
\cite{codecomp} concatenated with a repetition code. A fundamental
difference is that pair-cat codes consist of infinite superpositions
of Fock states while $\chi^{(2)}$ and noon codes are finite-dimensional.
In group theory jargon, cat and pair-cat codes live in irreducible
subspaces of the \textit{non-compact} group $SU(1,1)$ generated by
two-photon loss and occupation number operators (see Sec.~\ref{subsec:Primer-on-pair-coherent}),
while $\chi^{(2)}$ and noon codes are similarly related to \textit{compact}
groups such as $SU(N)$ associated with a $\chi^{(2)}$ Hamiltonian
\cite{niu1} and beam-splitter transformations \cite{Bergmann2016a},
respectively. As a result, only a finite number of photons can be
lost for $\chi^{(2)}$ and noon codes while pair-cat codes have a
nonzero (albeit exponentially vanishing) probability of losing an
arbitrary number of photons. None of the $\chi^{(2)}$ codes correct
against more than one individual loss event in each mode, but the
two- and three-mode $\chi^{(2)}$-BC codes can correct more than one
loss if one also knows the total number of photons lost.$^{\ref{fn:If-we-change}}$
Due to concatenation, noon codes require at least four modes to correct
single loss events. Generalized two- or higher-mode pair-cat codes
with $S>0$ {[}see Eq.~(\ref{eq:genpaircat}){]} can detect (and,
in the $\g\rightarrow\infty$ limit, correct) up to $S$ loss errors
in each mode using only knowledge given from error syndromes. Most
importantly, $S=0$ higher-mode pair-cat codes can detect all single-mode
losses \textit{and} gains, something that none of the other codes
can do. However, the two mode $\chi^{(2)}$-BC code can correct dephasing
errors $\ph^{\ell}$ exactly up to $\ell\leq N$, while pair-cat codes
correct dephasing approximately (see. Sec.~\ref{subsec:Pair-cat-code-error}).
It would be interesting to extend the analysis of Ref.~\cite{codecomp}
to two modes to determine the theoretically possible performance of
these codes against photon loss.

\begin{figure}
\centering \includegraphics[width=0.9\columnwidth]{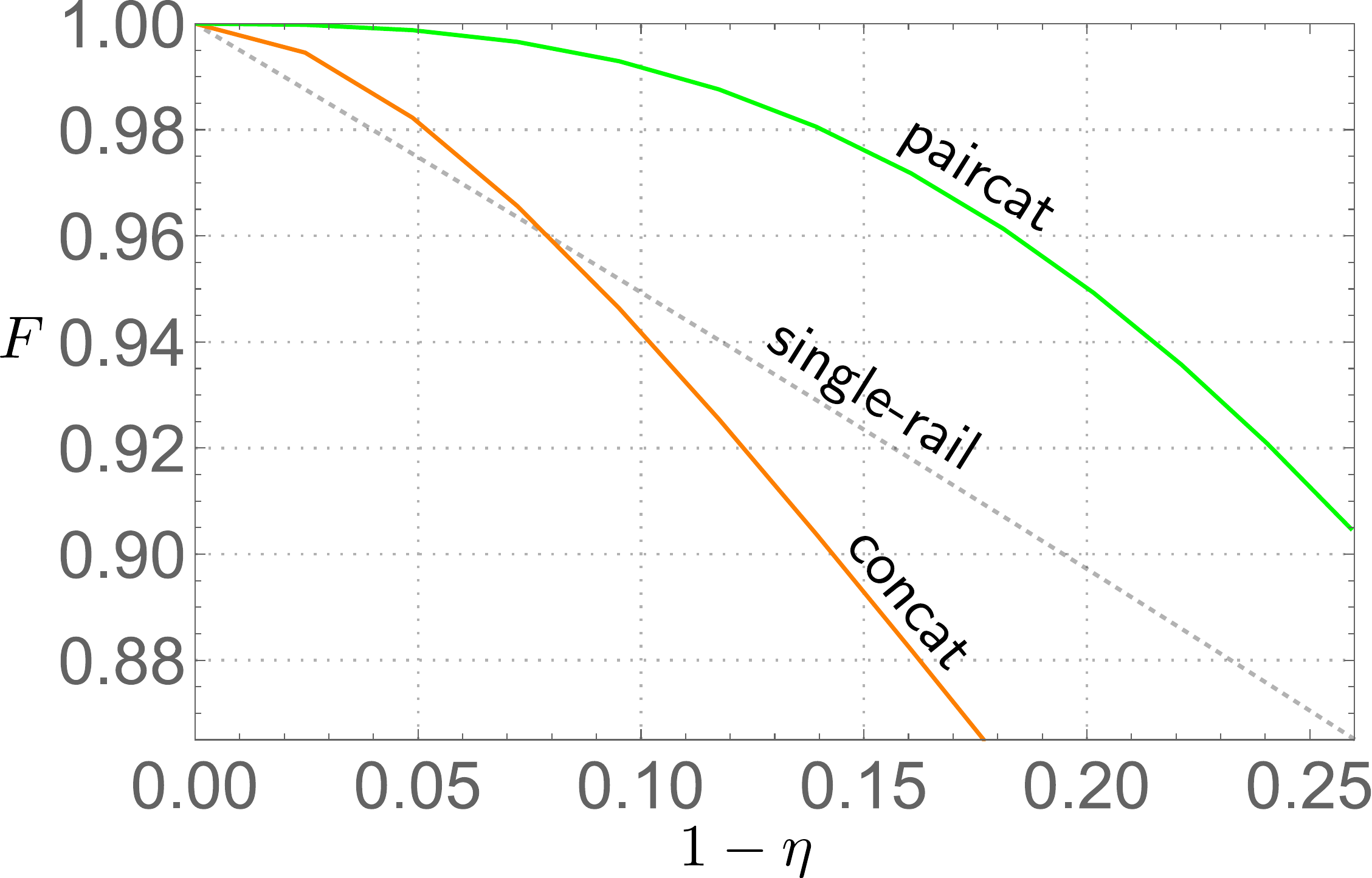}\caption{\label{fig:Plot-comparing-the}Plot comparing the entanglement fidelity
$F$ (\ref{eq:entfid}) of our three-mode code, Eq. (\ref{eq:3mcode})
for $M=3$, with the concatenated cat code (con-cat) from Eq. (\ref{eq:concat})
and the single-mode encoding into Fock states $\{|0\protect\ket,|1\protect\ket\}$
(single-rail). The horizontal axis is the loss rate $1-\eta$, written
in terms of the transmissivity $\eta=e^{-\protect\k t}$ (\ref{eq:transmissivity})
of the loss channel (assuming equal decay rates for each mode). This
result does not provide a full-fledged comparison for two reasons:
(1) the average photon number per mode is set to $\approx1.08$ for
both codes and (2) The fidelity is calculated assuming the transpose
recovery operation, which is a factor of two away from the optimal
recovery procedure \cite{Ng2010}. }
\end{figure}

\subsubsection{Concatenated cat code}

One can consider taking single-mode codes and concatenating with multi-qubit
codes. The simplest cat-code $\{|\a_{\Pi=0}\ket,|\a_{\Pi=1}\ket\}$
cannot correct against photon loss events, so scheme I uses a different
set of code states (see Table \ref{tab:1}). However, given that all
cat-codes suppress dephasing errors for sufficiently large $\a$,
one can concatenate the simplest cat code with another code that corrects
against loss. Loss errors cause a bit-flip within the logical subspace
$\{|\a_{\Pi=0}\ket,|\a_{\Pi=1}\ket\}$, so concatenating that code
with a repetition code yields a code \cite{Munro2002} with logical
states ($\m\in\{0,1\}$)
\begin{equation}
|\m_{\a}^{\cc}\ket=|\a_{\Pi=\m}\ket^{\otimes3}\label{eq:concat}
\end{equation}
that can correct both leading-order loss and (for sufficiently large
$\a$) dephasing errors in all three modes. This \textit{con}catenated
\textit{cat}-code (con-cat) is a candidate for a future bosonic logical
qubit {[}\citealp{cohenthesis}, Sec. 4.3{]}. Although a full comparison
between con-cat and our three-mode code {[}Eq. (\ref{eq:3mcode})
for $M=3${]} is outside the scope of this work, we have reason to
believe that pair-cat outperforms con-cat in one-photon-per-mode regime.

Recall that both single- and multi-mode cat codes suppress dephasing
errors as $\a$ and $\g$ increase, respectively. However, both codes
I and II also have the ability to suppress dephasing at optimally
configured ``sweet spots'' $\a$, $\g$. In Sec. \ref{subsec:Dephasing-errors-1},
we showed that the two-mode pair-cat code can protect against lowest-order
dephasing at the optimal value of $\nb\approx1.3$ photons per mode
($\g\approx1.3$). Our three-mode code allows for the same protection
at $\nb\approx1.08$ photons per mode ($\g\approx1.2$). While the
single-mode cat code I also allows for such beneficial fine-tuning,
the con-cat code does not because it consists of coherent states $|\pm\a\ket$
whose overlap does not oscillate with increasing $\a$. Therefore,
con-cat does not have a sweet spot and requires a larger $\a$, and
thus a larger number of photons, to protect against dephasing.

To corroborate this observation, we calculated a lower bound on the
ultimate performance of con-cat and pair-cat, both set at $\nb\approx1.08$
photons. We calculated the entanglement fidelity of both codes, assuming
photon loss and the transpose recovery operation. The procedure consists
of starting with an initial maximally entangled state $|\varPsi\ket$
of two qubits, encoding one of the qubits in either the con-cat or
pair-cat encoding via the isometry ${\cal S}$, applying the photon
loss channel $e^{\k t\DD[\aa]}$ with Kraus operators (\ref{eq:loss})
and equal decay rates $\k_{\aa}=\k_{\bb}=\k_{c}$ to that encoded
qubit, recovering via the transpose recovery ${\cal R}$, and then
decoding via the reverse isometry. The entanglement fidelity $F$
is the overlap between the state after recovery with the initial state,
\begin{equation}
F=\bra\varPsi|[{\cal S}^{-1}{\cal R}e^{\k t\DD[\aa]}{\cal S}\otimes{\cal I}](|\varPsi\ket\bra\varPsi|)|\varPsi\ket\,,\label{eq:entfid}
\end{equation}
where ${\cal I}$ is the identity channel. This is identical to a
single-mode code comparison {[}\citealp{codecomp}, Sec. I.B{]}, with
the exception that the recovery used now is not optimal and $\nb$
is fixed to 1.08.\footnote{The three-mode Hilbert space we use has at most $8$ photons per mode,
yielding dimension $(8+1)^{3}=729$. The transpose-recovery calculation
took several days on an above-average desktop computer, and calculating
the optimal recovery for such a space is intractable. Comparing the
transpose recovery fidelity to the optimal one {[}\citealp{codecomp},
Fig. 1{]} is unfortunately not fair.} However, the transpose recovery is guaranteed to yield a fidelity
at most a factor of two from the fidelity of the optimal recovery
procedure \cite{Ng2010}.

The result is shown in Fig. \ref{fig:Plot-comparing-the}; one can
see that pair-cat outperforms con-cat for all visible values. In a
circuit QED experimental setting, $\k\lesssim1$ kHz, and we would
prefer to correct ten times more often, i.e., at $10$ kHz. This yields
a $1-\eta\approx0.02$, and we observe that pair-cat outperforms con-cat
in that regime. While this is only a bound whose infidelity is guaranteed
to be within a factor of 1/2 from the optimal result, the improvement
of pair-cat over con-cat is more than that, e.g., $\half(1-\F_{\cc})\approx2.6\times10^{-3}$
while $1-\F_{\pc}\approx0.2\times10^{-3}$ at $1-\eta\approx0.025$.
In fact, pair-cat even reaches a fidelity of 99\% at the large loss
rate of $0.1$, which is in the regime of applicability to quantum
repeater architectures. This is evidence that pair-cat has a substantial
advantage in this low photon regime. However, this does not suggest
that pair-cat always outperforms con-cat since increasing $\nb$ for
both codes leads to further suppression of dephasing errors in con-cat.
Unfortunately, we cannot compare the codes at larger values of $\nb$
because the Hilbert space required to house the states becomes too
large to be computationally tractable.

\section{Realizing continuous QEC against dephasing\label{sec:Continuous-vs-loss}}

\begin{figure}
\centering \includegraphics[width=1\columnwidth]{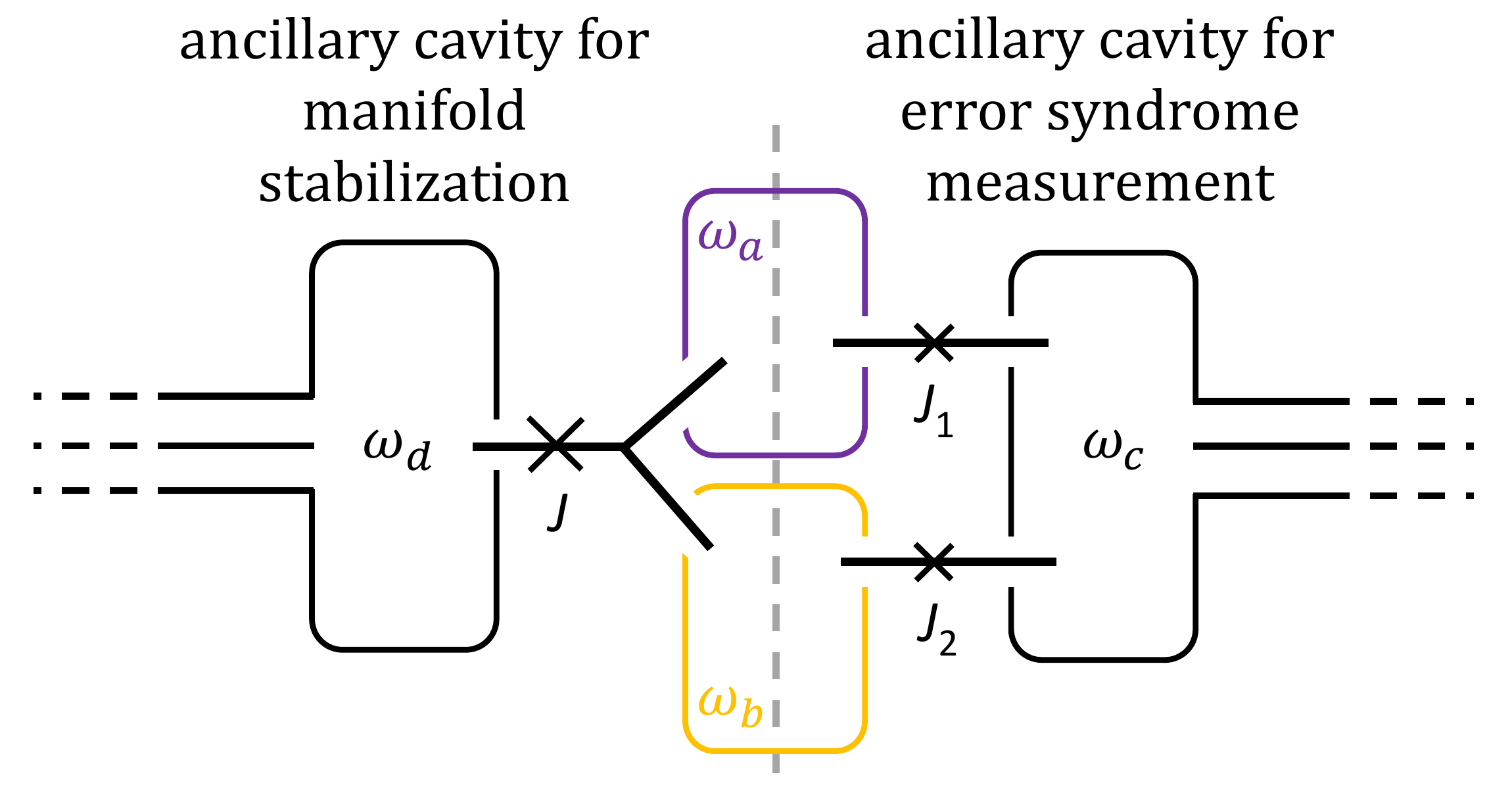} \caption{Proposed experimental setup. Two high-Q cavities at frequency $\omega_{a}/(2\pi)$
(purple) and $\omega_{b}/(2\pi)$ (orange) are coupled to three junction-modes.
The left half of the setup implements a driven-dissipative process
of the form $\protect\kt\protect\lt=\protect\kt\mathcal{D}\left[a^{2}b^{2}-\gamma^{4}\right]$
(\ref{eq:maindiss}) by cascading two four-wave mixing processes using
the junction mode labeled $J$, which is in turn coupled to a low-Q
cavity ($\omega_{d}$) facilitating entropy extraction. The right
half of the setup is used to perform measurement of the error syndrome
$\protect\di=\protect\bb^{\protect\dg}\protect\bb-\protect\aa^{\protect\dg}\protect\aa$
(\ref{eq:diff}) --- the photon number difference in the high-Q cavities.
Here, both high-Q cavities are coupled to individual junction modes
$J_{1,2}$, which each couple to a shared low-Q cavity ($\protect\o_{c}$).
Under appropriate pumping, these junction modes realize a displacement
of the low-Q cavity that is proportional to $\protect\di$. \label{fig:complete_setup}}
\end{figure}

In this section, we propose a realization of a driven-dissipative
process $\kt\lt$ corresponding to the left side of Fig.~\ref{fig:complete_setup}
by cascading a pair of two-photon exchange processes using a Raman
transition \citep{Mundhada2017}. The sub-system under consideration
consists of two high-Q cavity modes coupled to a Josephson junction
mode denoted by $J$ whose first three states are $|g\ket$, $|e\ket$,
and $|f\ket$. The junction mode is in turn coupled to a low-Q resonator
$d$ for the purpose of entropy extraction. One can engineer an exchange
of either of the cavities coupled to the $g\leftrightarrow e$ or
$e\leftrightarrow f$ transitions of the junction mode. Figure~\ref{fig:cascading_explanation}(a)
shows the schematic of cascading two such two-photon exchange processes
to get a simultaneous exchange of two photons of each of the cavities
with the $g\leftrightarrow f$ transition of the junction mode. Subsequent
decay of the junction mode translates to the loss of two-photons on
both cavities. The reverse process of exciting both cavities simultaneously
with two photons each is also possible by exciting the junction mode
to the $f$ state and then swapping the junction excitation into the
cavities.

\subsubsection{Setting up the Hamiltonian}

Consider a Hamiltonian consisting of the two high-Q cavities (with
lowering operators $\aa,\bb$ and frequencies $\o_{a,b}$) and a Josephson
junction mode (with lowering operator $J$, frequency $\o_{J}$, and
Josephson energy $E_{J}$) driven by a time-dependent drive $\hbar\varepsilon(t)J+\text{h.c.}$.
Let 
\begin{equation}
\frac{H_{0}}{\hbar}=\omega_{a}a^{\dagger}a+\omega_{b}b^{\dagger}b+\omega_{J}J^{\dagger}J
\end{equation}
consist of the harmonic portion of the full Hamiltonian. The anharmonic
portion of the junction is then $-E_{J}(\frac{\hat{\varphi}^{2}}{2}+\cos\hat{\varphi})$,
where the phase difference across the junction is
\begin{equation}
\hat{\varphi}=\phi_{a}a+\phi_{b}\bb+\phi_{J}J+\mathrm{\mathrm{h.c.}}\,.
\end{equation}
Here, $\phi_{a,b,J}=\phi_{\mathrm{ZPF},(a,b,J)}/\phi_{0}$ denote
the amplitude participation ratios of the respective modes in the
junction, with $\phi_{\mathrm{ZPF},(a,b,J)}$ corresponding to the
zero point fluctuations of the respective modes as seen by the junction
and $\phi_{0}=\frac{\hbar}{2e}$ being the reduced superconducting
flux quantum \cite{girvinbook}. Combining the harmonic and anharmonic
portions with the drive term and assuming that $\hbar\o_{a,b,J},|\hbar\varepsilon(t)|\ll E_{J}$
(for all $t$) and that all mode frequencies are non-commensurate,
we expand the cosine to fourth order \cite{girvinbook} and obtain
our the Hamiltonian
\begin{equation}
H^{\prime}=H_{0}-{\textstyle \frac{1}{24}}E_{J}\hat{\varphi}^{4}+\hbar\left(\varepsilon(t)J+\varepsilon^{\star}(t)J^{\dg}\right)\,.
\end{equation}

We consider a three-tone drive,
\begin{equation}
\varepsilon(t)=\sum_{k=1}^{3}\epsilon_{pk}\exp(i\omega_{pk}t)\,.
\end{equation}
and apply a sequence of transformations which absorbs, one tone at
a time, the entire $\varepsilon$ drive into $\hat{\varphi}^{4}$,
the anharmonic part of the junction (see Ref.~\cite{Leghtas2014},
Supplementary Materials). Let us consider tone 1 and go into the rotating
frame defined by $J\rightarrow J\exp(-i\o_{p1}t)$. The drive term
corresponding to tone 1 is now time-independent, so let us displace
$J\rightarrow J-\frac{\epsilon_{p1}^{\star}}{\o_{J}-\o_{p1}}$ in
order to move that term into $\hat{\varphi}^{4}$. Finally, we move
out of the interaction picture using $J\rightarrow J\exp(i\o_{p1}t)$.
Due to the displacement, the other two tones $k\in\{2,3\}$ produce
time-dependent offset terms which are proportional to the identity;
we ignore such terms from now on. This procedure is then performed
sequentially for tones 2 and 3, yielding 
\begin{align}
H & =H_{0}-{\textstyle \frac{1}{24}}E_{J}[\hat{\Phi}(t)]^{4}\,,\label{eq:displaced_1}
\end{align}
where the new time-dependent phase difference is
\begin{equation}
\hat{\Phi}(t)=\phi_{a}a+\phi_{b}b+\phi_{J}\left(J+\sum_{k=1}^{3}\xi_{pk}\exp(i\omega_{pk}t)\right)+\text{h.c.}
\end{equation}
and $\xi_{pk}\propto\epsilon_{pk}$ is the displacement of the junction
mode due to the $k$th drive.

\begin{figure}[t]
\centering \includegraphics[width=0.8\columnwidth]{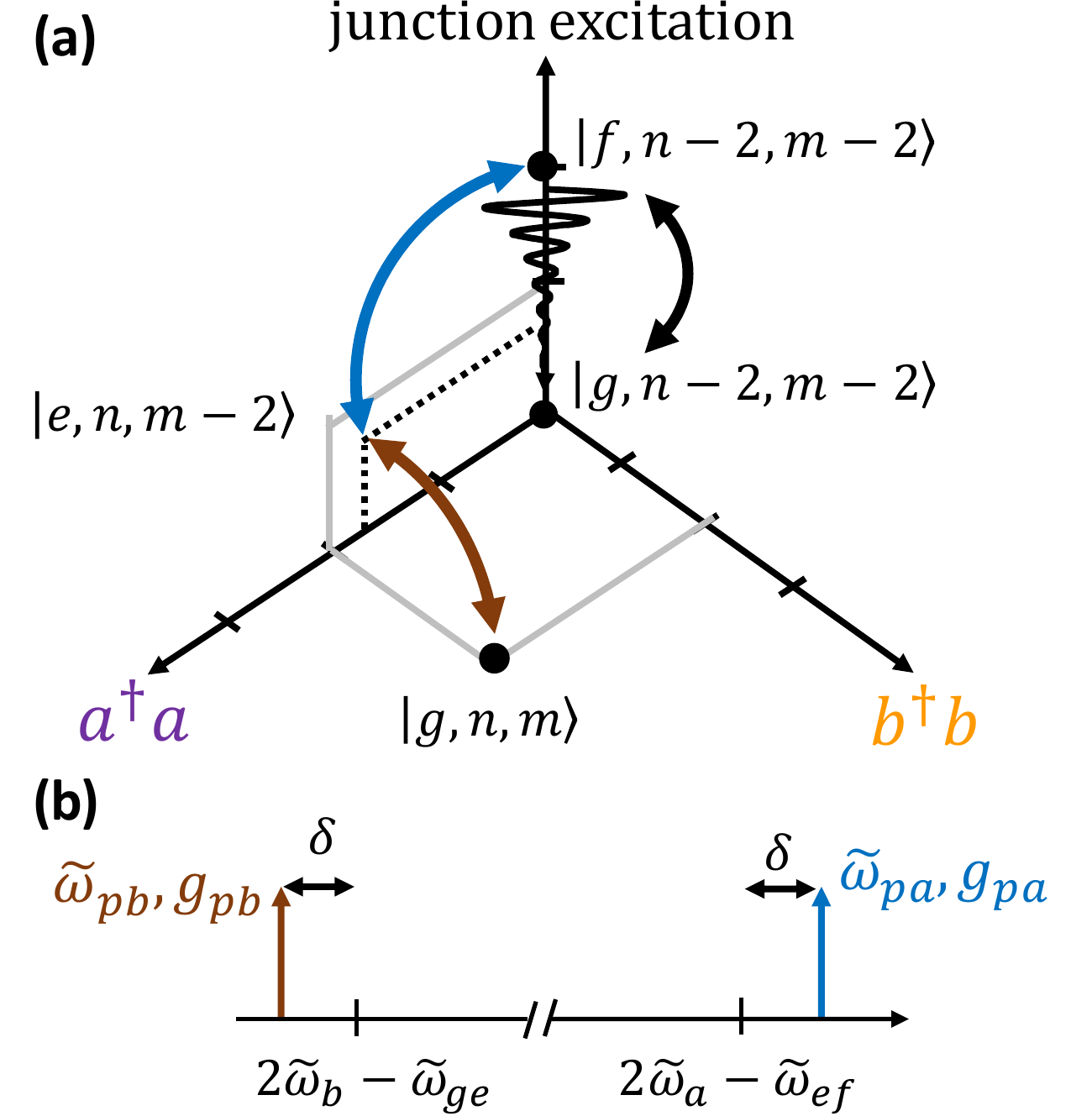} \caption{Schematic description of the cascading process. \textbf{(a)} The two-photon
exchange drives in frequency domain. The drives shown are detuned
from the respective resonance condition by a frequency $\delta$;
see Eqs.~(\ref{eq:detuning}-c). \textbf{(b)} Explanation of the
cascading process using a three-dimensional energy level description
of the system. The Fock-states of the high-Q cavities are denoted
by numbers and the lowest three eigenstates of the junction mode are
denoted by letters $g$, $e$ and $f$. The initial state is taken
to be $|g,n,m\protect\ket$. The first pump (brown) connects this
state with a virtual state detuned from the state $|{e,n,m-2}\protect\ket$
by $\delta$ (dashed line). The second pump (blue) connects this virtual
state with the state $|{f,n-2,m-2}\protect\ket$. Thus, a pair of
two-photon exchanges are combined to create a transition from $|{g,n,m}\protect\ket$
to $|{f,n-2,m-2}\protect\ket$ exchanging two-photons of each cavity
with the junction excitation. The effective two-photon dissipation
on each cavity is implemented by resetting the junction mode from
$|{f,n-2,m-2}\protect\ket$ to $|{g,n-2,m-2}\protect\ket$ (wavy arrow).
The simultaneous two-photon drive on both cavities comes from the
inverse process (black arrow), where a $gf/2$ drive on the junction
mode excites it from $|{g,n-2,m-2}\protect\ket$ to $|{f,n-2,m-2}\protect\ket$.
The off-resonant pumps then bring this state to $|{g,n,m}\protect\ket$.
The combination of these two processes yields the desired driven-dissipative
process $\protect\kt\mathcal{D}\left[a^{2}b^{2}-\gamma^{4}\right]$.\label{fig:cascading_explanation} }
\end{figure}

We now expand the $\Phi^{4}$ term in order to eventually tune the
drives $\{\o_{pk}\}_{k=1}^{3}$ such that our desired terms are selected
in a particular rotating frame. Normal-ordering the $\hat{\Phi}^{4}$-term
(Lamb- and Stark-)shifts the cavity frequencies $\o_{a,b}$ to new
frequencies $\tilde{\o}_{a,b}$ (which are here very different from
$\o_{a,b}$), so the rotating frame we pick is with respect to the
new frequencies. Regarding the junction, we consider only its first
three states $\{|g\ket,|e\ket,|f\ket\}$, defining transition frequencies
$\tilde{\omega}_{ge}$ ($\tilde{\omega}_{ef}$) between $|g\ket$
and $|e\ket$ ($|e\ket$ and $|f\ket$). We can absorb the cavity
shifts as well as any self-energy terms describing the junction's
first three levels into a noninteracting part 
\begin{equation}
\frac{\tilde{H}_{0}}{\hbar}=\tilde{\omega}_{a}a^{\dagger}a+\tilde{\omega}_{b}b^{\dagger}b+\tilde{\omega}_{ge}\hat{\sigma}_{ee}+\left(\tilde{\omega}_{ge}+\tilde{\omega}_{ef}\right)\hat{\sigma}_{ff}\,,\label{eq:hnt}
\end{equation}
where $\hat{\sigma}_{kl}=|{l}\ket\bra{k}|$ and $|{k,l}\ket$ are
junction states. 

Let us consider going into the rotating frame with respect to $\tilde{H}_{0}$
in order to select the desired terms
\begin{equation}
\{b^{\dg2}\hat{\sigma}_{ge},\,\aa^{\dg2}\hat{\sigma}_{ef}\,,\hat{\sigma}_{gf}\}\label{eq:desire}
\end{equation}
in the anharmonic term $\hat{\Phi}^{4}$. The drive $\o_{p1}$ ($\o_{p2}$)
is used to introduce an exchange of two photons of cavity $b$ ($a$)
with the excitation of the junction mode from the $g$ ($e$) to the
$e$ ($f$) state. The drive frequencies are thus\begin{subequations}
\begin{align}
\omega_{p1} & =2\tilde{\omega}_{b}-\tilde{\omega}_{ge}-\delta\label{eq:detuning}\\
\omega_{p2} & =2\tilde{\omega}_{a}-\tilde{\omega}_{ef}+\delta\\
\omega_{p3} & ={\textstyle \half}(\tilde{\omega}_{ge}+\tilde{\omega}_{ef})\,.
\end{align}
\end{subequations}The frequencies are detuned by $\pm\d$ {[}dotted
line in Fig.~\ref{fig:cascading_explanation}(a){]} such that together
they produce an exchange of two photons in each cavity with the $ef$
excitation of the junction. The third drive $\o_{p3}$ selects the
term $\hat{\s}_{gf}$ that, in presence of dissipation, will translate
into a simultaneous two-photon drive on both cavities and produce
$\ft$ with a nonzero $\g$. The rest of the junction levels are ignored
under the assumption that the anharmonicity of the junction mode is
much greater than the detuning,
\begin{equation}
\tilde{\omega}_{ge}-\tilde{\omega}_{ef}\gg\delta\,.\label{eq:approx}
\end{equation}
A sketch of all this is shown in Fig.~\ref{fig:cascading_explanation}(b).

Assuming the above approximations and the values of the drive tones,
we keep only the diagonal terms and our desired two-photon exchange
terms (\ref{eq:desire}) in $H$ from Eq.~(\ref{eq:displaced_1}).
In the rotating frame of $\tilde{H}_{0}-\d\hat{\s}_{ee}$, this yields
our time-independent system Hamiltonian
\begin{align}
\frac{H_{\mathrm{sys}}}{\hbar} & =\begin{pmatrix}0 & g_{1}^{\star}b^{\dg2} & \epsilon_{gf}^{\star}\\
g_{1}b^{2} & \d & g_{2}^{\star}a^{\dg2}\\
\epsilon_{gf} & g_{2}a^{2} & 0
\end{pmatrix}-\frac{H_{\text{anhrm}}}{\hbar}\,,\label{eq:sys_hamiltonian}
\end{align}
where the $3\times3$ matrix acts on the junction basis $\{|g\ket,|e\ket,|f\ket\}$.
The remaining piece $H_{\text{anhrm}}$ contains all the non-rotating
anharmonic terms of $H$,
\begin{align}
\frac{H_{\text{anhrm}}}{\hbar} & ={\textstyle \half}(\chi_{aa}a^{\dagger2}a^{2}+\chi_{bb}b^{\dagger2}b^{2})+\chi_{ab}a^{\dagger}ab^{\dagger}b\nonumber \\
 & +(\chi_{aJ}a^{\dagger}a+\chi_{bJ}b^{\dagger}b)(\hat{\sigma}_{ee}+2\hat{\sigma}_{ff})\,,
\end{align}
where cavity self- and cross-Kerr terms are $\chi_{qq}=\frac{E_{J}}{2\hbar}\phi_{q}^{4}$
and $\chi_{pq}=\frac{E_{J}}{\hbar}\phi_{p}^{2}\phi_{q}^{2}$ for $p\neq q$,
respectively. (The $\sqrt{2}$ difference between $g_{1,2}$ comes
from the differing strengths of the $ge$ and $ef$ transitions.)
The new drive strengths are\begin{subequations}
\begin{align}
g_{1}= & -\frac{E_{J}}{2\hbar}\phi_{b}^{2}\phi_{J}^{2}\xi_{p1}\\
g_{2}= & -\frac{E_{J}}{\sqrt{2}\hbar}\phi_{a}^{2}\phi_{J}^{2}\xi_{p2}\\
\epsilon_{gf}= & \frac{E_{J}}{\sqrt{2}\hbar}\phi_{J}^{4}\,.
\end{align}
\end{subequations}

\subsubsection{Eliminating the junction}

Here we show that an effective $f\rightarrow g$ transition through
the detuned $|e\ket$ state comes at the price of the four photon
loss $\aa^{2}\bb^{2}$, as desired. This is already hinted in Eq.~(\ref{eq:sys_hamiltonian}).
We first eliminate the $|e\ket$ state and then, with the help of
dissipation, the entire junction. We use second-order perturbation
theory for the first part and adiabatic elimination for the second,
but note that both parts can also be done either entirely using adiabatic
elimination or using a generalization of the RWA \cite{Mirrahimi2015}
(similar to the analogous realization of scheme I \cite{Mundhada2017}).
Since the two parts are sequential and not simultaneous, we have to
perform the second part --- adiabatic elimination --- on the timescales
$t\gtrsim\nicefrac{\d}{|g_{1,2}|^{2}},\nicefrac{\d}{|\e_{gf}|^{2}}$
during which the perturbation theory is valid.

For the first part, we perform degenerate perturbation theory on the
$\{|g\ket,|f\ket\}$ subspace. Let $-\d|e\ket\bra e|-H_{\text{anhrm}}$
be the unperturbed part of $H_{\mathrm{sys}}$ (\ref{eq:sys_hamiltonian}),
with the remaining parts $V$ constituting the perturbation. Letting
$P=\hat{\s}_{gg}+\hat{\s}_{ee}$ and adding the first-order ($PVP$)
and second-order ($PVH^{-1}VP$) corrections to the $\{|g\ket,|f\ket\}$
subspace yields
\begin{align}
\!\!H_{\text{pt}} & =\begin{pmatrix}0 & \epsilon_{gf}^{\star}\\
\epsilon_{gf} & 0
\end{pmatrix}+\frac{1}{\d}\begin{pmatrix}|g_{1}|^{2}\bb^{2}\bb^{\dg2} & g_{1}^{\star}g_{2}^{\star}\aa^{\dg2}\bb^{\dg2}\\
g_{1}g_{2}\aa^{2}\bb^{2} & |g_{2}|^{2}\aa^{2}\aa^{\dg2}
\end{pmatrix}-H_{\text{anhrm}},
\end{align}
where $H^{-1}=\d^{-1}|e\ket\bra e|$ is a pseudoinverse and the $2\times2$
matrix acts on the $\{|g\ket,|f\ket\}$ subspace. The $\aa^{2}\bb^{2}\hat{\s}_{gf}$
term gives the expected simultaneous two-photon exchange coupled to
the $g\leftrightarrow f$ transition of the junction mode.

The second part uses the junction's intrinsic dissipation, which we
assume is of Lindblad form. Within the $\{|g\ket,|f\ket\}$ subspace,
we have
\begin{equation}
{\cal L}_{gf}(\r)=-i[H_{\text{pt}},\r]+\G_{fg}\DD[\hat{\s}_{fg}](\r)\,.
\end{equation}
We proceed to adiabatically eliminate the $f$-state by the standard
procedure \{e.g., Ref.~\cite{Leghtas2014}, Supplementary Materials;
see also \cite{carmichael2,Verstraete2009}\}. In other words, we
turn the Hamiltonian $F\hat{\s}_{gf}+\text{h.c.}$, where here $F=\frac{g_{1}g_{2}}{\d}\aa^{2}\bb^{2}+\epsilon_{gf}$
is an operator on the two cavities, into a dissipator with jump operator
$F$. We assume the junction is lossy, i.e., $\G_{fg}$ is much greater
than all of the other parameters in $H_{\text{pt}}$, and derive the
effective dynamics of the two cavities under the assumption that the
junction is perturbed away from $|g\ket$ by a small parameter. This
yields the two-cavity Lindbladian 
\begin{equation}
{\cal L}_{\text{cav}}(\r)=-i[H_{\text{cav}},\r]+\kt\lt(\r)\label{eq:eom}
\end{equation}
with Hamiltonian
\begin{equation}
H_{\text{cav}}=\left({\textstyle \frac{1}{\d}|g_{1}|^{2}-\half\chi_{bb}}\right)b^{\dagger2}b^{2}-{\textstyle \half}\chi_{aa}a^{\dagger2}a^{2}-\chi_{ab}a^{\dagger}ab^{\dagger}b\label{eq:hcav}
\end{equation}
and dissipator parameters
\begin{equation}
\kt=\frac{4|g_{1}g_{2}|^{2}}{\G_{fg}\d^{2}}\,\,\,\,\,\,\,\,\,\,\,\,\,\text{and}\,\,\,\,\,\,\,\,\,\,\,\,\,\g=\left(-\frac{\e_{gf}\d}{g_{1}g_{2}}\right)^{1/4}\,.
\end{equation}

\subsubsection{Leading-order error processes}

While we have obtained our dissipator above, the Hamiltonian $H_{\text{cav}}$
(\ref{eq:hcav}) unfortunately carries undesirable anharmonic terms.
However, we have the ability to cancel the anharmonicity of the $b$
mode by adjusting the parameters to achieve $|g_{1}|^{2}/\delta=\chi_{bb}/2$.
Note that the anharmonicity of the $a$ mode remains unchanged.

The above procedure eliminating the junction unfortunately carries
with it one more leading-order error, which we have omitted previously
for simplicity. Physically, this corresponds to the ability of the
junction state to decay from $|e\ket$ back to $|g\ket$ instead of
following through the virtual transition to $|f\ket$ {[}see Fig.~\ref{fig:cascading_explanation}(a){]}.
After elimination of $|e\ket$, this induces a two-photon loss in
the $b$-mode. This process was not accounted for in our previous
derivations because we had not introduced dissipation until after
we eliminated $|e\ket$. If we include the dissipation $\Gamma_{eg}\DD[\hat{\sigma}_{eg}]$
and perform adiabatic elimination of $|e\ket$, we find that ${\cal L}_{\text{cav}}$
(\ref{eq:eom}) gains the term
\begin{equation}
{\cal L}_{\text{err}}(\r)=\frac{|g_{1}|^{2}}{\d^{2}}\G_{eg}\DD[\bb^{2}](\r)\,.\label{eq:errror}
\end{equation}
However, this two-photon dissipation can be corrected if we engineer
a device that can measure at least five distinct values of $\D$ (see
Fig.~\ref{fig:displacement_readout} and the next Subsection). This
is a key difference between the analogous experimental realization
of scheme I \cite{Mundhada2017} and the design here. While the analogous
leading-order dissipative error leads to uncorrectable logical errors
for cat codes, here such an error can in principle be corrected.

In terms of the additional $\aa$-mode anharmonicity in $H_{\text{cav}}$
(\ref{eq:hcav}) and the inherited $\bb$-mode two-photon dissipation
${\cal L}_{\text{err}}$ (\ref{eq:errror}), the $a$ and $b$ modes
are not on equal footing. This asymmetry has been built into the dynamics
owing to the fact that $b^{2}$ couples to the $ge$ transition and
$a^{2}$ couples to the $ef$ transition of the junction mode. However,
by carefully canceling one of the anharmonicities and discrete error-correction,
we have shown that the undesirable effect of this asymmetry can be
minimized.

\section{Realizing discrete QEC against photon loss\label{sec:Discrete-QEC-against}}

\begin{figure}[t]
\centering \includegraphics[width=0.8\columnwidth]{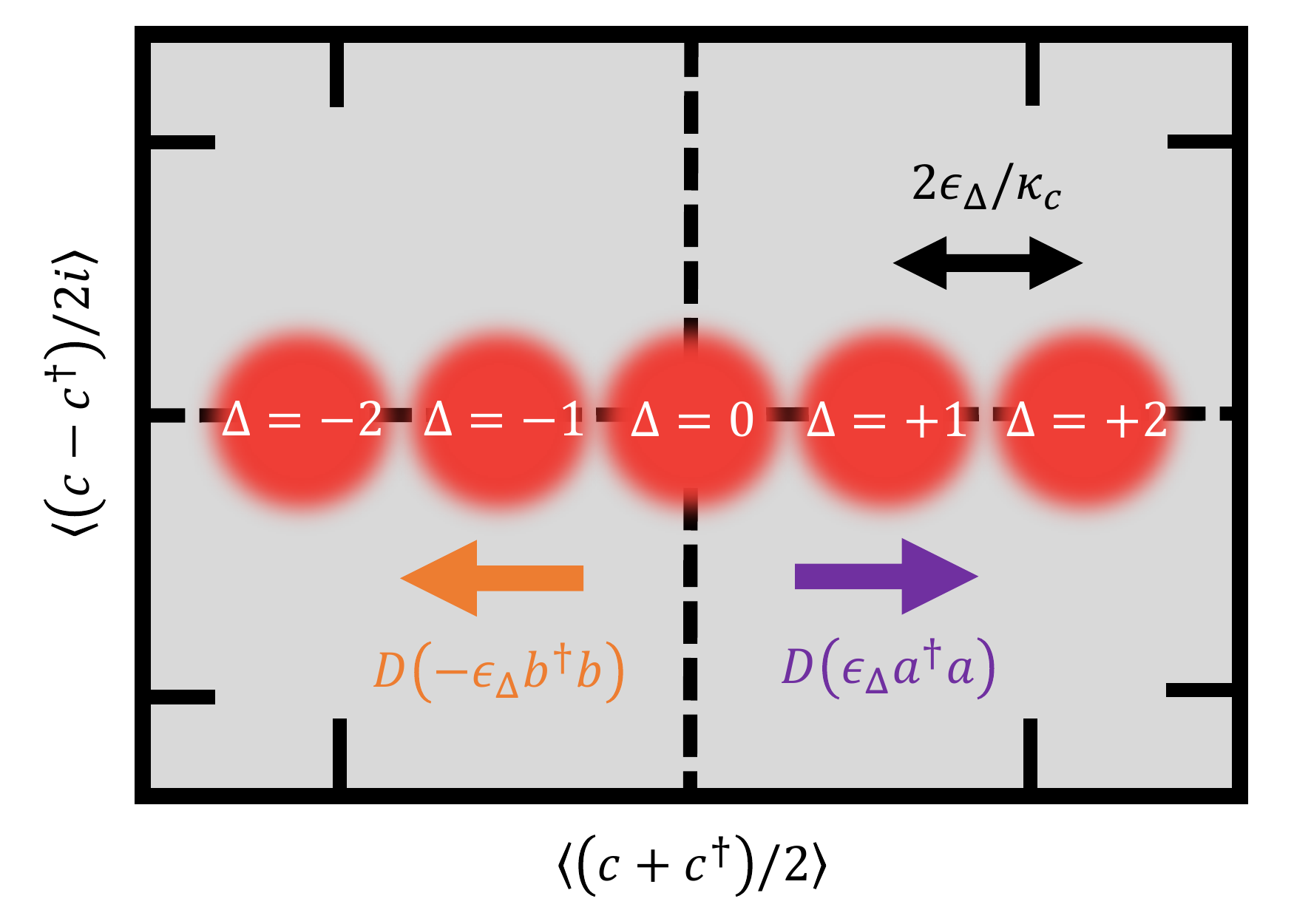} \caption{Principle of the error syndrome measurement. Pumping both the junction
modes independently at the resonant frequency of the low-Q readout
cavity $c$, we make the terms $g_{a}a^{\dagger}a(c+c^{\dagger})$
and $g_{b}b^{\dagger}b(c+c^{\dagger})$ resonant in the effective
system Hamiltonian. Here, the couplings $g_{a}$ and $g_{b}$ depend
on the physical parameters of the system and on the applied pumps
(see text). This exerts two displacement forces on the readout cavity
(purple arrow for $a$ and orange arrow for $b$) which are respectively
proportional to the photon numbers in the high-Q cavities. Adjusting
the magnitudes and phases of the pumps so that $g_{a}=-g_{b}$ results
in a total $a,b$-cavity-dependent displacement on cavity $c$, allowing
for direct measurement of the error syndrome $\protect\di$. Note
that this is only a sketch since we have ignored the $\chi$-, $g_{s}$-,
and $g_{c}$-dependent terms in Eq.~(\ref{eq:final2}).\label{fig:displacement_readout} }
\end{figure}

In this section, we propose a way to realize discrete QEC against
photon loss. The proposal involves using the four-wave mixing capabilities
of two Josephson junction modes to link the displacement of a low-Q
resonator mode to the photon number difference between the two high-Q
modes.

As shown in Fig.~\ref{fig:displacement_readout}, we have two junction
modes $J_{1,2}$ coupling the two cavities $\aa,\bb$ to a low-Q readout
cavity $c$. It is assumed that cavity $a$ couples only to junction
$J_{1}$ and cavity $b$ couples only to junction $J_{2}$. Both junctions
couple to cavity $c$. It is assumed that the two junction modes are
isolated from each other and can be driven independently. The two
junctions and cavity $c$ are driven with drives parameterized by
$\e_{1,2,c}$, respectively. Assuming $|\hbar\e_{1,2,c}|\ll E_{J1,2}$
and expanding the anharmonic parts of the two junctions yields
\begin{align}
H & =H_{0}-{\textstyle \frac{1}{24}}E_{J1}\hat{\varphi}_{1}^{4}-{\textstyle \frac{1}{24}}E_{J2}\hat{\varphi}_{2}^{4}\\
 & +\hbar\exp(i\tilde{\omega}_{c}t)\left(\epsilon_{c}c+\epsilon_{1}J_{1}+\epsilon_{2}J_{2}\right)+\text{h.c.}\,,\nonumber 
\end{align}
where $\frac{H_{0}}{\hbar}=\omega_{a}a^{\dagger}a+\omega_{b}b^{\dagger}b+\omega_{c}c^{\dagger}c+\omega_{J1}J_{1}^{\dagger}J_{1}+\omega_{J2}J_{2}^{\dagger}J_{2}$
is the harmonic part, the phase differences across the junctions $k\in\{1,2\}$
are\begin{subequations}
\begin{align}
\hat{\varphi}_{1}= & \phi_{a1}a+\phi_{c1}c+\phi_{1}J_{1}+\mathrm{\mathrm{h.c.}}\,,\\
\hat{\varphi}_{2}= & \phi_{b2}a+\phi_{c2}c+\phi_{2}J_{2}+\mathrm{\mathrm{h.c.}}\,,
\end{align}
\end{subequations}and $\phi$ are the amplitude participation ratios.
The two junctions are both independently driven at the frequency $\tilde{\omega}_{c}$
of the low-Q cavity, which we set to be the shifted frequency of mode
$c$ after normal ordering. We also apply a direct resonant drive
on the low-Q cavity of strength $\epsilon_{c}$; the importance of
this drive will be clear in the next couple of steps.

We absorb the $J_{1,2}$-drives on the junction modes in the respective
anharmonicities {[}similar to Eq.~(\ref{eq:displaced_1}){]}, but
leave the $c$-cavity drive untouched. This yields
\begin{align}
H' & =H_{0}-{\textstyle \frac{1}{24}}E_{J1}[\hat{\Phi}_{1}(t)]^{4}-{\textstyle \frac{1}{24}}E_{J2}[\hat{\Phi}_{2}(t)]^{4}\\
 & +\hbar\epsilon_{c}\exp(i\tilde{\omega}_{c}t)c+\text{h.c.}\,,\nonumber 
\end{align}
where the time-dependent phase differences are\begin{subequations}
\begin{align}
\hat{\Phi}_{1}(t)= & \phi_{a1}a+\phi_{c1}c+\phi_{1}J_{1}+\phi_{1}\xi_{1}\exp(i\tilde{\omega}_{c}t)+\mathrm{\mathrm{h.c.}}\\
\hat{\Phi}_{2}(t)= & \phi_{b2}b+\phi_{c2}c+\phi_{2}J_{2}+\phi_{2}\xi_{2}\exp(i\tilde{\omega}_{c}t)+\mathrm{\mathrm{h.c.}}.
\end{align}
\end{subequations}and $\xi_{k}$ are the displacements of the junction
modes due to the respective drives. Finally, we normal-order the anharmonicities,
go into a rotating frame with respect to the Lamb- and Stark-shifted
shifted harmonic part, 
\begin{align}
\frac{\tilde{H}_{0}}{\hbar} & =\tilde{\omega}_{a}a^{\dagger}a+\tilde{\omega}_{b}b^{\dagger}b+\tilde{\omega}_{c}c^{\dagger}c+\tilde{\omega}_{J1}J_{1}^{\dagger}J_{1}+\tilde{\omega}_{J2}J_{2}^{\dagger}J_{2}\,,
\end{align}
and keep only the non-rotating terms. Since the only drive frequency
is $\tilde{\o}_{c}$, the only off-diagonal time-independent terms
are those for which the number of $c$ terms is equal to the number
of $\xi_{1,2}^{\star}$ terms plus the number of $c^{\dg}$ terms
(and their Hermitian conjugates). We also assume that the junction
modes $J_{1,2}$ are never resonantly driven and hence are never populated.
Therefore, for the sake of compactness, we drop all the diagonal terms
involving the $J_{k}^{\dagger}J_{k}$ operator. The system Hamiltonian
becomes 
\begin{align}
\frac{H_{\mathrm{sys}}}{\hbar}= & -\chi_{ac}a^{\dagger}ac^{\dagger}c-\chi_{bc}b^{\dagger}bc^{\dagger}c-\sum_{r=a,b,c}\frac{\chi_{rr}}{2}r^{\dagger2}r^{2}\nonumber \\
 & -\left(\epsilon_{c}+g_{\mathrm{dir}}+g_{s}c+\sum_{r=a,b,c}g_{r}r^{\dagger}r\right)c+\mathrm{\mathrm{h.c.}}\,,\label{eq:final}
\end{align}
where the couplings are\begin{subequations}
\begin{align}
g_{\mathrm{dir}} & =\frac{1}{2\hbar}\sum_{k=1,2}E_{Jk}\phi_{k}^{3}\phi_{ck}|\xi_{k}|^{2}\xi_{k}\label{eq:aa}\\
g_{s} & =\frac{1}{4\hbar}\sum_{k=1,2}E_{Jk}\phi_{ck}^{2}\phi_{k}^{2}\xi_{k}^{2}\\
g_{a} & =\frac{E_{J1}}{\hbar}\phi_{a1}^{2}\phi_{c1}\phi_{1}\xi_{1}\\
g_{b} & =\frac{E_{J2}}{\hbar}\phi_{b2}^{2}\phi_{c2}\phi_{2}\xi_{2}\\
g_{c} & =\frac{1}{2\hbar}\sum_{k=1,2}E_{Jk}\phi_{ck}^{3}\phi_{k}\xi_{k}\label{eq:cc}
\end{align}
\end{subequations}

The remaining step now is to tune the second line of Eq.~(\ref{eq:final})
such that we obtain the term $\di c+\text{h.c.}$. We can adjust the
amplitude and the phase of $\xi_{1,2}$ such that
\begin{equation}
|g_{a}|=|g_{b}|=\epsilon_{\Delta}\,\,\,\mbox{ and }\,\,\,\mathrm{arg}(g_{a})=\mathrm{arg}(g_{b})+\pi=0\,.\label{eq:matching_condition}
\end{equation}
For $\phi_{a1}\approx\phi_{b2}$, $\phi_{c1}\approx\phi_{c2}$, $E_{J1}\approx E_{J2}$,
and $\phi_{1}\approx\phi_{2}$, the magnitude of the terms in Eqs.~(\ref{eq:aa})
and (\ref{eq:cc}) becomes minimal. The remaining $g_{\mathrm{dir}}$
can be canceled by setting $\epsilon_{c}=-g_{\mathrm{dir}}$, yielding
\begin{align}
\frac{H_{\mathrm{sys}}}{\hbar}= & -\chi_{ac}a^{\dagger}ac^{\dagger}c-\chi_{bc}b^{\dagger}bc^{\dagger}c-\sum_{r=a,b,c}\frac{\chi_{rr}}{2}r^{\dagger2}r^{2}\nonumber \\
 & -\left(g_{s}c+g_{c}c^{\dagger}c+\e_{\D}\di\right)c+\mathrm{\mathrm{h.c.}}\label{eq:final2}
\end{align}
Hence we can engineer the displacement term of the low-Q cavity mode
$c$ to be proportional to the error syndrome measurement operator
$\hat{\Delta}=\bb^{\dg}\bb-\aa^{\dg}\aa$ (\ref{eq:diff}).

In an idealized scenario (with the $\chi$-, $g_{s}$-, and $g_{c}$-dependent
terms ignored) and in presence of dissipation with rate $\kappa_{c}$,
the steady state of the low-Q cavity $c$ is a coherent state $|{\nu_{\D}}\ket$,
where $\nu_{\D}=2\epsilon_{\Delta}\Delta/\kappa_{c}$ is the complex
amplitude of the coherent state. We sketch a simplified IQ phase diagram
in Fig.~\ref{fig:displacement_readout}, showing this state for different
values of $\Delta$. If $\epsilon_{\Delta}\ge\kappa_{c}$, the average
photon number occupation of the steady state is given by $|\nu_{\D}|^{2}\ge4\Delta^{2}$.
This should enable us to perform single-shot measurements of the error
syndrome using typical heterodyne detection of the signal coming out
of the low-Q cavity $c$ by employing a quantum limited amplifier,
e.g., the Josephson Parametric Converter \citep{Hatridge2013}. After
measurement, we can continue tracking the shifts (in the spirit of
Pauli frames) and take it into account when decoding.

We conclude by commenting on the $\chi$-, $g_{s}$-, and $g_{c}$-dependent
terms in Eq.~(\ref{eq:final2}). Such terms will necessarily distort
the idealized signal, as the full linear part of the cavity $c$ Hamiltonian
corresponds to an oscillator displaced by $\e_{\D}\di$ and squeezed
by $g_{s}$ and there are several nonlinearities in the system. While
these corrections will make the states corresponding to different
values of $\di$ harder to resolve, it will nevertheless be possible
since the states manifestly occupy different portions of phase space.
We thus leave further optimization of this scheme to future work.

\begin{figure}[t]
\centering \includegraphics[width=0.9\columnwidth]{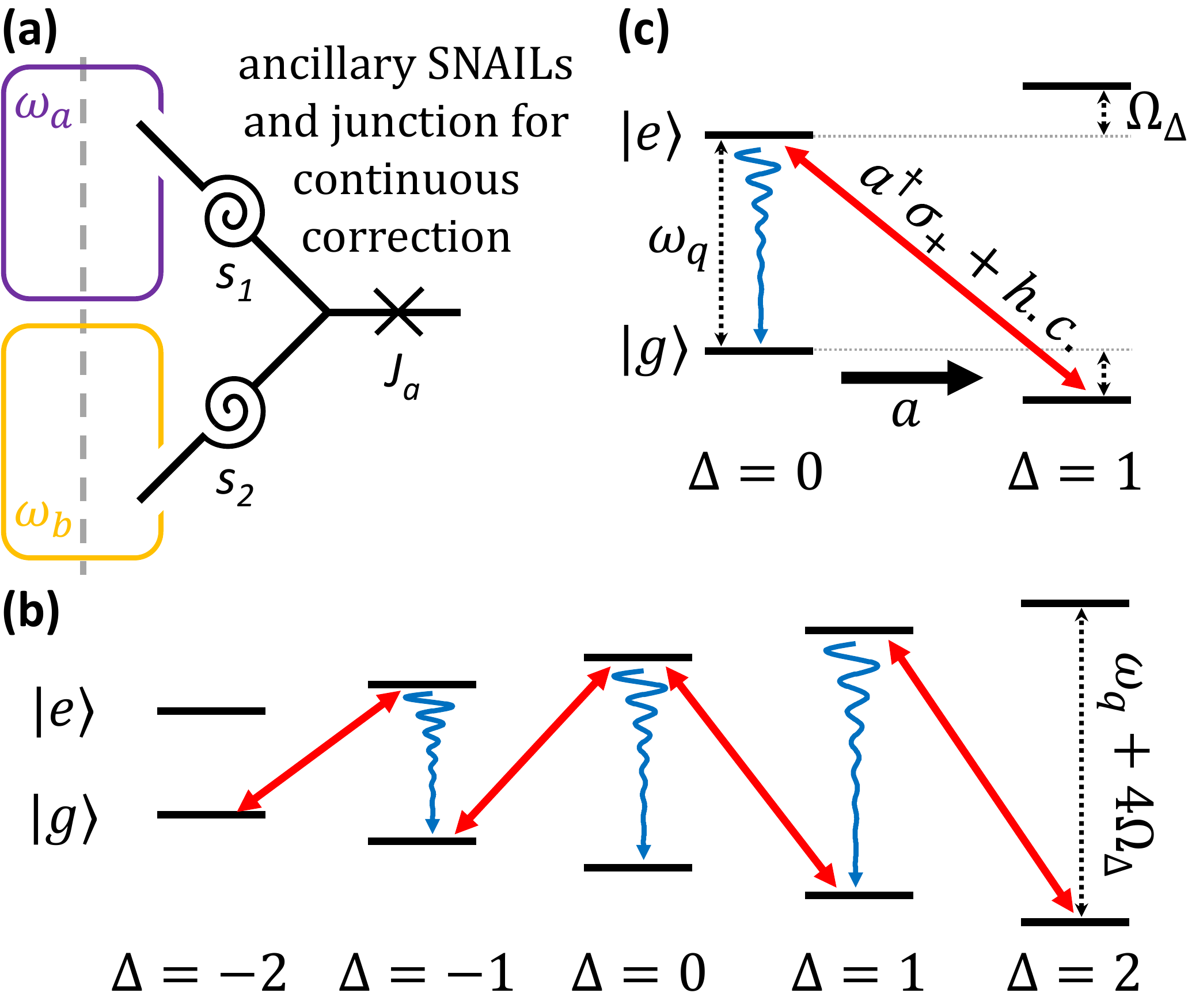} \caption{\label{fig:autonomous}\textbf{(a)} Sketch of the continuous QEC circuit,
which substitutes the discrete QEC circuit to the right the two cavities
$\protect\o_{a},\protect\o_{b}$ in Fig.~\ref{fig:complete_setup}(a).
The spiral circuit elements are SNAILs \cite{Frattini2017}, which
are three-wave mixers that allow one to couple the cavities of the
junction $J_{a}$ without any undesired Kerr nonlinearities. \textbf{(b)}
The telescope of levels due to the engineered cross-Kerr interaction
term $\Omega_{\protect\D}\protect\di\protect\s_{z}$ in addition to
the usual junction term $\protect\half\protect\o_{q}\protect\s_{z}$
(see Sec.~\ref{sec:Continuous-QEC-against}), which induces a $\protect\di$-dependent
junction frequency. For each pair of levels, the two cavities are
in the subspace of fixed $\protect\D$ while the junction is either
in $|g\protect\ket$ for the bottom level or in $|e\protect\ket$
for the top. \textbf{(c)} Sketch of the step correcting a single loss
event in cavity $\protect\aa$. Upon the event (thick arrow), the
logical qubit stored in the two cavities is transferred into the subspace
$\protect\tp_{\protect\D=1}$ while the junction remains in $|g\protect\ket$.
A pulse (two-headed arrow) then drives the junction to its excited
state $|e\protect\ket$ while simultaneously applying $\protect\aa^{\protect\dg}$
to the cavity system, thereby returning the logical state back to
$\protect\tp_{\protect\D=0}$. The junction then decays from $|e\protect\ket$
to $|g\protect\ket$ (wavy arrow) to complete the process.}
\end{figure}

\section{Realizing continuous QEC against photon loss\label{sec:Continuous-QEC-against}}

In this section, we propose a way to continuously correct against
photon loss. Since we only need one form of QEC against loss, this
proposal is meant to substitute the discrete QEC proposal of the previous
section. Instead of coupling the cavities storing the pair-cat qubit
to an ancillary cavity, we couple them to an ancillary junction mode
through two three-wave mixers dubbed Superconducting Nonlinear Asymmetric
Inductive eLements (SNAILs) \cite{Frattini2017}. Schematically, we
substitute the circuit to the right of the two cavities $\o_{a},\o_{b}$
in Fig.~\ref{fig:complete_setup}(a) with the circuit in Fig.~\ref{fig:autonomous}(a).
The main idea is to compensate single photon losses in either cavity
by adding photon gain jump operators that are conditional on $\D=\pm1$.
For mode $\aa$, the jump operator is
\begin{equation}
F(1)=\aa^{\dg}\tp_{\D=1}\,,\label{eq:autojumpplus1}
\end{equation}
and similarly $F(-1)=\bb^{\dg}\tp_{\D=-1}$ for mode $\bb$. A difference
between these jumps and the ideal continuous QEC jumps (\ref{eq:loss2})
with $\D=\pm1$ is the extra raising operator. Even though this recovery
implements first-order dephasing errors $\aa\aa^{\dg}$ or $\bb\bb^{\dg}$
when combined with the preceding loss event, these have negligible
effect on the code states for sufficiently large $\g$ (see Subsec.~\ref{subsec:Pair-cat-code-error}).
We now sketch this proposal, focusing on correction for $\Delta=1$
(mode $\aa$ loss) only {[}sketched in Fig.~\ref{fig:complete_setup}(b){]}.
Note that this proposal is similar in spirit to a continuous QEC proposal
for scheme I {[}\citealp{cohenthesis}, Sec.~4.2.2{]} and is extendable
to higher $\D$ in natural fashion {[}see Fig.~\ref{fig:autonomous}(b){]}.
In the exposition below, we apply perturbation theory sequentially.
However, all terms can be introduced simultaneously in a more involved
calculation that yields the same low-order result with higher order
corrections.

The SNAIL three-wave mixer provides Jaynes-Cummings type couplings
between the junction and the cavities without any additional Kerr
nonlinearities. We thus begin with a two-mode driven Jaynes-Cummings
Hamiltonian
\begin{align}
H & =\o_{a}\aa^{\dg}\aa+\o_{b}\bb^{\dg}\bb+{\textstyle \frac{1}{2}\o_{q}}\s_{z}\\
 & +g(\aa\s_{+}e^{-i\o_{pa}t}+b\s_{+}e^{-i\o_{pb}t}+h.c.)\nonumber 
\end{align}
with cavity frequencies $\o_{a,b}$, Josephson junction frequency
$\o_{q}$, real pump drive $g$, and pump frequencies $\o_{pa},\o_{pb}$.
Setting $\o_{pa}=\o_{a}-\o_{q}+\d$ and $\o_{pb}=\o_{b}-\o_{q}-\d$
and going into the rotating frame with respect to $H_{0}=\o_{a}\aa^{\dg}\aa+\o_{b}\bb^{\dg}\bb+{\textstyle \frac{1}{2}\o_{q}}\s_{z}+\d\di$
for $\d>0$. The pump strengths are then set so we are in the dispersive
regime: $g\bra k^{\dg}k\ket\ll\d$ with $\bra k^{\dg}k\ket$ being
the average occupation number in cavity $k\in\{a,b\}$ {[}\citealp{scully_book},
Sec.~19.3{]}. In the dispersive limit, the Hamiltonian becomes $\frac{g^{2}}{\d}\di\s_{z}\equiv\Omega_{\D}\di\s_{z}$.
This $\di$-dependent junction frequency is responsible for the telescope
of levels in Fig.~\ref{fig:autonomous}(b). For each pair of levels,
the two cavities are in the subspace of fixed $\D$ while the junction
is either in $|g\ket$ for the bottom level or in $|e\ket$ for the
top. Note that we can use this Hamiltonian to perform error syndrome
readout, but this scheme is slower than the one from Sec.~\ref{sec:Discrete-QEC-against}
since the readout here would have to be performed bit by bit. Instead
of utilizing this for discrete readout, we now describe how to continuously
perform the correction operation.

Since the junction frequency depends on the error syndrome, we can
add photons to the cavity selectively depending on $\di$. In particular,
for $\Delta=1$, we utilize the SNAILs to couple the junction to cavity
$\aa$ via the (counter-rotating) term $\aa^{\dg}\s_{+}\exp(-i\o_{CR}t)$
with frequency $\o_{CR}$ in the rotating frame with respect to $H_{0}$.
We set $\o_{CR}=3\O_{\D}$, the frequency of the effective two-level
system at $\D=1$ from Fig.~\ref{fig:autonomous}(b). The Hamiltonian
is then
\begin{equation}
H_{1}=\Omega_{\D}\di\s_{z}+g_{\D}(\aa^{\dg}\s_{+}e^{-3i\Omega_{\D}t}+h.c.)\,,
\end{equation}
with $g_{\D}\ll\O_{\D}\ll\d$. Going into the rotating frame with
respect to $\Omega_{\D}\di\s_{z}$ yields the desired transition
\begin{equation}
\tilde{H}=g_{\D}\left(\aa^{\dg}\s_{+}e^{2i\Omega_{\D}(\di-1)t}+h.c.\right)\,.\label{eq:shift}
\end{equation}
Furthermore, assuming the junction decays with jump operator $\sqrt{\G}\s_{-}$
and adiabatically eliminating the junction yields a dissipator with
jump operator $\frac{4g_{\D}^{2}}{\G}F(1)$ (\ref{eq:autojumpplus1}).
Thus, we have our desired result to first-order in the RWA.

\section{Conclusion\label{sec:Conclusion}}

In a non-trivial extension of cat-codes to multiple modes, we introduce
a family of two-mode continuous-variable codes based on pair-coherent
states (also known as Barut-Girardello states). We analyze which errors
the code can correct and extend single-mode cat-code gates to this
paradigm. We provide several experimental realizations of the full
error-correction scheme associated with this code, including continuous
error-correction based on reservoir engineering and discrete (i.e.,
measurement-based) error-correction based on measurements of the occupation
number difference between the two modes. We introduce ways to completely
visualize certain two-mode states in a two-dimensional complex plane,
avoiding the need to take cross-sections of the states' four-dimensional
Wigner functions. An extension of the codes to multiple modes makes
contact with the stabilizer formalism from multi-qubit error correction
and yields codes which can simultaneously correct against single-mode
losses and gains.

\phfnoteRestoreDefs{origcmds}
\begin{acknowledgments}
The authors thank Mazyar Mirrahimi and Kyungjoo Noh for fruitful discussions.
We acknowledge support from the ARL-CDQI, ARO (Grants No. W911NF-14-1-0011
and No. W911NF-14-1-0563), ARO MURI (W911NF-16-1-0349), NSF (EFMA-1640959),
AFOSR MURI (FA9550-14-1-0052 and No. FA9550-15-1-0015), the Alfred
P. Sloan Foundation (BR2013-049), the Packard Foundation (2013-39273),
and the Walter Burke Institute for Theoretical Physics at Caltech.
\end{acknowledgments}

\bibliographystyle{apsrev4-1t}
\bibliography{C:/Users/russi/Documents/library}

\begin{thebibliography}{123}%
\makeatletter
\providecommand \@ifxundefined [1]{%
 \@ifx{#1\undefined}
}%
\providecommand \@ifnum [1]{%
 \ifnum #1\expandafter \@firstoftwo
 \else \expandafter \@secondoftwo
 \fi
}%
\providecommand \@ifx [1]{%
 \ifx #1\expandafter \@firstoftwo
 \else \expandafter \@secondoftwo
 \fi
}%
\providecommand \natexlab [1]{#1}%
\providecommand \enquote  [1]{``#1''}%
\providecommand \bibnamefont  [1]{#1}%
\providecommand \bibfnamefont [1]{#1}%
\providecommand \citenamefont [1]{#1}%
\providecommand \href@noop [0]{\@secondoftwo}%
\providecommand \href [0]{\begingroup \@sanitize@url \@href}%
\providecommand \@href[1]{\@@startlink{#1}\@@href}%
\providecommand \@@href[1]{\endgroup#1\@@endlink}%
\providecommand \@sanitize@url [0]{\catcode `\\12\catcode `\$12\catcode
  `\&12\catcode `\#12\catcode `\^12\catcode `\_12\catcode `\%12\relax}%
\providecommand \@@startlink[1]{}%
\providecommand \@@endlink[0]{}%
\providecommand \url  [0]{\begingroup\@sanitize@url \@url }%
\providecommand \@url [1]{\endgroup\@href {#1}{\urlprefix }}%
\providecommand \urlprefix  [0]{URL }%
\providecommand \Eprint [0]{\href }%
\providecommand \doibase [0]{http://dx.doi.org/}%
\providecommand \selectlanguage [0]{\@gobble}%
\providecommand \bibinfo  [0]{\@secondoftwo}%
\providecommand \bibfield  [0]{\@secondoftwo}%
\providecommand \translation [1]{[#1]}%
\providecommand \BibitemOpen [0]{}%
\providecommand \bibitemStop [0]{}%
\providecommand \bibitemNoStop [0]{.\EOS\space}%
\providecommand \EOS [0]{\spacefactor3000\relax}%
\providecommand \BibitemShut  [1]{\csname bibitem#1\endcsname}%
\let\auto@bib@innerbib\@empty
%</preamble>
\bibitem [{\citenamefont {Braunstein}\ and\ \citenamefont {van
  Loock}(2005)}]{Braunstein2005}%
  \BibitemOpen
  \bibfield  {author} {\bibinfo {author} {\bibfnamefont {S.~L.}\ \bibnamefont
  {Braunstein}}\ and\ \bibinfo {author} {\bibfnamefont {P.}~\bibnamefont {van
  Loock}},\ }\bibfield  {{Quantum information with continuous variables}}
  {\emph {\bibinfo {title} {{Quantum information with continuous variables}},\
  }}\href {http://journals.aps.org/rmp/abstract/10.1103/RevModPhys.77.513}
  {\bibfield  {journal} {\bibinfo  {journal} {Rev. Mod. Phys.}\ }\textbf
  {\bibinfo {volume} {77}} (\bibinfo {year} {2005})}\BibitemShut {NoStop}%
\bibitem [{\citenamefont {Weedbrook}\ \emph {et~al.}(2012)\citenamefont
  {Weedbrook}, \citenamefont {Pirandola}, \citenamefont
  {Garc{\'{i}}a-Patr{\'{o}}n}, \citenamefont {Cerf}, \citenamefont {Ralph},
  \citenamefont {Shapiro},\ and\ \citenamefont {Lloyd}}]{Weedbrook2012}%
  \BibitemOpen
  \bibfield  {author} {\bibinfo {author} {\bibfnamefont {C.}~\bibnamefont
  {Weedbrook}}, \bibinfo {author} {\bibfnamefont {S.}~\bibnamefont
  {Pirandola}}, \bibinfo {author} {\bibfnamefont {R.}~\bibnamefont
  {Garc{\'{i}}a-Patr{\'{o}}n}}, \bibinfo {author} {\bibfnamefont {N.~J.}\
  \bibnamefont {Cerf}}, \bibinfo {author} {\bibfnamefont {T.~C.}\ \bibnamefont
  {Ralph}}, \bibinfo {author} {\bibfnamefont {J.~H.}\ \bibnamefont {Shapiro}},
  \ and\ \bibinfo {author} {\bibfnamefont {S.}~\bibnamefont {Lloyd}},\
  }\bibfield  {{Gaussian quantum information}} {\emph {\bibinfo {title}
  {{Gaussian quantum information}},\ }}\href {\doibase
  10.1103/RevModPhys.84.621} {\bibfield  {journal} {\bibinfo  {journal} {Rev.
  Mod. Phys.}\ }\textbf {\bibinfo {volume} {84}},\ \bibinfo {pages} {621}
  (\bibinfo {year} {2012})}\BibitemShut {NoStop}%
\bibitem [{\citenamefont {Cerf}\ \emph {et~al.}(2007)\citenamefont {Cerf},
  \citenamefont {Leuchs},\ and\ \citenamefont {Polzik}}]{cvbook}%
  \BibitemOpen
  \bibfield  {author} {\bibinfo {author} {\bibfnamefont {N.~J.}\ \bibnamefont
  {Cerf}}, \bibinfo {author} {\bibfnamefont {G.}~\bibnamefont {Leuchs}}, \ and\
  \bibinfo {author} {\bibfnamefont {E.~S.}\ \bibnamefont {Polzik}},\ }\href
  {\doibase 10.1142/9781860948169} {\emph {\bibinfo {title} {{Quantum
  Information with Continuous Variables of Atoms and Light}}}}\ (\bibinfo
  {publisher} {World Scientific},\ \bibinfo {address} {London},\ \bibinfo
  {year} {2007})\BibitemShut {NoStop}%
\bibitem [{\citenamefont {Serafini}(2017)}]{serafinibook}%
  \BibitemOpen
  \bibfield  {author} {\bibinfo {author} {\bibfnamefont {A.}~\bibnamefont
  {Serafini}},\ }\href
  {https://www.crcpress.com/Quantum-Continuous-Variables-A-Primer-of-Theoretical-Methods/Serafini/p/book/9781482246346}
  {\emph {\bibinfo {title} {{Quantum Continuous Variables: A Primer of
  Theoretical Methods}}}}\ (\bibinfo  {publisher} {CRC Press},\ \bibinfo
  {address} {Boca Raton FL},\ \bibinfo {year} {2017})\BibitemShut {NoStop}%
\bibitem [{\citenamefont {Chuang}\ and\ \citenamefont
  {Yamamoto}(1995)}]{Chuang1995}%
  \BibitemOpen
  \bibfield  {author} {\bibinfo {author} {\bibfnamefont {I.~L.}\ \bibnamefont
  {Chuang}}\ and\ \bibinfo {author} {\bibfnamefont {Y.}~\bibnamefont
  {Yamamoto}},\ }\bibfield  {{Simple quantum computer}} {\emph {\bibinfo
  {title} {{Simple quantum computer}},\ }}\href {\doibase
  10.1103/PhysRevA.52.3489} {\bibfield  {journal} {\bibinfo  {journal} {Phys.
  Rev. A}\ }\textbf {\bibinfo {volume} {52}},\ \bibinfo {pages} {3489}
  (\bibinfo {year} {1995})}\BibitemShut {NoStop}%
\bibitem [{\citenamefont {Chuang}\ \emph {et~al.}(1997)\citenamefont {Chuang},
  \citenamefont {Leung},\ and\ \citenamefont {Yamamoto}}]{Chuang1997}%
  \BibitemOpen
  \bibfield  {author} {\bibinfo {author} {\bibfnamefont {I.~L.}\ \bibnamefont
  {Chuang}}, \bibinfo {author} {\bibfnamefont {D.~W.}\ \bibnamefont {Leung}}, \
  and\ \bibinfo {author} {\bibfnamefont {Y.}~\bibnamefont {Yamamoto}},\
  }\bibfield  {{Bosonic quantum codes for amplitude damping}} {\emph {\bibinfo
  {title} {{Bosonic quantum codes for amplitude damping}},\ }}\href {\doibase
  10.1103/PhysRevA.56.1114} {\bibfield  {journal} {\bibinfo  {journal} {Phys.
  Rev. A}\ }\textbf {\bibinfo {volume} {56}},\ \bibinfo {pages} {1114}
  (\bibinfo {year} {1997})}\BibitemShut {NoStop}%
\bibitem [{\citenamefont {Knill}\ \emph {et~al.}(2001)\citenamefont {Knill},
  \citenamefont {Laflamme},\ and\ \citenamefont {Milburn}}]{Knill2001}%
  \BibitemOpen
  \bibfield  {author} {\bibinfo {author} {\bibfnamefont {E.}~\bibnamefont
  {Knill}}, \bibinfo {author} {\bibfnamefont {R.}~\bibnamefont {Laflamme}}, \
  and\ \bibinfo {author} {\bibfnamefont {G.~J.}\ \bibnamefont {Milburn}},\
  }\bibfield  {{A scheme for efficient quantum computation with linear optics}}
  {\emph {\bibinfo {title} {{A scheme for efficient quantum computation with
  linear optics}},\ }}\href {\doibase 10.1038/35051009} {\bibfield  {journal}
  {\bibinfo  {journal} {Nature}\ }\textbf {\bibinfo {volume} {409}},\ \bibinfo
  {pages} {46} (\bibinfo {year} {2001})}\BibitemShut {NoStop}%
\bibitem [{\citenamefont {Ralph}\ \emph {et~al.}(2005)\citenamefont {Ralph},
  \citenamefont {Hayes},\ and\ \citenamefont {Gilchrist}}]{Ralph2005}%
  \BibitemOpen
  \bibfield  {author} {\bibinfo {author} {\bibfnamefont {T.~C.}\ \bibnamefont
  {Ralph}}, \bibinfo {author} {\bibfnamefont {A.~J.~F.}\ \bibnamefont {Hayes}},
  \ and\ \bibinfo {author} {\bibfnamefont {A.}~\bibnamefont {Gilchrist}},\
  }\bibfield  {{Loss-Tolerant Optical Qubits}} {\emph {\bibinfo {title}
  {{Loss-Tolerant Optical Qubits}},\ }}\href {\doibase
  10.1103/PhysRevLett.95.100501} {\bibfield  {journal} {\bibinfo  {journal}
  {Phys. Rev. Lett.}\ }\textbf {\bibinfo {volume} {95}},\ \bibinfo {pages}
  {100501} (\bibinfo {year} {2005})}\BibitemShut {NoStop}%
\bibitem [{\citenamefont {Wasilewski}\ and\ \citenamefont
  {Banaszek}(2007)}]{Wasilewski2007}%
  \BibitemOpen
  \bibfield  {author} {\bibinfo {author} {\bibfnamefont {W.}~\bibnamefont
  {Wasilewski}}\ and\ \bibinfo {author} {\bibfnamefont {K.}~\bibnamefont
  {Banaszek}},\ }\bibfield  {{Protecting an optical qubit against photon loss}}
  {\emph {\bibinfo {title} {{Protecting an optical qubit against photon
  loss}},\ }}\href {\doibase 10.1103/PhysRevA.75.042316} {\bibfield  {journal}
  {\bibinfo  {journal} {Phys. Rev. A}\ }\textbf {\bibinfo {volume} {75}},\
  \bibinfo {pages} {042316} (\bibinfo {year} {2007})}\BibitemShut {NoStop}%
\bibitem [{\citenamefont {Bergmann}\ and\ \citenamefont {van
  Loock}(2016{\natexlab{a}})}]{Bergmann2016a}%
  \BibitemOpen
  \bibfield  {author} {\bibinfo {author} {\bibfnamefont {M.}~\bibnamefont
  {Bergmann}}\ and\ \bibinfo {author} {\bibfnamefont {P.}~\bibnamefont {van
  Loock}},\ }\bibfield  {{Quantum error correction against photon loss using
  NOON states}} {\emph {\bibinfo {title} {{Quantum error correction against
  photon loss using NOON states}},\ }}\href {\doibase
  10.1103/PhysRevA.94.012311} {\bibfield  {journal} {\bibinfo  {journal} {Phys.
  Rev. A}\ }\textbf {\bibinfo {volume} {94}},\ \bibinfo {pages} {012311}
  (\bibinfo {year} {2016}{\natexlab{a}})}\BibitemShut {NoStop}%
\bibitem [{\citenamefont {Michael}\ \emph {et~al.}(2016)\citenamefont
  {Michael}, \citenamefont {Silveri}, \citenamefont {Brierley}, \citenamefont
  {Albert}, \citenamefont {Salmilehto}, \citenamefont {Jiang},\ and\
  \citenamefont {Girvin}}]{bin}%
  \BibitemOpen
  \bibfield  {author} {\bibinfo {author} {\bibfnamefont {M.~H.}\ \bibnamefont
  {Michael}}, \bibinfo {author} {\bibfnamefont {M.}~\bibnamefont {Silveri}},
  \bibinfo {author} {\bibfnamefont {R.~T.}\ \bibnamefont {Brierley}}, \bibinfo
  {author} {\bibfnamefont {V.~V.}\ \bibnamefont {Albert}}, \bibinfo {author}
  {\bibfnamefont {J.}~\bibnamefont {Salmilehto}}, \bibinfo {author}
  {\bibfnamefont {L.}~\bibnamefont {Jiang}}, \ and\ \bibinfo {author}
  {\bibfnamefont {S.~M.}\ \bibnamefont {Girvin}},\ }\bibfield  {{New Class of
  Quantum Error-Correcting Codes for a Bosonic Mode}} {\emph {\bibinfo {title}
  {{New Class of Quantum Error-Correcting Codes for a Bosonic Mode}},\ }}\href
  {\doibase 10.1103/PhysRevX.6.031006} {\bibfield  {journal} {\bibinfo
  {journal} {Phys. Rev. X}\ }\textbf {\bibinfo {volume} {6}},\ \bibinfo {pages}
  {031006} (\bibinfo {year} {2016})}\BibitemShut {NoStop}%
\bibitem [{\citenamefont {Niu}\ \emph {et~al.}(2018{\natexlab{a}})\citenamefont
  {Niu}, \citenamefont {Chuang},\ and\ \citenamefont {Shapiro}}]{Niu2017}%
  \BibitemOpen
  \bibfield  {author} {\bibinfo {author} {\bibfnamefont {M.~Y.}\ \bibnamefont
  {Niu}}, \bibinfo {author} {\bibfnamefont {I.~L.}\ \bibnamefont {Chuang}}, \
  and\ \bibinfo {author} {\bibfnamefont {J.~H.}\ \bibnamefont {Shapiro}},\
  }\bibfield  {{Hardware-efficient bosonic quantum error-correcting codes based
  on symmetry operators}} {\emph {\bibinfo {title} {{Hardware-efficient bosonic
  quantum error-correcting codes based on symmetry operators}},\ }}\href
  {\doibase 10.1103/PhysRevA.97.032323} {\bibfield  {journal} {\bibinfo
  {journal} {Phys. Rev. A}\ }\textbf {\bibinfo {volume} {97}},\ \bibinfo
  {pages} {032323} (\bibinfo {year} {2018}{\natexlab{a}})}\BibitemShut
  {NoStop}%
\bibitem [{\citenamefont {Lloyd}\ and\ \citenamefont
  {Slotine}(1998)}]{Lloyd1998}%
  \BibitemOpen
  \bibfield  {author} {\bibinfo {author} {\bibfnamefont {S.}~\bibnamefont
  {Lloyd}}\ and\ \bibinfo {author} {\bibfnamefont {J.-J.~E.}\ \bibnamefont
  {Slotine}},\ }\bibfield  {{Analog Quantum Error Correction}} {\emph {\bibinfo
  {title} {{Analog Quantum Error Correction}},\ }}\href {\doibase
  10.1103/PhysRevLett.80.4088} {\bibfield  {journal} {\bibinfo  {journal}
  {Phys. Rev. Lett.}\ }\textbf {\bibinfo {volume} {80}},\ \bibinfo {pages}
  {4088} (\bibinfo {year} {1998})}\BibitemShut {NoStop}%
\bibitem [{\citenamefont {Braunstein}(1998)}]{Braunstein1998}%
  \BibitemOpen
  \bibfield  {author} {\bibinfo {author} {\bibfnamefont {S.~L.}\ \bibnamefont
  {Braunstein}},\ }\bibfield  {{Error Correction for Continuous Quantum
  Variables}} {\emph {\bibinfo {title} {{Error Correction for Continuous
  Quantum Variables}},\ }}\href {\doibase 10.1103/PhysRevLett.80.4084}
  {\bibfield  {journal} {\bibinfo  {journal} {Phys. Rev. Lett.}\ }\textbf
  {\bibinfo {volume} {80}},\ \bibinfo {pages} {4084} (\bibinfo {year}
  {1998})}\BibitemShut {NoStop}%
\bibitem [{\citenamefont {Gottesman}\ \emph {et~al.}(2001)\citenamefont
  {Gottesman}, \citenamefont {{Yu. Kitaev}},\ and\ \citenamefont
  {Preskill}}]{Gottesman2001}%
  \BibitemOpen
  \bibfield  {author} {\bibinfo {author} {\bibfnamefont {D.}~\bibnamefont
  {Gottesman}}, \bibinfo {author} {\bibfnamefont {A.}~\bibnamefont {{Yu.
  Kitaev}}}, \ and\ \bibinfo {author} {\bibfnamefont {J.}~\bibnamefont
  {Preskill}},\ }\bibfield  {{Encoding a qubit in an oscillator}} {\emph
  {\bibinfo {title} {{Encoding a qubit in an oscillator}},\ }}\href {\doibase
  10.1103/PhysRevA.64.012310} {\bibfield  {journal} {\bibinfo  {journal} {Phys.
  Rev. A}\ }\textbf {\bibinfo {volume} {64}},\ \bibinfo {pages} {012310}
  (\bibinfo {year} {2001})}\BibitemShut {NoStop}%
\bibitem [{\citenamefont {Menicucci}(2014)}]{Menicucci2014}%
  \BibitemOpen
  \bibfield  {author} {\bibinfo {author} {\bibfnamefont {N.~C.}\ \bibnamefont
  {Menicucci}},\ }\bibfield  {{Fault-Tolerant Measurement-Based Quantum
  Computing with Continuous-Variable Cluster States}} {\emph {\bibinfo {title}
  {{Fault-Tolerant Measurement-Based Quantum Computing with Continuous-Variable
  Cluster States}},\ }}\href {\doibase 10.1103/PhysRevLett.112.120504}
  {\bibfield  {journal} {\bibinfo  {journal} {Phys. Rev. Lett.}\ }\textbf
  {\bibinfo {volume} {112}},\ \bibinfo {pages} {120504} (\bibinfo {year}
  {2014})}\BibitemShut {NoStop}%
\bibitem [{\citenamefont {Hayden}\ \emph {et~al.}(2016)\citenamefont {Hayden},
  \citenamefont {Nezami}, \citenamefont {Salton},\ and\ \citenamefont
  {Sanders}}]{Hayden2016}%
  \BibitemOpen
  \bibfield  {author} {\bibinfo {author} {\bibfnamefont {P.}~\bibnamefont
  {Hayden}}, \bibinfo {author} {\bibfnamefont {S.}~\bibnamefont {Nezami}},
  \bibinfo {author} {\bibfnamefont {G.}~\bibnamefont {Salton}}, \ and\ \bibinfo
  {author} {\bibfnamefont {B.~C.}\ \bibnamefont {Sanders}},\ }\bibfield
  {{Spacetime replication of continuous variable quantum information}} {\emph
  {\bibinfo {title} {{Spacetime replication of continuous variable quantum
  information}},\ }}\href {\doibase 10.1088/1367-2630/18/8/083043} {\bibfield
  {journal} {\bibinfo  {journal} {New J. Phys.}\ }\textbf {\bibinfo {volume}
  {18}},\ \bibinfo {pages} {083043} (\bibinfo {year} {2016})}\BibitemShut
  {NoStop}%
\bibitem [{\citenamefont {Ketterer}\ \emph {et~al.}(2016)\citenamefont
  {Ketterer}, \citenamefont {Keller}, \citenamefont {Walborn}, \citenamefont
  {Coudreau},\ and\ \citenamefont {Milman}}]{Ketterer2016}%
  \BibitemOpen
  \bibfield  {author} {\bibinfo {author} {\bibfnamefont {A.}~\bibnamefont
  {Ketterer}}, \bibinfo {author} {\bibfnamefont {A.}~\bibnamefont {Keller}},
  \bibinfo {author} {\bibfnamefont {S.~P.}\ \bibnamefont {Walborn}}, \bibinfo
  {author} {\bibfnamefont {T.}~\bibnamefont {Coudreau}}, \ and\ \bibinfo
  {author} {\bibfnamefont {P.}~\bibnamefont {Milman}},\ }\bibfield  {{Quantum
  information processing in phase space: A modular variables approach}} {\emph
  {\bibinfo {title} {{Quantum information processing in phase space: A modular
  variables approach}},\ }}\href {\doibase 10.1103/PhysRevA.94.022325}
  {\bibfield  {journal} {\bibinfo  {journal} {Phys. Rev. A}\ }\textbf {\bibinfo
  {volume} {94}},\ \bibinfo {pages} {022325} (\bibinfo {year}
  {2016})}\BibitemShut {NoStop}%
\bibitem [{\citenamefont {Cochrane}\ \emph {et~al.}(1999)\citenamefont
  {Cochrane}, \citenamefont {Milburn},\ and\ \citenamefont
  {Munro}}]{Cochrane1999}%
  \BibitemOpen
  \bibfield  {author} {\bibinfo {author} {\bibfnamefont {P.~T.}\ \bibnamefont
  {Cochrane}}, \bibinfo {author} {\bibfnamefont {G.~J.}\ \bibnamefont
  {Milburn}}, \ and\ \bibinfo {author} {\bibfnamefont {W.~J.}\ \bibnamefont
  {Munro}},\ }\bibfield  {{Macroscopically distinct quantum-superposition
  states as a bosonic code for amplitude damping}} {\emph {\bibinfo {title}
  {{Macroscopically distinct quantum-superposition states as a bosonic code for
  amplitude damping}},\ }}\href {\doibase 10.1103/PhysRevA.59.2631} {\bibfield
  {journal} {\bibinfo  {journal} {Phys. Rev. A}\ }\textbf {\bibinfo {volume}
  {59}},\ \bibinfo {pages} {2631} (\bibinfo {year} {1999})}\BibitemShut
  {NoStop}%
\bibitem [{\citenamefont {Niset}\ \emph {et~al.}(2008)\citenamefont {Niset},
  \citenamefont {Andersen},\ and\ \citenamefont {Cerf}}]{Niset2008}%
  \BibitemOpen
  \bibfield  {author} {\bibinfo {author} {\bibfnamefont {J.}~\bibnamefont
  {Niset}}, \bibinfo {author} {\bibfnamefont {U.~L.}\ \bibnamefont {Andersen}},
  \ and\ \bibinfo {author} {\bibfnamefont {N.~J.}\ \bibnamefont {Cerf}},\
  }\bibfield  {{Experimentally Feasible Quantum Erasure-Correcting Code for
  Continuous Variables}} {\emph {\bibinfo {title} {{Experimentally Feasible
  Quantum Erasure-Correcting Code for Continuous Variables}},\ }}\href
  {\doibase 10.1103/PhysRevLett.101.130503} {\bibfield  {journal} {\bibinfo
  {journal} {Phys. Rev. Lett.}\ }\textbf {\bibinfo {volume} {101}},\ \bibinfo
  {pages} {130503} (\bibinfo {year} {2008})}\BibitemShut {NoStop}%
\bibitem [{\citenamefont {Leghtas}\ \emph {et~al.}(2013)\citenamefont
  {Leghtas}, \citenamefont {Kirchmair}, \citenamefont {Vlastakis},
  \citenamefont {Schoelkopf}, \citenamefont {Devoret},\ and\ \citenamefont
  {Mirrahimi}}]{Leghtas2013b}%
  \BibitemOpen
  \bibfield  {author} {\bibinfo {author} {\bibfnamefont {Z.}~\bibnamefont
  {Leghtas}}, \bibinfo {author} {\bibfnamefont {G.}~\bibnamefont {Kirchmair}},
  \bibinfo {author} {\bibfnamefont {B.}~\bibnamefont {Vlastakis}}, \bibinfo
  {author} {\bibfnamefont {R.~J.}\ \bibnamefont {Schoelkopf}}, \bibinfo
  {author} {\bibfnamefont {M.~H.}\ \bibnamefont {Devoret}}, \ and\ \bibinfo
  {author} {\bibfnamefont {M.}~\bibnamefont {Mirrahimi}},\ }\bibfield
  {{Hardware-Efficient Autonomous Quantum Memory Protection}} {\emph {\bibinfo
  {title} {{Hardware-Efficient Autonomous Quantum Memory Protection}},\ }}\href
  {\doibase 10.1103/PhysRevLett.111.120501} {\bibfield  {journal} {\bibinfo
  {journal} {Phys. Rev. Lett.}\ }\textbf {\bibinfo {volume} {111}},\ \bibinfo
  {pages} {120501} (\bibinfo {year} {2013})}\BibitemShut {NoStop}%
\bibitem [{\citenamefont {Lacerda}\ \emph {et~al.}(2016)\citenamefont
  {Lacerda}, \citenamefont {Renes},\ and\ \citenamefont
  {Scholz}}]{Lacerda2016}%
  \BibitemOpen
  \bibfield  {author} {\bibinfo {author} {\bibfnamefont {F.}~\bibnamefont
  {Lacerda}}, \bibinfo {author} {\bibfnamefont {J.~M.}\ \bibnamefont {Renes}},
  \ and\ \bibinfo {author} {\bibfnamefont {V.~B.}\ \bibnamefont {Scholz}},\
  }\bibfield  {{Coherent state constellations for Bosonic Gaussian channels}}
  {\emph {\bibinfo {title} {{Coherent state constellations for Bosonic Gaussian
  channels}},\ }}\href {\doibase 10.1109/ISIT.2016.7541749} {\bibfield
  {journal} {\bibinfo  {journal} {2016 IEEE Int. Symp. Inf. Theory}\ ,\
  \bibinfo {pages} {2499}} (\bibinfo {year} {2016})}\BibitemShut {NoStop}%
\bibitem [{\citenamefont {Lee}\ and\ \citenamefont {Jeong}(2013)}]{Lee2013}%
  \BibitemOpen
  \bibfield  {author} {\bibinfo {author} {\bibfnamefont {S.-W.}\ \bibnamefont
  {Lee}}\ and\ \bibinfo {author} {\bibfnamefont {H.}~\bibnamefont {Jeong}},\
  }\bibfield  {{Near-deterministic quantum teleportation and resource-efficient
  quantum computation using linear optics and hybrid qubits}} {\emph {\bibinfo
  {title} {{Near-deterministic quantum teleportation and resource-efficient
  quantum computation using linear optics and hybrid qubits}},\ }}\href
  {\doibase 10.1103/PhysRevA.87.022326} {\bibfield  {journal} {\bibinfo
  {journal} {Phys. Rev. A}\ }\textbf {\bibinfo {volume} {87}},\ \bibinfo
  {pages} {022326} (\bibinfo {year} {2013})}\BibitemShut {NoStop}%
\bibitem [{\citenamefont {Kapit}(2016)}]{Kapit2016}%
  \BibitemOpen
  \bibfield  {author} {\bibinfo {author} {\bibfnamefont {E.}~\bibnamefont
  {Kapit}},\ }\bibfield  {{Hardware-Efficient and Fully Autonomous Quantum
  Error Correction in Superconducting Circuits}} {\emph {\bibinfo {title}
  {{Hardware-Efficient and Fully Autonomous Quantum Error Correction in
  Superconducting Circuits}},\ }}\href {\doibase
  10.1103/PhysRevLett.116.150501} {\bibfield  {journal} {\bibinfo  {journal}
  {Phys. Rev. Lett.}\ }\textbf {\bibinfo {volume} {116}},\ \bibinfo {pages}
  {150501} (\bibinfo {year} {2016})}\BibitemShut {NoStop}%
\bibitem [{\citenamefont {Girvin}(2015)}]{girvinbook}%
  \BibitemOpen
  \bibfield  {author} {\bibinfo {author} {\bibfnamefont {S.~M.}\ \bibnamefont
  {Girvin}},\ }in\ \href {\doibase 10.1080/00107514.2015.1005684} {\emph
  {\bibinfo {booktitle} {Quantum Mach. Meas. Control Eng. quantum Syst.}}},\
  \bibinfo {editor} {edited by\ \bibinfo {editor} {\bibfnamefont {M.~H.}\
  \bibnamefont {Devoret}}, \bibinfo {editor} {\bibfnamefont {B.}~\bibnamefont
  {Huard}}, \bibinfo {editor} {\bibfnamefont {R.~J.}\ \bibnamefont
  {Schoelkopf}}, \ and\ \bibinfo {editor} {\bibfnamefont {L.~F.}\ \bibnamefont
  {Cugliandolo}}}\ (\bibinfo  {publisher} {Oxford University Press},\ \bibinfo
  {address} {Oxford},\ \bibinfo {year} {2015})\ Chap.~\bibinfo {chapter}
  {3}\BibitemShut {NoStop}%
\bibitem [{\citenamefont {Gu}\ \emph {et~al.}(2017)\citenamefont {Gu},
  \citenamefont {Kockum}, \citenamefont {Miranowicz}, \citenamefont {Liu},\
  and\ \citenamefont {Nori}}]{Gu2017}%
  \BibitemOpen
  \bibfield  {author} {\bibinfo {author} {\bibfnamefont {X.}~\bibnamefont
  {Gu}}, \bibinfo {author} {\bibfnamefont {A.~F.}\ \bibnamefont {Kockum}},
  \bibinfo {author} {\bibfnamefont {A.}~\bibnamefont {Miranowicz}}, \bibinfo
  {author} {\bibfnamefont {Y.-x.}\ \bibnamefont {Liu}}, \ and\ \bibinfo
  {author} {\bibfnamefont {F.}~\bibnamefont {Nori}},\ }\bibfield  {{Microwave
  photonics with superconducting quantum circuits}} {\emph {\bibinfo {title}
  {{Microwave photonics with superconducting quantum circuits}},\ }}\href
  {\doibase 10.1016/j.physrep.2017.10.002} {\bibfield  {journal} {\bibinfo
  {journal} {Phys. Rep.}\ }\textbf {\bibinfo {volume} {718-719}},\ \bibinfo
  {pages} {1} (\bibinfo {year} {2017})}\BibitemShut {NoStop}%
\bibitem [{\citenamefont {Mirrahimi}\ \emph {et~al.}(2014)\citenamefont
  {Mirrahimi}, \citenamefont {Leghtas}, \citenamefont {Albert}, \citenamefont
  {Touzard}, \citenamefont {Schoelkopf}, \citenamefont {Jiang},\ and\
  \citenamefont {Devoret}}]{cats}%
  \BibitemOpen
  \bibfield  {author} {\bibinfo {author} {\bibfnamefont {M.}~\bibnamefont
  {Mirrahimi}}, \bibinfo {author} {\bibfnamefont {Z.}~\bibnamefont {Leghtas}},
  \bibinfo {author} {\bibfnamefont {V.~V.}\ \bibnamefont {Albert}}, \bibinfo
  {author} {\bibfnamefont {S.}~\bibnamefont {Touzard}}, \bibinfo {author}
  {\bibfnamefont {R.~J.}\ \bibnamefont {Schoelkopf}}, \bibinfo {author}
  {\bibfnamefont {L.}~\bibnamefont {Jiang}}, \ and\ \bibinfo {author}
  {\bibfnamefont {M.~H.}\ \bibnamefont {Devoret}},\ }\bibfield  {{Dynamically
  protected cat-qubits: a new paradigm for universal quantum computation}}
  {\emph {\bibinfo {title} {{Dynamically protected cat-qubits: a new paradigm
  for universal quantum computation}},\ }}\href
  {http://arxiv-web3.library.cornell.edu/abs/1312.2017} {\bibfield  {journal}
  {\bibinfo  {journal} {New J. Phys.}\ }\textbf {\bibinfo {volume} {16}},\
  \bibinfo {pages} {045014} (\bibinfo {year} {2014})}\BibitemShut {NoStop}%
\bibitem [{\citenamefont {Albert}\ \emph
  {et~al.}(2016{\natexlab{a}})\citenamefont {Albert}, \citenamefont {Shu},
  \citenamefont {Krastanov}, \citenamefont {Shen}, \citenamefont {Liu},
  \citenamefont {Yang}, \citenamefont {Schoelkopf}, \citenamefont {Mirrahimi},
  \citenamefont {Devoret},\ and\ \citenamefont {Jiang}}]{Albert2015}%
  \BibitemOpen
  \bibfield  {author} {\bibinfo {author} {\bibfnamefont {V.~V.}\ \bibnamefont
  {Albert}}, \bibinfo {author} {\bibfnamefont {C.}~\bibnamefont {Shu}},
  \bibinfo {author} {\bibfnamefont {S.}~\bibnamefont {Krastanov}}, \bibinfo
  {author} {\bibfnamefont {C.}~\bibnamefont {Shen}}, \bibinfo {author}
  {\bibfnamefont {R.-B.}\ \bibnamefont {Liu}}, \bibinfo {author} {\bibfnamefont
  {Z.-B.}\ \bibnamefont {Yang}}, \bibinfo {author} {\bibfnamefont {R.~J.}\
  \bibnamefont {Schoelkopf}}, \bibinfo {author} {\bibfnamefont
  {M.}~\bibnamefont {Mirrahimi}}, \bibinfo {author} {\bibfnamefont {M.~H.}\
  \bibnamefont {Devoret}}, \ and\ \bibinfo {author} {\bibfnamefont
  {L.}~\bibnamefont {Jiang}},\ }\bibfield  {{Holonomic Quantum Control with
  Continuous Variable Systems}} {\emph {\bibinfo {title} {{Holonomic Quantum
  Control with Continuous Variable Systems}},\ }}\href {\doibase
  10.1103/PhysRevLett.116.140502} {\bibfield  {journal} {\bibinfo  {journal}
  {Phys. Rev. Lett.}\ }\textbf {\bibinfo {volume} {116}},\ \bibinfo {pages}
  {140502} (\bibinfo {year} {2016}{\natexlab{a}})}\BibitemShut {NoStop}%
\bibitem [{\citenamefont {Bergmann}\ and\ \citenamefont {van
  Loock}(2016{\natexlab{b}})}]{Bergmann2016}%
  \BibitemOpen
  \bibfield  {author} {\bibinfo {author} {\bibfnamefont {M.}~\bibnamefont
  {Bergmann}}\ and\ \bibinfo {author} {\bibfnamefont {P.}~\bibnamefont {van
  Loock}},\ }\bibfield  {{Quantum error correction against photon loss using
  multicomponent cat states}} {\emph {\bibinfo {title} {{Quantum error
  correction against photon loss using multicomponent cat states}},\ }}\href
  {\doibase 10.1103/PhysRevA.94.042332} {\bibfield  {journal} {\bibinfo
  {journal} {Phys. Rev. A}\ }\textbf {\bibinfo {volume} {94}},\ \bibinfo
  {pages} {042332} (\bibinfo {year} {2016}{\natexlab{b}})}\BibitemShut
  {NoStop}%
\bibitem [{\citenamefont {Li}\ \emph {et~al.}(2017)\citenamefont {Li},
  \citenamefont {Zou}, \citenamefont {Albert}, \citenamefont {Muralidharan},
  \citenamefont {Girvin},\ and\ \citenamefont {Jiang}}]{Li2016}%
  \BibitemOpen
  \bibfield  {author} {\bibinfo {author} {\bibfnamefont {L.}~\bibnamefont
  {Li}}, \bibinfo {author} {\bibfnamefont {C.-l.}\ \bibnamefont {Zou}},
  \bibinfo {author} {\bibfnamefont {V.~V.}\ \bibnamefont {Albert}}, \bibinfo
  {author} {\bibfnamefont {S.}~\bibnamefont {Muralidharan}}, \bibinfo {author}
  {\bibfnamefont {S.~M.}\ \bibnamefont {Girvin}}, \ and\ \bibinfo {author}
  {\bibfnamefont {L.}~\bibnamefont {Jiang}},\ }\bibfield  {{Cat Codes with
  Optimal Decoherence Suppression for a Lossy Bosonic Channel}} {\emph
  {\bibinfo {title} {{Cat Codes with Optimal Decoherence Suppression for a
  Lossy Bosonic Channel}},\ }}\href {\doibase 10.1103/PhysRevLett.119.030502}
  {\bibfield  {journal} {\bibinfo  {journal} {Phys. Rev. Lett.}\ }\textbf
  {\bibinfo {volume} {119}},\ \bibinfo {pages} {030502} (\bibinfo {year}
  {2017})}\BibitemShut {NoStop}%
\bibitem [{\citenamefont {Leghtas}\ \emph {et~al.}(2015)\citenamefont
  {Leghtas}, \citenamefont {Touzard}, \citenamefont {Pop}, \citenamefont {Kou},
  \citenamefont {Vlastakis}, \citenamefont {Petrenko}, \citenamefont {Sliwa},
  \citenamefont {Narla}, \citenamefont {Shankar}, \citenamefont {Hatridge},
  \citenamefont {Reagor}, \citenamefont {Frunzio}, \citenamefont {Schoelkopf},
  \citenamefont {Mirrahimi},\ and\ \citenamefont {Devoret}}]{Leghtas2014}%
  \BibitemOpen
  \bibfield  {author} {\bibinfo {author} {\bibfnamefont {Z.}~\bibnamefont
  {Leghtas}}, \bibinfo {author} {\bibfnamefont {S.}~\bibnamefont {Touzard}},
  \bibinfo {author} {\bibfnamefont {I.~M.}\ \bibnamefont {Pop}}, \bibinfo
  {author} {\bibfnamefont {A.}~\bibnamefont {Kou}}, \bibinfo {author}
  {\bibfnamefont {B.}~\bibnamefont {Vlastakis}}, \bibinfo {author}
  {\bibfnamefont {A.}~\bibnamefont {Petrenko}}, \bibinfo {author}
  {\bibfnamefont {K.~M.}\ \bibnamefont {Sliwa}}, \bibinfo {author}
  {\bibfnamefont {A.}~\bibnamefont {Narla}}, \bibinfo {author} {\bibfnamefont
  {S.}~\bibnamefont {Shankar}}, \bibinfo {author} {\bibfnamefont {M.~J.}\
  \bibnamefont {Hatridge}}, \bibinfo {author} {\bibfnamefont {M.}~\bibnamefont
  {Reagor}}, \bibinfo {author} {\bibfnamefont {L.}~\bibnamefont {Frunzio}},
  \bibinfo {author} {\bibfnamefont {R.~J.}\ \bibnamefont {Schoelkopf}},
  \bibinfo {author} {\bibfnamefont {M.}~\bibnamefont {Mirrahimi}}, \ and\
  \bibinfo {author} {\bibfnamefont {M.~H.}\ \bibnamefont {Devoret}},\
  }\bibfield  {{Confining the state of light to a quantum manifold by
  engineered two-photon loss}} {\emph {\bibinfo {title} {{Confining the state
  of light to a quantum manifold by engineered two-photon loss}},\ }}\href
  {\doibase 10.1126/science.aaa2085} {\bibfield  {journal} {\bibinfo  {journal}
  {Science (80-. ).}\ }\textbf {\bibinfo {volume} {347}},\ \bibinfo {pages}
  {853} (\bibinfo {year} {2015})}\BibitemShut {NoStop}%
\bibitem [{\citenamefont {Ofek}\ \emph {et~al.}(2016)\citenamefont {Ofek},
  \citenamefont {Petrenko}, \citenamefont {Heeres}, \citenamefont {Reinhold},
  \citenamefont {Leghtas}, \citenamefont {Vlastakis}, \citenamefont {Liu},
  \citenamefont {Frunzio}, \citenamefont {Girvin}, \citenamefont {Jiang},
  \citenamefont {Mirrahimi}, \citenamefont {Devoret},\ and\ \citenamefont
  {Schoelkopf}}]{Ofek2016}%
  \BibitemOpen
  \bibfield  {author} {\bibinfo {author} {\bibfnamefont {N.}~\bibnamefont
  {Ofek}}, \bibinfo {author} {\bibfnamefont {A.}~\bibnamefont {Petrenko}},
  \bibinfo {author} {\bibfnamefont {R.}~\bibnamefont {Heeres}}, \bibinfo
  {author} {\bibfnamefont {P.}~\bibnamefont {Reinhold}}, \bibinfo {author}
  {\bibfnamefont {Z.}~\bibnamefont {Leghtas}}, \bibinfo {author} {\bibfnamefont
  {B.}~\bibnamefont {Vlastakis}}, \bibinfo {author} {\bibfnamefont
  {Y.}~\bibnamefont {Liu}}, \bibinfo {author} {\bibfnamefont {L.}~\bibnamefont
  {Frunzio}}, \bibinfo {author} {\bibfnamefont {S.~M.}\ \bibnamefont {Girvin}},
  \bibinfo {author} {\bibfnamefont {L.}~\bibnamefont {Jiang}}, \bibinfo
  {author} {\bibfnamefont {M.}~\bibnamefont {Mirrahimi}}, \bibinfo {author}
  {\bibfnamefont {M.~H.}\ \bibnamefont {Devoret}}, \ and\ \bibinfo {author}
  {\bibfnamefont {R.~J.}\ \bibnamefont {Schoelkopf}},\ }\bibfield  {{Extending
  the lifetime of a quantum bit with error correction in superconducting
  circuits}} {\emph {\bibinfo {title} {{Extending the lifetime of a quantum bit
  with error correction in superconducting circuits}},\ }}\href {\doibase
  10.1038/nature18949} {\bibfield  {journal} {\bibinfo  {journal} {Nature}\
  }\textbf {\bibinfo {volume} {536}},\ \bibinfo {pages} {441} (\bibinfo {year}
  {2016})}\BibitemShut {NoStop}%
\bibitem [{\citenamefont {Heeres}\ \emph {et~al.}(2017)\citenamefont {Heeres},
  \citenamefont {Reinhold}, \citenamefont {Ofek}, \citenamefont {Frunzio},
  \citenamefont {Jiang}, \citenamefont {Devoret},\ and\ \citenamefont
  {Schoelkopf}}]{Heeres2016}%
  \BibitemOpen
  \bibfield  {author} {\bibinfo {author} {\bibfnamefont {R.~W.}\ \bibnamefont
  {Heeres}}, \bibinfo {author} {\bibfnamefont {P.}~\bibnamefont {Reinhold}},
  \bibinfo {author} {\bibfnamefont {N.}~\bibnamefont {Ofek}}, \bibinfo {author}
  {\bibfnamefont {L.}~\bibnamefont {Frunzio}}, \bibinfo {author} {\bibfnamefont
  {L.}~\bibnamefont {Jiang}}, \bibinfo {author} {\bibfnamefont {M.~H.}\
  \bibnamefont {Devoret}}, \ and\ \bibinfo {author} {\bibfnamefont {R.~J.}\
  \bibnamefont {Schoelkopf}},\ }\bibfield  {{Implementing a universal gate set
  on a logical qubit encoded in an oscillator}} {\emph {\bibinfo {title}
  {{Implementing a universal gate set on a logical qubit encoded in an
  oscillator}},\ }}\href {\doibase 10.1038/s41467-017-00045-1} {\bibfield
  {journal} {\bibinfo  {journal} {Nat. Commun.}\ }\textbf {\bibinfo {volume}
  {8}},\ \bibinfo {pages} {94} (\bibinfo {year} {2017})}\BibitemShut {NoStop}%
\bibitem [{\citenamefont {Arrangoiz-Arriola}\ \emph {et~al.}(2018)\citenamefont
  {Arrangoiz-Arriola}, \citenamefont {Wollack}, \citenamefont {Pechal},
  \citenamefont {Witmer}, \citenamefont {Hill},\ and\ \citenamefont
  {Safavi-Naeini}}]{Arrangoiz-Arriola2018}%
  \BibitemOpen
  \bibfield  {author} {\bibinfo {author} {\bibfnamefont {P.}~\bibnamefont
  {Arrangoiz-Arriola}}, \bibinfo {author} {\bibfnamefont {E.~A.}\ \bibnamefont
  {Wollack}}, \bibinfo {author} {\bibfnamefont {M.}~\bibnamefont {Pechal}},
  \bibinfo {author} {\bibfnamefont {J.~D.}\ \bibnamefont {Witmer}}, \bibinfo
  {author} {\bibfnamefont {J.~T.}\ \bibnamefont {Hill}}, \ and\ \bibinfo
  {author} {\bibfnamefont {A.~H.}\ \bibnamefont {Safavi-Naeini}},\ }\bibfield
  {{Coupling a Superconducting Quantum Circuit to a Phononic Crystal Defect
  Cavity}} {\emph {\bibinfo {title} {{Coupling a Superconducting Quantum
  Circuit to a Phononic Crystal Defect Cavity}},\ }}\href {\doibase
  10.1103/PhysRevX.8.031007} {\bibfield  {journal} {\bibinfo  {journal} {Phys.
  Rev. X}\ }\textbf {\bibinfo {volume} {8}},\ \bibinfo {pages} {031007}
  (\bibinfo {year} {2018})}\BibitemShut {NoStop}%
\bibitem [{\citenamefont {Chu}\ \emph {et~al.}(2017)\citenamefont {Chu},
  \citenamefont {Kharel}, \citenamefont {Renninger}, \citenamefont {Burkhart},
  \citenamefont {Frunzio}, \citenamefont {Rakich},\ and\ \citenamefont
  {Schoelkopf}}]{Chu2017}%
  \BibitemOpen
  \bibfield  {author} {\bibinfo {author} {\bibfnamefont {Y.}~\bibnamefont
  {Chu}}, \bibinfo {author} {\bibfnamefont {P.}~\bibnamefont {Kharel}},
  \bibinfo {author} {\bibfnamefont {W.~H.}\ \bibnamefont {Renninger}}, \bibinfo
  {author} {\bibfnamefont {L.~D.}\ \bibnamefont {Burkhart}}, \bibinfo {author}
  {\bibfnamefont {L.}~\bibnamefont {Frunzio}}, \bibinfo {author} {\bibfnamefont
  {P.~T.}\ \bibnamefont {Rakich}}, \ and\ \bibinfo {author} {\bibfnamefont
  {R.~J.}\ \bibnamefont {Schoelkopf}},\ }\bibfield  {{Quantum acoustics with
  superconducting qubits}} {\emph {\bibinfo {title} {{Quantum acoustics with
  superconducting qubits}},\ }}\href {\doibase 10.1126/science.aao1511}
  {\bibfield  {journal} {\bibinfo  {journal} {Science (80-. ).}\ }\textbf
  {\bibinfo {volume} {358}},\ \bibinfo {pages} {199} (\bibinfo {year}
  {2017})}\BibitemShut {NoStop}%
\bibitem [{\citenamefont {Poyatos}\ \emph {et~al.}(1996)\citenamefont
  {Poyatos}, \citenamefont {Cirac},\ and\ \citenamefont
  {Zoller}}]{Poyatos1996}%
  \BibitemOpen
  \bibfield  {author} {\bibinfo {author} {\bibfnamefont {J.~F.}\ \bibnamefont
  {Poyatos}}, \bibinfo {author} {\bibfnamefont {J.~I.}\ \bibnamefont {Cirac}},
  \ and\ \bibinfo {author} {\bibfnamefont {P.}~\bibnamefont {Zoller}},\
  }\bibfield  {{Quantum Reservoir Engineering with Laser Cooled Trapped Ions}}
  {\emph {\bibinfo {title} {{Quantum Reservoir Engineering with Laser Cooled
  Trapped Ions}},\ }}\href {\doibase 10.1103/PhysRevLett.77.4728} {\bibfield
  {journal} {\bibinfo  {journal} {Phys. Rev. Lett.}\ }\textbf {\bibinfo
  {volume} {77}},\ \bibinfo {pages} {4728} (\bibinfo {year}
  {1996})}\BibitemShut {NoStop}%
\bibitem [{\citenamefont {Belavin}\ \emph {et~al.}(1969)\citenamefont
  {Belavin}, \citenamefont {{Ya. Zel'dovich}}, \citenamefont {Perelomov},\ and\
  \citenamefont {Popov}}]{Belavin1969}%
  \BibitemOpen
  \bibfield  {author} {\bibinfo {author} {\bibfnamefont {A.~A.}\ \bibnamefont
  {Belavin}}, \bibinfo {author} {\bibfnamefont {B.}~\bibnamefont {{Ya.
  Zel'dovich}}}, \bibinfo {author} {\bibfnamefont {A.~M.}\ \bibnamefont
  {Perelomov}}, \ and\ \bibinfo {author} {\bibfnamefont {V.~S.}\ \bibnamefont
  {Popov}},\ }\bibfield  {{Relaxation of Quantum Systems with Equidistant
  Spectra}} {\emph {\bibinfo {title} {{Relaxation of Quantum Systems with
  Equidistant Spectra}},\ }}\href
  {http://www.jetp.ac.ru/cgi-bin/e/index/e/29/1/p145?a=list} {\bibfield
  {journal} {\bibinfo  {journal} {Sov. Phys. JETP-USSR}\ }\textbf {\bibinfo
  {volume} {29}},\ \bibinfo {pages} {145} (\bibinfo {year} {1969})}\BibitemShut
  {NoStop}%
\bibitem [{\citenamefont {Lindblad}(1976)}]{Lindblad1976}%
  \BibitemOpen
  \bibfield  {author} {\bibinfo {author} {\bibfnamefont {G.}~\bibnamefont
  {Lindblad}},\ }\bibfield  {{On the generators of quantum dynamical
  semigroups}} {\emph {\bibinfo {title} {{On the generators of quantum
  dynamical semigroups}},\ }}\href
  {http://projecteuclid.org/euclid.cmp/1103899849} {\bibfield  {journal}
  {\bibinfo  {journal} {Commun. Math. Phys.}\ }\textbf {\bibinfo {volume}
  {48}},\ \bibinfo {pages} {119} (\bibinfo {year} {1976})}\BibitemShut
  {NoStop}%
\bibitem [{\citenamefont {Gorini}\ \emph {et~al.}(1976)\citenamefont {Gorini},
  \citenamefont {Kossakowski},\ and\ \citenamefont {Sudarshan}}]{Gorini1976a}%
  \BibitemOpen
  \bibfield  {author} {\bibinfo {author} {\bibfnamefont {V.}~\bibnamefont
  {Gorini}}, \bibinfo {author} {\bibfnamefont {A.}~\bibnamefont {Kossakowski}},
  \ and\ \bibinfo {author} {\bibfnamefont {E.~C.~G.}\ \bibnamefont
  {Sudarshan}},\ }\bibfield  {{Completely positive dynamical semigroups of
  N-level systems}} {\emph {\bibinfo {title} {{Completely positive dynamical
  semigroups of N-level systems}},\ }}\href {\doibase 10.1063/1.522979}
  {\bibfield  {journal} {\bibinfo  {journal} {J. Math. Phys.}\ }\textbf
  {\bibinfo {volume} {17}},\ \bibinfo {pages} {821} (\bibinfo {year}
  {1976})}\BibitemShut {NoStop}%
\bibitem [{\citenamefont {Duan}\ and\ \citenamefont {Guo}(1997)}]{Duan1997}%
  \BibitemOpen
  \bibfield  {author} {\bibinfo {author} {\bibfnamefont {L.-M.}\ \bibnamefont
  {Duan}}\ and\ \bibinfo {author} {\bibfnamefont {G.-C.}\ \bibnamefont {Guo}},\
  }\bibfield  {{Preserving Coherence in Quantum Computation by Pairing Quantum
  Bits}} {\emph {\bibinfo {title} {{Preserving Coherence in Quantum Computation
  by Pairing Quantum Bits}},\ }}\href {\doibase 10.1103/PhysRevLett.79.1953}
  {\bibfield  {journal} {\bibinfo  {journal} {Phys. Rev. Lett.}\ }\textbf
  {\bibinfo {volume} {79}},\ \bibinfo {pages} {1953} (\bibinfo {year}
  {1997})}\BibitemShut {NoStop}%
\bibitem [{\citenamefont {Zanardi}\ and\ \citenamefont
  {Rasetti}(1997)}]{Zanardi1997}%
  \BibitemOpen
  \bibfield  {author} {\bibinfo {author} {\bibfnamefont {P.}~\bibnamefont
  {Zanardi}}\ and\ \bibinfo {author} {\bibfnamefont {M.}~\bibnamefont
  {Rasetti}},\ }\bibfield  {{Noiseless Quantum Codes}} {\emph {\bibinfo {title}
  {{Noiseless Quantum Codes}},\ }}\href {\doibase 10.1103/PhysRevLett.79.3306}
  {\bibfield  {journal} {\bibinfo  {journal} {Phys. Rev. Lett.}\ }\textbf
  {\bibinfo {volume} {79}},\ \bibinfo {pages} {3306} (\bibinfo {year}
  {1997})}\BibitemShut {NoStop}%
\bibitem [{\citenamefont {Lidar}\ \emph {et~al.}(1998)\citenamefont {Lidar},
  \citenamefont {Chuang},\ and\ \citenamefont {Whaley}}]{Lidar1998}%
  \BibitemOpen
  \bibfield  {author} {\bibinfo {author} {\bibfnamefont {D.~A.}\ \bibnamefont
  {Lidar}}, \bibinfo {author} {\bibfnamefont {I.~L.}\ \bibnamefont {Chuang}}, \
  and\ \bibinfo {author} {\bibfnamefont {K.~B.}\ \bibnamefont {Whaley}},\
  }\bibfield  {{Decoherence-Free Subspaces for Quantum Computation}} {\emph
  {\bibinfo {title} {{Decoherence-Free Subspaces for Quantum Computation}},\
  }}\href {\doibase 10.1103/PhysRevLett.81.2594} {\bibfield  {journal}
  {\bibinfo  {journal} {Phys. Rev. Lett.}\ }\textbf {\bibinfo {volume} {81}},\
  \bibinfo {pages} {2594} (\bibinfo {year} {1998})}\BibitemShut {NoStop}%
\bibitem [{\citenamefont {Terhal}(2015)}]{Terhal2015}%
  \BibitemOpen
  \bibfield  {author} {\bibinfo {author} {\bibfnamefont {B.~M.}\ \bibnamefont
  {Terhal}},\ }\bibfield  {{Quantum error correction for quantum memories}}
  {\emph {\bibinfo {title} {{Quantum error correction for quantum memories}},\
  }}\href {\doibase 10.1103/RevModPhys.87.307} {\bibfield  {journal} {\bibinfo
  {journal} {Rev. Mod. Phys.}\ }\textbf {\bibinfo {volume} {87}},\ \bibinfo
  {pages} {307} (\bibinfo {year} {2015})}\BibitemShut {NoStop}%
\bibitem [{\citenamefont {Paz}\ and\ \citenamefont {Zurek}(1998)}]{Paz1998}%
  \BibitemOpen
  \bibfield  {author} {\bibinfo {author} {\bibfnamefont {J.~P.}\ \bibnamefont
  {Paz}}\ and\ \bibinfo {author} {\bibfnamefont {W.~H.}\ \bibnamefont
  {Zurek}},\ }\bibfield  {{Continuous error correction}} {\emph {\bibinfo
  {title} {{Continuous error correction}},\ }}\href
  {http://rspa.royalsocietypublishing.org/content/454/1969/355} {\bibfield
  {journal} {\bibinfo  {journal} {Proc. R. Soc. A}\ }\textbf {\bibinfo {volume}
  {454}} (\bibinfo {year} {1998})}\BibitemShut {NoStop}%
\bibitem [{\citenamefont {Barnes}\ and\ \citenamefont
  {Warren}(2000)}]{Barnes2000}%
  \BibitemOpen
  \bibfield  {author} {\bibinfo {author} {\bibfnamefont {J.~P.}\ \bibnamefont
  {Barnes}}\ and\ \bibinfo {author} {\bibfnamefont {W.~S.}\ \bibnamefont
  {Warren}},\ }\bibfield  {{Automatic Quantum Error Correction}} {\emph
  {\bibinfo {title} {{Automatic Quantum Error Correction}},\ }}\href {\doibase
  10.1103/PhysRevLett.85.856} {\bibfield  {journal} {\bibinfo  {journal} {Phys.
  Rev. Lett.}\ }\textbf {\bibinfo {volume} {85}},\ \bibinfo {pages} {856}
  (\bibinfo {year} {2000})}\BibitemShut {NoStop}%
\bibitem [{\citenamefont {Ahn}\ \emph {et~al.}(2002)\citenamefont {Ahn},
  \citenamefont {Doherty},\ and\ \citenamefont {Landahl}}]{Ahn2002}%
  \BibitemOpen
  \bibfield  {author} {\bibinfo {author} {\bibfnamefont {C.}~\bibnamefont
  {Ahn}}, \bibinfo {author} {\bibfnamefont {A.~C.}\ \bibnamefont {Doherty}}, \
  and\ \bibinfo {author} {\bibfnamefont {A.~J.}\ \bibnamefont {Landahl}},\
  }\bibfield  {{Continuous quantum error correction via quantum feedback
  control}} {\emph {\bibinfo {title} {{Continuous quantum error correction via
  quantum feedback control}},\ }}\href {\doibase 10.1103/PhysRevA.65.042301}
  {\bibfield  {journal} {\bibinfo  {journal} {Phys. Rev. A}\ }\textbf {\bibinfo
  {volume} {65}},\ \bibinfo {pages} {042301} (\bibinfo {year}
  {2002})}\BibitemShut {NoStop}%
\bibitem [{\citenamefont {Sarovar}\ and\ \citenamefont
  {Milburn}(2005)}]{Sarovar2005}%
  \BibitemOpen
  \bibfield  {author} {\bibinfo {author} {\bibfnamefont {M.}~\bibnamefont
  {Sarovar}}\ and\ \bibinfo {author} {\bibfnamefont {G.~J.}\ \bibnamefont
  {Milburn}},\ }\bibfield  {{Continuous quantum error correction by cooling}}
  {\emph {\bibinfo {title} {{Continuous quantum error correction by cooling}},\
  }}\href {\doibase 10.1103/PhysRevA.72.012306} {\bibfield  {journal} {\bibinfo
   {journal} {Phys. Rev. A}\ }\textbf {\bibinfo {volume} {72}},\ \bibinfo
  {pages} {012306} (\bibinfo {year} {2005})}\BibitemShut {NoStop}%
\bibitem [{\citenamefont {Oreshkov}\ and\ \citenamefont
  {Brun}(2007)}]{Oreshkov2007}%
  \BibitemOpen
  \bibfield  {author} {\bibinfo {author} {\bibfnamefont {O.}~\bibnamefont
  {Oreshkov}}\ and\ \bibinfo {author} {\bibfnamefont {T.~A.}\ \bibnamefont
  {Brun}},\ }\bibfield  {{Continuous quantum error correction for non-Markovian
  decoherence}} {\emph {\bibinfo {title} {{Continuous quantum error correction
  for non-Markovian decoherence}},\ }}\href {\doibase
  10.1103/PhysRevA.76.022318} {\bibfield  {journal} {\bibinfo  {journal} {Phys.
  Rev. A}\ }\textbf {\bibinfo {volume} {76}},\ \bibinfo {pages} {022318}
  (\bibinfo {year} {2007})}\BibitemShut {NoStop}%
\bibitem [{\citenamefont {Kerckhoff}\ \emph {et~al.}(2010)\citenamefont
  {Kerckhoff}, \citenamefont {Nurdin}, \citenamefont {Pavlichin},\ and\
  \citenamefont {Mabuchi}}]{Kerckhoff2010}%
  \BibitemOpen
  \bibfield  {author} {\bibinfo {author} {\bibfnamefont {J.}~\bibnamefont
  {Kerckhoff}}, \bibinfo {author} {\bibfnamefont {H.~I.}\ \bibnamefont
  {Nurdin}}, \bibinfo {author} {\bibfnamefont {D.~S.}\ \bibnamefont
  {Pavlichin}}, \ and\ \bibinfo {author} {\bibfnamefont {H.}~\bibnamefont
  {Mabuchi}},\ }\bibfield  {{Designing Quantum Memories with Embedded Control:
  Photonic Circuits for Autonomous Quantum Error Correction}} {\emph {\bibinfo
  {title} {{Designing Quantum Memories with Embedded Control: Photonic Circuits
  for Autonomous Quantum Error Correction}},\ }}\href {\doibase
  10.1103/PhysRevLett.105.040502} {\bibfield  {journal} {\bibinfo  {journal}
  {Phys. Rev. Lett.}\ }\textbf {\bibinfo {volume} {105}},\ \bibinfo {pages}
  {040502} (\bibinfo {year} {2010})}\BibitemShut {NoStop}%
\bibitem [{\citenamefont {Kerckhoff}\ \emph {et~al.}(2011)\citenamefont
  {Kerckhoff}, \citenamefont {Pavlichin}, \citenamefont {Chalabi},\ and\
  \citenamefont {Mabuchi}}]{Kerckhoff2011}%
  \BibitemOpen
  \bibfield  {author} {\bibinfo {author} {\bibfnamefont {J.}~\bibnamefont
  {Kerckhoff}}, \bibinfo {author} {\bibfnamefont {D.~S.}\ \bibnamefont
  {Pavlichin}}, \bibinfo {author} {\bibfnamefont {H.}~\bibnamefont {Chalabi}},
  \ and\ \bibinfo {author} {\bibfnamefont {H.}~\bibnamefont {Mabuchi}},\
  }\bibfield  {{Design of nanophotonic circuits for autonomous subsystem
  quantum error correction}} {\emph {\bibinfo {title} {{Design of nanophotonic
  circuits for autonomous subsystem quantum error correction}},\ }}\href
  {\doibase 10.1088/1367-2630/13/5/055022} {\bibfield  {journal} {\bibinfo
  {journal} {New J. Phys.}\ }\textbf {\bibinfo {volume} {13}},\ \bibinfo
  {pages} {055022} (\bibinfo {year} {2011})}\BibitemShut {NoStop}%
\bibitem [{\citenamefont {Sarma}\ and\ \citenamefont
  {Mabuchi}(2013)}]{Sarma2013}%
  \BibitemOpen
  \bibfield  {author} {\bibinfo {author} {\bibfnamefont {G.}~\bibnamefont
  {Sarma}}\ and\ \bibinfo {author} {\bibfnamefont {H.}~\bibnamefont
  {Mabuchi}},\ }\bibfield  {{Gauge subsystems, separability and robustness in
  autonomous quantum memories}} {\emph {\bibinfo {title} {{Gauge subsystems,
  separability and robustness in autonomous quantum memories}},\ }}\href
  {\doibase 10.1088/1367-2630/15/3/035014} {\bibfield  {journal} {\bibinfo
  {journal} {New J. Phys.}\ }\textbf {\bibinfo {volume} {15}},\ \bibinfo
  {pages} {035014} (\bibinfo {year} {2013})}\BibitemShut {NoStop}%
\bibitem [{\citenamefont {Lihm}\ \emph {et~al.}(2018)\citenamefont {Lihm},
  \citenamefont {Noh},\ and\ \citenamefont {Fischer}}]{Jae-MoLihmKyungjooNoh}%
  \BibitemOpen
  \bibfield  {author} {\bibinfo {author} {\bibfnamefont {J.-M.}\ \bibnamefont
  {Lihm}}, \bibinfo {author} {\bibfnamefont {K.}~\bibnamefont {Noh}}, \ and\
  \bibinfo {author} {\bibfnamefont {U.~R.}\ \bibnamefont {Fischer}},\
  }\bibfield  {{Implementation-independent sufficient condition of the
  Knill-Laflamme type for the autonomous protection of logical qudits by strong
  engineered dissipation}} {\emph {\bibinfo {title}
  {{Implementation-independent sufficient condition of the Knill-Laflamme type
  for the autonomous protection of logical qudits by strong engineered
  dissipation}},\ }}\href {\doibase 10.1103/PhysRevA.98.012317} {\bibfield
  {journal} {\bibinfo  {journal} {Phys. Rev. A}\ }\textbf {\bibinfo {volume}
  {98}},\ \bibinfo {pages} {012317} (\bibinfo {year} {2018})}\BibitemShut
  {NoStop}%
\bibitem [{\citenamefont {Cohen}(2017)}]{cohenthesis}%
  \BibitemOpen
  \bibfield  {author} {\bibinfo {author} {\bibfnamefont {J.}~\bibnamefont
  {Cohen}},\ }\emph {\bibinfo {title} {{Autonomous quantum error correction
  with superconducting qubits}}},\ \href
  {https://tel.archives-ouvertes.fr/tel-01545186} {Ph.D. thesis},\ \bibinfo
  {school} {Ecole Normale Superieure} (\bibinfo {year} {2017})\BibitemShut
  {NoStop}%
\bibitem [{\citenamefont {Sun}\ \emph {et~al.}(2014)\citenamefont {Sun},
  \citenamefont {Petrenko}, \citenamefont {Leghtas}, \citenamefont {Vlastakis},
  \citenamefont {Kirchmair}, \citenamefont {Sliwa}, \citenamefont {Narla},
  \citenamefont {Hatridge}, \citenamefont {Shankar}, \citenamefont {Blumoff},
  \citenamefont {Frunzio}, \citenamefont {Mirrahimi}, \citenamefont {Devoret},\
  and\ \citenamefont {Schoelkopf}}]{Sun2014}%
  \BibitemOpen
  \bibfield  {author} {\bibinfo {author} {\bibfnamefont {L.}~\bibnamefont
  {Sun}}, \bibinfo {author} {\bibfnamefont {A.}~\bibnamefont {Petrenko}},
  \bibinfo {author} {\bibfnamefont {Z.}~\bibnamefont {Leghtas}}, \bibinfo
  {author} {\bibfnamefont {B.}~\bibnamefont {Vlastakis}}, \bibinfo {author}
  {\bibfnamefont {G.}~\bibnamefont {Kirchmair}}, \bibinfo {author}
  {\bibfnamefont {K.~M.}\ \bibnamefont {Sliwa}}, \bibinfo {author}
  {\bibfnamefont {A.}~\bibnamefont {Narla}}, \bibinfo {author} {\bibfnamefont
  {M.}~\bibnamefont {Hatridge}}, \bibinfo {author} {\bibfnamefont
  {S.}~\bibnamefont {Shankar}}, \bibinfo {author} {\bibfnamefont
  {J.}~\bibnamefont {Blumoff}}, \bibinfo {author} {\bibfnamefont
  {L.}~\bibnamefont {Frunzio}}, \bibinfo {author} {\bibfnamefont
  {M.}~\bibnamefont {Mirrahimi}}, \bibinfo {author} {\bibfnamefont {M.~H.}\
  \bibnamefont {Devoret}}, \ and\ \bibinfo {author} {\bibfnamefont {R.~J.}\
  \bibnamefont {Schoelkopf}},\ }\bibfield  {{Tracking photon jumps with
  repeated quantum non-demolition parity measurements.}} {\emph {\bibinfo
  {title} {{Tracking photon jumps with repeated quantum non-demolition parity
  measurements.}}\ }}\href {\doibase 10.1038/nature13436} {\bibfield  {journal}
  {\bibinfo  {journal} {Nature}\ }\textbf {\bibinfo {volume} {511}},\ \bibinfo
  {pages} {444} (\bibinfo {year} {2014})}\BibitemShut {NoStop}%
\bibitem [{\citenamefont {Cohen}\ \emph {et~al.}(2017)\citenamefont {Cohen},
  \citenamefont {Smith}, \citenamefont {Devoret},\ and\ \citenamefont
  {Mirrahimi}}]{Cohen2016}%
  \BibitemOpen
  \bibfield  {author} {\bibinfo {author} {\bibfnamefont {J.}~\bibnamefont
  {Cohen}}, \bibinfo {author} {\bibfnamefont {W.~C.}\ \bibnamefont {Smith}},
  \bibinfo {author} {\bibfnamefont {M.~H.}\ \bibnamefont {Devoret}}, \ and\
  \bibinfo {author} {\bibfnamefont {M.}~\bibnamefont {Mirrahimi}},\ }\bibfield
  {{Degeneracy-Preserving Quantum Nondemolition Measurement of Parity-Type
  Observables for Cat Qubits}} {\emph {\bibinfo {title} {{Degeneracy-Preserving
  Quantum Nondemolition Measurement of Parity-Type Observables for Cat
  Qubits}},\ }}\href {\doibase 10.1103/PhysRevLett.119.060503} {\bibfield
  {journal} {\bibinfo  {journal} {Phys. Rev. Lett.}\ }\textbf {\bibinfo
  {volume} {119}},\ \bibinfo {pages} {060503} (\bibinfo {year}
  {2017})}\BibitemShut {NoStop}%
\bibitem [{\citenamefont {Frattini}\ \emph {et~al.}(2017)\citenamefont
  {Frattini}, \citenamefont {Vool}, \citenamefont {Shankar}, \citenamefont
  {Narla}, \citenamefont {Sliwa},\ and\ \citenamefont
  {Devoret}}]{Frattini2017}%
  \BibitemOpen
  \bibfield  {author} {\bibinfo {author} {\bibfnamefont {N.~E.}\ \bibnamefont
  {Frattini}}, \bibinfo {author} {\bibfnamefont {U.}~\bibnamefont {Vool}},
  \bibinfo {author} {\bibfnamefont {S.}~\bibnamefont {Shankar}}, \bibinfo
  {author} {\bibfnamefont {A.}~\bibnamefont {Narla}}, \bibinfo {author}
  {\bibfnamefont {K.~M.}\ \bibnamefont {Sliwa}}, \ and\ \bibinfo {author}
  {\bibfnamefont {M.~H.}\ \bibnamefont {Devoret}},\ }\bibfield  {{3-wave mixing
  Josephson dipole element}} {\emph {\bibinfo {title} {{3-wave mixing Josephson
  dipole element}},\ }}\href {\doibase 10.1063/1.4984142} {\bibfield  {journal}
  {\bibinfo  {journal} {Appl. Phys. Lett.}\ }\textbf {\bibinfo {volume}
  {110}},\ \bibinfo {pages} {222603} (\bibinfo {year} {2017})}\BibitemShut
  {NoStop}%
\bibitem [{\citenamefont {Mundadha}\ \emph {et~al.}(2017)\citenamefont
  {Mundadha}, \citenamefont {Grimm}, \citenamefont {Touzard}, \citenamefont
  {Vool}, \citenamefont {Shankar}, \citenamefont {Devoret},\ and\ \citenamefont
  {Mirrahimi}}]{Mundhada2017}%
  \BibitemOpen
  \bibfield  {author} {\bibinfo {author} {\bibfnamefont {S.~O.}\ \bibnamefont
  {Mundadha}}, \bibinfo {author} {\bibfnamefont {A.}~\bibnamefont {Grimm}},
  \bibinfo {author} {\bibfnamefont {S.}~\bibnamefont {Touzard}}, \bibinfo
  {author} {\bibfnamefont {U.}~\bibnamefont {Vool}}, \bibinfo {author}
  {\bibfnamefont {S.}~\bibnamefont {Shankar}}, \bibinfo {author} {\bibfnamefont
  {M.~H.}\ \bibnamefont {Devoret}}, \ and\ \bibinfo {author} {\bibfnamefont
  {M.}~\bibnamefont {Mirrahimi}},\ }\bibfield  {{Generating higher-order
  quantum dissipation from lower-order parametric processes}} {\emph {\bibinfo
  {title} {{Generating higher-order quantum dissipation from lower-order
  parametric processes}},\ }}\href
  {http://stacks.iop.org/2058-9565/2/i=2/a=024005} {\bibfield  {journal}
  {\bibinfo  {journal} {Quantum Sci. Technol.}\ }\textbf {\bibinfo {volume}
  {2}},\ \bibinfo {pages} {24005} (\bibinfo {year} {2017})}\BibitemShut
  {NoStop}%
\bibitem [{\citenamefont {Munro}\ \emph {et~al.}(2002)\citenamefont {Munro},
  \citenamefont {Nemoto}, \citenamefont {Milburn},\ and\ \citenamefont
  {Braunstein}}]{Munro2002}%
  \BibitemOpen
  \bibfield  {author} {\bibinfo {author} {\bibfnamefont {W.~J.}\ \bibnamefont
  {Munro}}, \bibinfo {author} {\bibfnamefont {K.}~\bibnamefont {Nemoto}},
  \bibinfo {author} {\bibfnamefont {G.~J.}\ \bibnamefont {Milburn}}, \ and\
  \bibinfo {author} {\bibfnamefont {S.~L.}\ \bibnamefont {Braunstein}},\
  }\bibfield  {{Weak-force detection with superposed coherent states}} {\emph
  {\bibinfo {title} {{Weak-force detection with superposed coherent states}},\
  }}\href {\doibase 10.1103/PhysRevA.66.023819} {\bibfield  {journal} {\bibinfo
   {journal} {Phys. Rev. A}\ }\textbf {\bibinfo {volume} {66}},\ \bibinfo
  {pages} {023819} (\bibinfo {year} {2002})}\BibitemShut {NoStop}%
\bibitem [{\citenamefont {Zhang}\ \emph {et~al.}(2017)\citenamefont {Zhang},
  \citenamefont {Zhao}, \citenamefont {Zheng}, \citenamefont {Yu},
  \citenamefont {Su},\ and\ \citenamefont {Yang}}]{zhang2017}%
  \BibitemOpen
  \bibfield  {author} {\bibinfo {author} {\bibfnamefont {Y.}~\bibnamefont
  {Zhang}}, \bibinfo {author} {\bibfnamefont {X.}~\bibnamefont {Zhao}},
  \bibinfo {author} {\bibfnamefont {Z.-F.}\ \bibnamefont {Zheng}}, \bibinfo
  {author} {\bibfnamefont {L.}~\bibnamefont {Yu}}, \bibinfo {author}
  {\bibfnamefont {Q.-P.}\ \bibnamefont {Su}}, \ and\ \bibinfo {author}
  {\bibfnamefont {C.-P.}\ \bibnamefont {Yang}},\ }\bibfield  {{Universal
  controlled-phase gate with cat-state qubits in circuit QED}} {\emph {\bibinfo
  {title} {{Universal controlled-phase gate with cat-state qubits in circuit
  QED}},\ }}\href {\doibase 10.1103/PhysRevA.96.052317} {\bibfield  {journal}
  {\bibinfo  {journal} {Phys. Rev. A}\ }\textbf {\bibinfo {volume} {96}},\
  \bibinfo {pages} {052317} (\bibinfo {year} {2017})}\BibitemShut {NoStop}%
\bibitem [{\citenamefont {Bennett}\ \emph {et~al.}(1996)\citenamefont
  {Bennett}, \citenamefont {DiVincenzo}, \citenamefont {Smolin},\ and\
  \citenamefont {Wootters}}]{Bennett1996}%
  \BibitemOpen
  \bibfield  {author} {\bibinfo {author} {\bibfnamefont {C.~H.}\ \bibnamefont
  {Bennett}}, \bibinfo {author} {\bibfnamefont {D.~P.}\ \bibnamefont
  {DiVincenzo}}, \bibinfo {author} {\bibfnamefont {J.~A.}\ \bibnamefont
  {Smolin}}, \ and\ \bibinfo {author} {\bibfnamefont {W.~K.}\ \bibnamefont
  {Wootters}},\ }\bibfield  {{Mixed-state entanglement and quantum error
  correction}} {\emph {\bibinfo {title} {{Mixed-state entanglement and quantum
  error correction}},\ }}\href {\doibase 10.1103/PhysRevA.54.3824} {\bibfield
  {journal} {\bibinfo  {journal} {Phys. Rev. A}\ }\textbf {\bibinfo {volume}
  {54}},\ \bibinfo {pages} {3824} (\bibinfo {year} {1996})}\BibitemShut
  {NoStop}%
\bibitem [{\citenamefont {Knill}\ and\ \citenamefont
  {Laflamme}(1997)}]{Knill1997}%
  \BibitemOpen
  \bibfield  {author} {\bibinfo {author} {\bibfnamefont {E.}~\bibnamefont
  {Knill}}\ and\ \bibinfo {author} {\bibfnamefont {R.}~\bibnamefont
  {Laflamme}},\ }\bibfield  {{Theory of quantum error-correcting codes}} {\emph
  {\bibinfo {title} {{Theory of quantum error-correcting codes}},\ }}\href
  {\doibase 10.1103/PhysRevA.55.900} {\bibfield  {journal} {\bibinfo  {journal}
  {Phys. Rev. A}\ }\textbf {\bibinfo {volume} {55}},\ \bibinfo {pages} {900}
  (\bibinfo {year} {1997})}\BibitemShut {NoStop}%
\bibitem [{\citenamefont {Nielsen}\ and\ \citenamefont
  {Chuang}(2011)}]{nielsen_chuang}%
  \BibitemOpen
  \bibfield  {author} {\bibinfo {author} {\bibfnamefont {M.~A.}\ \bibnamefont
  {Nielsen}}\ and\ \bibinfo {author} {\bibfnamefont {I.~L.}\ \bibnamefont
  {Chuang}},\ }\href
  {http://www.amazon.com/Quantum-Computation-Information-Anniversary-Edition/dp/1107002176}
  {\emph {\bibinfo {title} {{Quantum Computation and Quantum Information}}}}\
  (\bibinfo  {publisher} {Cambridge University Press},\ \bibinfo {address}
  {Cambridge},\ \bibinfo {year} {2011})\BibitemShut {NoStop}%
\bibitem [{\citenamefont {Ueda}(1989)}]{Ueda1989}%
  \BibitemOpen
  \bibfield  {author} {\bibinfo {author} {\bibfnamefont {M.}~\bibnamefont
  {Ueda}},\ }\bibfield  {{Probability-density-functional description of quantum
  photodetection processes}} {\emph {\bibinfo {title}
  {{Probability-density-functional description of quantum photodetection
  processes}},\ }}\href {\doibase 10.1088/0954-8998/1/2/005} {\bibfield
  {journal} {\bibinfo  {journal} {Quantum Opt.}\ }\textbf {\bibinfo {volume}
  {1}},\ \bibinfo {pages} {131} (\bibinfo {year} {1989})}\BibitemShut {NoStop}%
\bibitem [{\citenamefont {Lee}(1994)}]{Lee1994}%
  \BibitemOpen
  \bibfield  {author} {\bibinfo {author} {\bibfnamefont {C.~T.}\ \bibnamefont
  {Lee}},\ }\bibfield  {{Superoperators and their implications in the hybrid
  model for photodetection}} {\emph {\bibinfo {title} {{Superoperators and
  their implications in the hybrid model for photodetection}},\ }}\href
  {\doibase 10.1103/PhysRevA.49.4888} {\bibfield  {journal} {\bibinfo
  {journal} {Phys. Rev. A}\ }\textbf {\bibinfo {volume} {49}},\ \bibinfo
  {pages} {4888} (\bibinfo {year} {1994})}\BibitemShut {NoStop}%
\bibitem [{\citenamefont {Klimov}\ and\ \citenamefont
  {Chumakov}(2009)}]{klimov_book}%
  \BibitemOpen
  \bibfield  {author} {\bibinfo {author} {\bibfnamefont {A.~B.}\ \bibnamefont
  {Klimov}}\ and\ \bibinfo {author} {\bibfnamefont {S.~M.}\ \bibnamefont
  {Chumakov}},\ }\href
  {http://www.wiley.com/WileyCDA/WileyTitle/productCd-3527408797.html} {\emph
  {\bibinfo {title} {{A Group-Theoretical Approach to Quantum Optics}}}}\
  (\bibinfo  {publisher} {Wiley},\ \bibinfo {address} {Weinheim},\ \bibinfo
  {year} {2009})\BibitemShut {NoStop}%
\bibitem [{\citenamefont {Albert}\ \emph {et~al.}(2018)\citenamefont {Albert},
  \citenamefont {Noh}, \citenamefont {Duivenvoorden}, \citenamefont {Young},
  \citenamefont {Brierley}, \citenamefont {Reinhold}, \citenamefont {Vuillot},
  \citenamefont {Li}, \citenamefont {Shen}, \citenamefont {Girvin},
  \citenamefont {Terhal},\ and\ \citenamefont {Jiang}}]{codecomp}%
  \BibitemOpen
  \bibfield  {author} {\bibinfo {author} {\bibfnamefont {V.~V.}\ \bibnamefont
  {Albert}}, \bibinfo {author} {\bibfnamefont {K.}~\bibnamefont {Noh}},
  \bibinfo {author} {\bibfnamefont {K.}~\bibnamefont {Duivenvoorden}}, \bibinfo
  {author} {\bibfnamefont {D.~J.}\ \bibnamefont {Young}}, \bibinfo {author}
  {\bibfnamefont {R.~T.}\ \bibnamefont {Brierley}}, \bibinfo {author}
  {\bibfnamefont {P.}~\bibnamefont {Reinhold}}, \bibinfo {author}
  {\bibfnamefont {C.}~\bibnamefont {Vuillot}}, \bibinfo {author} {\bibfnamefont
  {L.}~\bibnamefont {Li}}, \bibinfo {author} {\bibfnamefont {C.}~\bibnamefont
  {Shen}}, \bibinfo {author} {\bibfnamefont {S.~M.}\ \bibnamefont {Girvin}},
  \bibinfo {author} {\bibfnamefont {B.~M.}\ \bibnamefont {Terhal}}, \ and\
  \bibinfo {author} {\bibfnamefont {L.}~\bibnamefont {Jiang}},\ }\bibfield
  {{Performance and structure of single-mode bosonic codes}} {\emph {\bibinfo
  {title} {{Performance and structure of single-mode bosonic codes}},\ }}\href
  {\doibase 10.1103/PhysRevA.97.032346} {\bibfield  {journal} {\bibinfo
  {journal} {Phys. Rev. A}\ }\textbf {\bibinfo {volume} {97}},\ \bibinfo
  {pages} {032346} (\bibinfo {year} {2018})}\BibitemShut {NoStop}%
\bibitem [{\citenamefont {Albert}\ \emph
  {et~al.}(2016{\natexlab{b}})\citenamefont {Albert}, \citenamefont {Bradlyn},
  \citenamefont {Fraas},\ and\ \citenamefont {Jiang}}]{ABFJ}%
  \BibitemOpen
  \bibfield  {author} {\bibinfo {author} {\bibfnamefont {V.~V.}\ \bibnamefont
  {Albert}}, \bibinfo {author} {\bibfnamefont {B.}~\bibnamefont {Bradlyn}},
  \bibinfo {author} {\bibfnamefont {M.}~\bibnamefont {Fraas}}, \ and\ \bibinfo
  {author} {\bibfnamefont {L.}~\bibnamefont {Jiang}},\ }\bibfield  {{Geometry
  and Response of Lindbladians}} {\emph {\bibinfo {title} {{Geometry and
  Response of Lindbladians}},\ }}\href {\doibase 10.1103/PhysRevX.6.041031}
  {\bibfield  {journal} {\bibinfo  {journal} {Phys. Rev. X}\ }\textbf {\bibinfo
  {volume} {6}},\ \bibinfo {pages} {041031} (\bibinfo {year}
  {2016}{\natexlab{b}})}\BibitemShut {NoStop}%
\bibitem [{\citenamefont {Zanardi}\ and\ \citenamefont {{Campos
  Venuti}}(2014)}]{Zanardi2014}%
  \BibitemOpen
  \bibfield  {author} {\bibinfo {author} {\bibfnamefont {P.}~\bibnamefont
  {Zanardi}}\ and\ \bibinfo {author} {\bibfnamefont {L.}~\bibnamefont {{Campos
  Venuti}}},\ }\bibfield  {{Coherent Quantum Dynamics in Steady-State Manifolds
  of Strongly Dissipative Systems}} {\emph {\bibinfo {title} {{Coherent Quantum
  Dynamics in Steady-State Manifolds of Strongly Dissipative Systems}},\
  }}\href {\doibase 10.1103/PhysRevLett.113.240406} {\bibfield  {journal}
  {\bibinfo  {journal} {Phys. Rev. Lett.}\ }\textbf {\bibinfo {volume} {113}},\
  \bibinfo {pages} {240406} (\bibinfo {year} {2014})}\BibitemShut {NoStop}%
\bibitem [{\citenamefont {Azouit}\ \emph {et~al.}(2016)\citenamefont {Azouit},
  \citenamefont {Sarlette},\ and\ \citenamefont {Rouchon}}]{Azouit2016}%
  \BibitemOpen
  \bibfield  {author} {\bibinfo {author} {\bibfnamefont {R.}~\bibnamefont
  {Azouit}}, \bibinfo {author} {\bibfnamefont {A.}~\bibnamefont {Sarlette}}, \
  and\ \bibinfo {author} {\bibfnamefont {P.}~\bibnamefont {Rouchon}},\ }in\
  \href {\doibase 10.1109/CDC.2016.7798963} {\emph {\bibinfo {booktitle} {2016
  IEEE 55th Conf. Decis. Control}}}\ (\bibinfo  {publisher} {IEEE},\ \bibinfo
  {year} {2016})\ pp.\ \bibinfo {pages} {4559--4565}\BibitemShut {NoStop}%
\bibitem [{\citenamefont {Facchi}\ and\ \citenamefont
  {Pascazio}(2002)}]{Facchi2002}%
  \BibitemOpen
  \bibfield  {author} {\bibinfo {author} {\bibfnamefont {P.}~\bibnamefont
  {Facchi}}\ and\ \bibinfo {author} {\bibfnamefont {S.}~\bibnamefont
  {Pascazio}},\ }\bibfield  {{Quantum Zeno Subspaces}} {\emph {\bibinfo {title}
  {{Quantum Zeno Subspaces}},\ }}\href {\doibase 10.1103/PhysRevLett.89.080401}
  {\bibfield  {journal} {\bibinfo  {journal} {Phys. Rev. Lett.}\ }\textbf
  {\bibinfo {volume} {89}},\ \bibinfo {pages} {080401} (\bibinfo {year}
  {2002})}\BibitemShut {NoStop}%
\bibitem [{\citenamefont {Arenz}\ \emph {et~al.}(2016)\citenamefont {Arenz},
  \citenamefont {Burgarth}, \citenamefont {Facchi}, \citenamefont
  {Giovannetti}, \citenamefont {Nakazato}, \citenamefont {Pascazio},\ and\
  \citenamefont {Yuasa}}]{Arenz2016}%
  \BibitemOpen
  \bibfield  {author} {\bibinfo {author} {\bibfnamefont {C.}~\bibnamefont
  {Arenz}}, \bibinfo {author} {\bibfnamefont {D.}~\bibnamefont {Burgarth}},
  \bibinfo {author} {\bibfnamefont {P.}~\bibnamefont {Facchi}}, \bibinfo
  {author} {\bibfnamefont {V.}~\bibnamefont {Giovannetti}}, \bibinfo {author}
  {\bibfnamefont {H.}~\bibnamefont {Nakazato}}, \bibinfo {author}
  {\bibfnamefont {S.}~\bibnamefont {Pascazio}}, \ and\ \bibinfo {author}
  {\bibfnamefont {K.}~\bibnamefont {Yuasa}},\ }\bibfield  {{Universal Control
  Induced by Noise}} {\emph {\bibinfo {title} {{Universal Control Induced by
  Noise}},\ }}\href {\doibase 10.1103/PhysRevA.93.062308} {\bibfield  {journal}
  {\bibinfo  {journal} {Phys. Rev. A}\ }\textbf {\bibinfo {volume} {93}},\
  \bibinfo {pages} {062308} (\bibinfo {year} {2016})}\BibitemShut {NoStop}%
\bibitem [{\citenamefont {Bu{\v{c}}a}\ and\ \citenamefont
  {Prosen}(2012)}]{prozen}%
  \BibitemOpen
  \bibfield  {author} {\bibinfo {author} {\bibfnamefont {B.}~\bibnamefont
  {Bu{\v{c}}a}}\ and\ \bibinfo {author} {\bibfnamefont {T.}~\bibnamefont
  {Prosen}},\ }\bibfield  {{A note on symmetry reductions of the Lindblad
  equation: transport in constrained open spin chains}} {\emph {\bibinfo
  {title} {{A note on symmetry reductions of the Lindblad equation: transport
  in constrained open spin chains}},\ }}\href {\doibase
  10.1088/1367-2630/14/7/073007} {\bibfield  {journal} {\bibinfo  {journal}
  {New J. Phys.}\ }\textbf {\bibinfo {volume} {14}},\ \bibinfo {pages} {073007}
  (\bibinfo {year} {2012})}\BibitemShut {NoStop}%
\bibitem [{\citenamefont {Albert}\ and\ \citenamefont {Jiang}(2014)}]{pub011}%
  \BibitemOpen
  \bibfield  {author} {\bibinfo {author} {\bibfnamefont {V.~V.}\ \bibnamefont
  {Albert}}\ and\ \bibinfo {author} {\bibfnamefont {L.}~\bibnamefont {Jiang}},\
  }\bibfield  {{Symmetries and conserved quantities in Lindblad master
  equations}} {\emph {\bibinfo {title} {{Symmetries and conserved quantities in
  Lindblad master equations}},\ }}\href {\doibase 10.1103/PhysRevA.89.022118}
  {\bibfield  {journal} {\bibinfo  {journal} {Phys. Rev. A}\ }\textbf {\bibinfo
  {volume} {89}},\ \bibinfo {pages} {022118} (\bibinfo {year}
  {2014})}\BibitemShut {NoStop}%
\bibitem [{\citenamefont {Dodonov}\ \emph {et~al.}(1974)\citenamefont
  {Dodonov}, \citenamefont {Malkin},\ and\ \citenamefont
  {Man'ko}}]{Dodonov1974}%
  \BibitemOpen
  \bibfield  {author} {\bibinfo {author} {\bibfnamefont {V.~V.}\ \bibnamefont
  {Dodonov}}, \bibinfo {author} {\bibfnamefont {I.}~\bibnamefont {Malkin}}, \
  and\ \bibinfo {author} {\bibfnamefont {V.}~\bibnamefont {Man'ko}},\
  }\bibfield  {{Even and odd coherent states and excitations of a singular
  oscillator}} {\emph {\bibinfo {title} {{Even and odd coherent states and
  excitations of a singular oscillator}},\ }}\href {\doibase
  10.1016/0031-8914(74)90215-8} {\bibfield  {journal} {\bibinfo  {journal}
  {Physica}\ }\textbf {\bibinfo {volume} {72}},\ \bibinfo {pages} {597}
  (\bibinfo {year} {1974})}\BibitemShut {NoStop}%
\bibitem [{\citenamefont {Vaughn}(2007)}]{vaughn_book}%
  \BibitemOpen
  \bibfield  {author} {\bibinfo {author} {\bibfnamefont {M.~T.}\ \bibnamefont
  {Vaughn}},\ }\href
  {http://ca.wiley.com/WileyCDA/WileyTitle/productCd-3527406271.html} {\emph
  {\bibinfo {title} {{Introduction to Mathematical Physics}}}}\ (\bibinfo
  {publisher} {Wiley},\ \bibinfo {address} {Weinheim},\ \bibinfo {year}
  {2007})\BibitemShut {NoStop}%
\bibitem [{\citenamefont {Albert}(2017)}]{thesis}%
  \BibitemOpen
  \bibfield  {author} {\bibinfo {author} {\bibfnamefont {V.~V.}\ \bibnamefont
  {Albert}},\ }\emph {\bibinfo {title} {{Lindbladians with multiple steady
  states: theory and applications}}},\ \href {https://arxiv.org/abs/1802.00010}
  {Ph.D. thesis},\ \bibinfo  {school} {Yale University} (\bibinfo {year}
  {2017})\BibitemShut {NoStop}%
\bibitem [{\citenamefont {Puri}\ \emph {et~al.}(2017)\citenamefont {Puri},
  \citenamefont {Boutin},\ and\ \citenamefont {Blais}}]{Puri2017}%
  \BibitemOpen
  \bibfield  {author} {\bibinfo {author} {\bibfnamefont {S.}~\bibnamefont
  {Puri}}, \bibinfo {author} {\bibfnamefont {S.}~\bibnamefont {Boutin}}, \ and\
  \bibinfo {author} {\bibfnamefont {A.}~\bibnamefont {Blais}},\ }\bibfield
  {{Engineering the quantum states of light in a Kerr-nonlinear resonator by
  two-photon driving}} {\emph {\bibinfo {title} {{Engineering the quantum
  states of light in a Kerr-nonlinear resonator by two-photon driving}},\
  }}\href {\doibase 10.1038/s41534-017-0019-1} {\bibfield  {journal} {\bibinfo
  {journal} {npj Quantum Inf.}\ }\textbf {\bibinfo {volume} {3}},\ \bibinfo
  {pages} {18} (\bibinfo {year} {2017})}\BibitemShut {NoStop}%
\bibitem [{\citenamefont {Ippoliti}\ \emph {et~al.}(2015)\citenamefont
  {Ippoliti}, \citenamefont {Mazza}, \citenamefont {Rizzi},\ and\ \citenamefont
  {Giovannetti}}]{Ippoliti2014}%
  \BibitemOpen
  \bibfield  {author} {\bibinfo {author} {\bibfnamefont {M.}~\bibnamefont
  {Ippoliti}}, \bibinfo {author} {\bibfnamefont {L.}~\bibnamefont {Mazza}},
  \bibinfo {author} {\bibfnamefont {M.}~\bibnamefont {Rizzi}}, \ and\ \bibinfo
  {author} {\bibfnamefont {V.}~\bibnamefont {Giovannetti}},\ }\bibfield
  {{Perturbative approach to continuous-time quantum error correction}} {\emph
  {\bibinfo {title} {{Perturbative approach to continuous-time quantum error
  correction}},\ }}\href {\doibase 10.1103/PhysRevA.91.042322} {\bibfield
  {journal} {\bibinfo  {journal} {Phys. Rev. A}\ }\textbf {\bibinfo {volume}
  {91}},\ \bibinfo {pages} {042322} (\bibinfo {year} {2015})}\BibitemShut
  {NoStop}%
\bibitem [{\citenamefont {Reiter}\ \emph {et~al.}(2017)\citenamefont {Reiter},
  \citenamefont {S{\o}rensen}, \citenamefont {Zoller},\ and\ \citenamefont
  {Muschik}}]{Reiter2017}%
  \BibitemOpen
  \bibfield  {author} {\bibinfo {author} {\bibfnamefont {F.}~\bibnamefont
  {Reiter}}, \bibinfo {author} {\bibfnamefont {A.~S.}\ \bibnamefont
  {S{\o}rensen}}, \bibinfo {author} {\bibfnamefont {P.}~\bibnamefont {Zoller}},
  \ and\ \bibinfo {author} {\bibfnamefont {C.~A.}\ \bibnamefont {Muschik}},\
  }\bibfield  {{Dissipative quantum error correction and application to quantum
  sensing with trapped ions}} {\emph {\bibinfo {title} {{Dissipative quantum
  error correction and application to quantum sensing with trapped ions}},\
  }}\href {\doibase 10.1038/s41467-017-01895-5} {\bibfield  {journal} {\bibinfo
   {journal} {Nat. Commun.}\ }\textbf {\bibinfo {volume} {8}},\ \bibinfo
  {pages} {1822} (\bibinfo {year} {2017})}\BibitemShut {NoStop}%
\bibitem [{\citenamefont {Cohen}\ and\ \citenamefont
  {Mirrahimi}(2014)}]{Cohen2014}%
  \BibitemOpen
  \bibfield  {author} {\bibinfo {author} {\bibfnamefont {J.}~\bibnamefont
  {Cohen}}\ and\ \bibinfo {author} {\bibfnamefont {M.}~\bibnamefont
  {Mirrahimi}},\ }\bibfield  {{Dissipation-induced continuous quantum error
  correction for superconducting circuits}} {\emph {\bibinfo {title}
  {{Dissipation-induced continuous quantum error correction for superconducting
  circuits}},\ }}\href {\doibase 10.1103/PhysRevA.90.062344} {\bibfield
  {journal} {\bibinfo  {journal} {Phys. Rev. A}\ }\textbf {\bibinfo {volume}
  {90}},\ \bibinfo {pages} {062344} (\bibinfo {year} {2014})}\BibitemShut
  {NoStop}%
\bibitem [{\citenamefont {Knill}(2005)}]{Knill2005}%
  \BibitemOpen
  \bibfield  {author} {\bibinfo {author} {\bibfnamefont {E.}~\bibnamefont
  {Knill}},\ }\bibfield  {{Quantum computing with realistically noisy devices}}
  {\emph {\bibinfo {title} {{Quantum computing with realistically noisy
  devices}},\ }}\href {\doibase 10.1038/nature03350} {\bibfield  {journal}
  {\bibinfo  {journal} {Nature}\ }\textbf {\bibinfo {volume} {434}},\ \bibinfo
  {pages} {39} (\bibinfo {year} {2005})}\BibitemShut {NoStop}%
\bibitem [{\citenamefont {Wilczek}\ and\ \citenamefont
  {Zee}(1984)}]{Wilczek1984}%
  \BibitemOpen
  \bibfield  {author} {\bibinfo {author} {\bibfnamefont {F.}~\bibnamefont
  {Wilczek}}\ and\ \bibinfo {author} {\bibfnamefont {A.}~\bibnamefont {Zee}},\
  }\bibfield  {{Appearance of Gauge Structure in Simple Dynamical Systems}}
  {\emph {\bibinfo {title} {{Appearance of Gauge Structure in Simple Dynamical
  Systems}},\ }}\href {\doibase 10.1103/PhysRevLett.52.2111} {\bibfield
  {journal} {\bibinfo  {journal} {Phys. Rev. Lett.}\ }\textbf {\bibinfo
  {volume} {52}},\ \bibinfo {pages} {2111} (\bibinfo {year}
  {1984})}\BibitemShut {NoStop}%
\bibitem [{\citenamefont {Krastanov}\ \emph {et~al.}(2015)\citenamefont
  {Krastanov}, \citenamefont {Albert}, \citenamefont {Shen}, \citenamefont
  {Zou}, \citenamefont {Heeres}, \citenamefont {Vlastakis}, \citenamefont
  {Schoelkopf},\ and\ \citenamefont {Jiang}}]{Krastanov2015}%
  \BibitemOpen
  \bibfield  {author} {\bibinfo {author} {\bibfnamefont {S.}~\bibnamefont
  {Krastanov}}, \bibinfo {author} {\bibfnamefont {V.~V.}\ \bibnamefont
  {Albert}}, \bibinfo {author} {\bibfnamefont {C.}~\bibnamefont {Shen}},
  \bibinfo {author} {\bibfnamefont {C.-L.}\ \bibnamefont {Zou}}, \bibinfo
  {author} {\bibfnamefont {R.}~\bibnamefont {Heeres}}, \bibinfo {author}
  {\bibfnamefont {B.}~\bibnamefont {Vlastakis}}, \bibinfo {author}
  {\bibfnamefont {R.}~\bibnamefont {Schoelkopf}}, \ and\ \bibinfo {author}
  {\bibfnamefont {L.}~\bibnamefont {Jiang}},\ }\bibfield  {{Universal control
  of an oscillator with dispersive coupling to a qubit}} {\emph {\bibinfo
  {title} {{Universal control of an oscillator with dispersive coupling to a
  qubit}},\ }}\href {\doibase 10.1103/PhysRevA.92.040303} {\bibfield  {journal}
  {\bibinfo  {journal} {Phys. Rev. A}\ }\textbf {\bibinfo {volume} {92}},\
  \bibinfo {pages} {040303(R)} (\bibinfo {year} {2015})}\BibitemShut {NoStop}%
\bibitem [{\citenamefont {Heeres}\ \emph {et~al.}(2015)\citenamefont {Heeres},
  \citenamefont {Vlastakis}, \citenamefont {Holland}, \citenamefont
  {Krastanov}, \citenamefont {Albert}, \citenamefont {Frunzio}, \citenamefont
  {Jiang},\ and\ \citenamefont {Schoelkopf}}]{Heeres2015}%
  \BibitemOpen
  \bibfield  {author} {\bibinfo {author} {\bibfnamefont {R.~W.}\ \bibnamefont
  {Heeres}}, \bibinfo {author} {\bibfnamefont {B.}~\bibnamefont {Vlastakis}},
  \bibinfo {author} {\bibfnamefont {E.}~\bibnamefont {Holland}}, \bibinfo
  {author} {\bibfnamefont {S.}~\bibnamefont {Krastanov}}, \bibinfo {author}
  {\bibfnamefont {V.~V.}\ \bibnamefont {Albert}}, \bibinfo {author}
  {\bibfnamefont {L.}~\bibnamefont {Frunzio}}, \bibinfo {author} {\bibfnamefont
  {L.}~\bibnamefont {Jiang}}, \ and\ \bibinfo {author} {\bibfnamefont {R.~J.}\
  \bibnamefont {Schoelkopf}},\ }\bibfield  {{Cavity State Manipulation Using
  Photon-Number Selective Phase Gates}} {\emph {\bibinfo {title} {{Cavity State
  Manipulation Using Photon-Number Selective Phase Gates}},\ }}\href {\doibase
  10.1103/PhysRevLett.115.137002} {\bibfield  {journal} {\bibinfo  {journal}
  {Phys. Rev. Lett.}\ }\textbf {\bibinfo {volume} {115}},\ \bibinfo {pages}
  {137002} (\bibinfo {year} {2015})}\BibitemShut {NoStop}%
\bibitem [{\citenamefont {Shen}\ \emph {et~al.}(2017)\citenamefont {Shen},
  \citenamefont {Noh}, \citenamefont {Albert}, \citenamefont {Krastanov},
  \citenamefont {Devoret}, \citenamefont {Schoelkopf}, \citenamefont {Girvin},\
  and\ \citenamefont {Jiang}}]{Shen2016}%
  \BibitemOpen
  \bibfield  {author} {\bibinfo {author} {\bibfnamefont {C.}~\bibnamefont
  {Shen}}, \bibinfo {author} {\bibfnamefont {K.}~\bibnamefont {Noh}}, \bibinfo
  {author} {\bibfnamefont {V.~V.}\ \bibnamefont {Albert}}, \bibinfo {author}
  {\bibfnamefont {S.}~\bibnamefont {Krastanov}}, \bibinfo {author}
  {\bibfnamefont {M.~H.}\ \bibnamefont {Devoret}}, \bibinfo {author}
  {\bibfnamefont {R.~J.}\ \bibnamefont {Schoelkopf}}, \bibinfo {author}
  {\bibfnamefont {S.~M.}\ \bibnamefont {Girvin}}, \ and\ \bibinfo {author}
  {\bibfnamefont {L.}~\bibnamefont {Jiang}},\ }\bibfield  {{Quantum channel
  construction with circuit quantum electrodynamics}} {\emph {\bibinfo {title}
  {{Quantum channel construction with circuit quantum electrodynamics}},\
  }}\href {\doibase 10.1103/PhysRevB.95.134501} {\bibfield  {journal} {\bibinfo
   {journal} {Phys. Rev. B}\ }\textbf {\bibinfo {volume} {95}},\ \bibinfo
  {pages} {134501} (\bibinfo {year} {2017})}\BibitemShut {NoStop}%
\bibitem [{\citenamefont {Lloyd}\ and\ \citenamefont
  {Viola}(2001)}]{Lloyd2001}%
  \BibitemOpen
  \bibfield  {author} {\bibinfo {author} {\bibfnamefont {S.}~\bibnamefont
  {Lloyd}}\ and\ \bibinfo {author} {\bibfnamefont {L.}~\bibnamefont {Viola}},\
  }\bibfield  {{Engineering quantum dynamics}} {\emph {\bibinfo {title}
  {{Engineering quantum dynamics}},\ }}\href {\doibase
  10.1103/PhysRevA.65.010101} {\bibfield  {journal} {\bibinfo  {journal} {Phys.
  Rev. A}\ }\textbf {\bibinfo {volume} {65}},\ \bibinfo {pages} {010101}
  (\bibinfo {year} {2001})}\BibitemShut {NoStop}%
\bibitem [{\citenamefont {Andersson}\ and\ \citenamefont
  {Oi}(2008)}]{Andersson2008}%
  \BibitemOpen
  \bibfield  {author} {\bibinfo {author} {\bibfnamefont {E.}~\bibnamefont
  {Andersson}}\ and\ \bibinfo {author} {\bibfnamefont {D.~K.~L.}\ \bibnamefont
  {Oi}},\ }\bibfield  {{Binary search trees for generalized measurements}}
  {\emph {\bibinfo {title} {{Binary search trees for generalized
  measurements}},\ }}\href {\doibase 10.1103/PhysRevA.77.052104} {\bibfield
  {journal} {\bibinfo  {journal} {Phys. Rev. A}\ }\textbf {\bibinfo {volume}
  {77}},\ \bibinfo {pages} {052104} (\bibinfo {year} {2008})}\BibitemShut
  {NoStop}%
\bibitem [{\citenamefont {Iten}\ \emph {et~al.}(2017)\citenamefont {Iten},
  \citenamefont {Colbeck},\ and\ \citenamefont {Christandl}}]{Iten2016}%
  \BibitemOpen
  \bibfield  {author} {\bibinfo {author} {\bibfnamefont {R.}~\bibnamefont
  {Iten}}, \bibinfo {author} {\bibfnamefont {R.}~\bibnamefont {Colbeck}}, \
  and\ \bibinfo {author} {\bibfnamefont {M.}~\bibnamefont {Christandl}},\
  }\bibfield  {{Quantum circuits for quantum channels}} {\emph {\bibinfo
  {title} {{Quantum circuits for quantum channels}},\ }}\href {\doibase
  10.1103/PhysRevA.95.052316} {\bibfield  {journal} {\bibinfo  {journal} {Phys.
  Rev. A}\ }\textbf {\bibinfo {volume} {95}},\ \bibinfo {pages} {052316}
  (\bibinfo {year} {2017})}\BibitemShut {NoStop}%
\bibitem [{\citenamefont {Barut}\ and\ \citenamefont
  {Girardello}(1971)}]{Barut1971}%
  \BibitemOpen
  \bibfield  {author} {\bibinfo {author} {\bibfnamefont {A.~O.}\ \bibnamefont
  {Barut}}\ and\ \bibinfo {author} {\bibfnamefont {L.}~\bibnamefont
  {Girardello}},\ }\bibfield  {{New ``coherent'' states associated with
  non-compact groups}} {\emph {\bibinfo {title} {{New ``coherent'' states
  associated with non-compact groups}},\ }}\href
  {http://projecteuclid.org/euclid.cmp/1103857258} {\bibfield  {journal}
  {\bibinfo  {journal} {Commun. Math. Phys.}\ }\textbf {\bibinfo {volume}
  {21}},\ \bibinfo {pages} {41} (\bibinfo {year} {1971})}\BibitemShut {NoStop}%
\bibitem [{\citenamefont {Agarwal}(1986)}]{Agarwal1986}%
  \BibitemOpen
  \bibfield  {author} {\bibinfo {author} {\bibfnamefont {G.~S.}\ \bibnamefont
  {Agarwal}},\ }\bibfield  {{Generation of Pair Coherent States and Squeezing
  via the Competition of Four-Wave Mixing and Amplified Spontaneous Emission}}
  {\emph {\bibinfo {title} {{Generation of Pair Coherent States and Squeezing
  via the Competition of Four-Wave Mixing and Amplified Spontaneous
  Emission}},\ }}\href {\doibase 10.1103/PhysRevLett.57.827} {\bibfield
  {journal} {\bibinfo  {journal} {Phys. Rev. Lett.}\ }\textbf {\bibinfo
  {volume} {57}},\ \bibinfo {pages} {827} (\bibinfo {year} {1986})}\BibitemShut
  {NoStop}%
\bibitem [{\citenamefont {Agarwal}(1988)}]{Agarwal1988}%
  \BibitemOpen
  \bibfield  {author} {\bibinfo {author} {\bibfnamefont {G.~S.}\ \bibnamefont
  {Agarwal}},\ }\bibfield  {{Nonclassical statistics of fields in pair coherent
  states}} {\emph {\bibinfo {title} {{Nonclassical statistics of fields in pair
  coherent states}},\ }}\href {\doibase 10.1364/JOSAB.5.001940} {\bibfield
  {journal} {\bibinfo  {journal} {J. Opt. Soc. Am. B}\ }\textbf {\bibinfo
  {volume} {5}},\ \bibinfo {pages} {1940} (\bibinfo {year} {1988})}\BibitemShut
  {NoStop}%
\bibitem [{\citenamefont {Perelomov}(1986)}]{perelomov_book}%
  \BibitemOpen
  \bibfield  {author} {\bibinfo {author} {\bibfnamefont {A.~M.}\ \bibnamefont
  {Perelomov}},\ }\href@noop {} {\emph {\bibinfo {title} {{Generalized Coherent
  States and Their Applications}}}}\ (\bibinfo  {publisher} {Springer},\
  \bibinfo {address} {Berlin},\ \bibinfo {year} {1986})\BibitemShut {NoStop}%
\bibitem [{\citenamefont {Gerry}\ and\ \citenamefont
  {Grobe}(1995)}]{Gerry1995}%
  \BibitemOpen
  \bibfield  {author} {\bibinfo {author} {\bibfnamefont {C.~C.}\ \bibnamefont
  {Gerry}}\ and\ \bibinfo {author} {\bibfnamefont {R.}~\bibnamefont {Grobe}},\
  }\bibfield  {{Nonclassical properties of correlated two-mode Schrodinger cat
  states}} {\emph {\bibinfo {title} {{Nonclassical properties of correlated
  two-mode Schrodinger cat states}},\ }}\href {\doibase
  10.1103/PhysRevA.51.1698} {\bibfield  {journal} {\bibinfo  {journal} {Phys.
  Rev. A}\ }\textbf {\bibinfo {volume} {51}},\ \bibinfo {pages} {1698}
  (\bibinfo {year} {1995})}\BibitemShut {NoStop}%
\bibitem [{\citenamefont {Gou}\ \emph {et~al.}(1996)\citenamefont {Gou},
  \citenamefont {Steinbach},\ and\ \citenamefont {Knight}}]{Gou1996}%
  \BibitemOpen
  \bibfield  {author} {\bibinfo {author} {\bibfnamefont {S.-C.}\ \bibnamefont
  {Gou}}, \bibinfo {author} {\bibfnamefont {J.}~\bibnamefont {Steinbach}}, \
  and\ \bibinfo {author} {\bibfnamefont {P.~L.}\ \bibnamefont {Knight}},\
  }\bibfield  {{Vibrational pair cat states}} {\emph {\bibinfo {title}
  {{Vibrational pair cat states}},\ }}\href {\doibase 10.1103/PhysRevA.54.4315}
  {\bibfield  {journal} {\bibinfo  {journal} {Phys. Rev. A}\ }\textbf {\bibinfo
  {volume} {54}},\ \bibinfo {pages} {4315} (\bibinfo {year}
  {1996})}\BibitemShut {NoStop}%
\bibitem [{\citenamefont {Liu}(2001)}]{Liu2001}%
  \BibitemOpen
  \bibfield  {author} {\bibinfo {author} {\bibfnamefont {X.-M.}\ \bibnamefont
  {Liu}},\ }\bibfield  {{Even and odd charge coherent states and their
  non-classical properties}} {\emph {\bibinfo {title} {{Even and odd charge
  coherent states and their non-classical properties}},\ }}\href {\doibase
  10.1016/S0375-9601(00)00803-3} {\bibfield  {journal} {\bibinfo  {journal}
  {Phys. Lett. A}\ }\textbf {\bibinfo {volume} {279}},\ \bibinfo {pages} {123}
  (\bibinfo {year} {2001})}\BibitemShut {NoStop}%
\bibitem [{\citenamefont {Choi}\ and\ \citenamefont {Yeon}(2008)}]{Choi2008}%
  \BibitemOpen
  \bibfield  {author} {\bibinfo {author} {\bibfnamefont {J.~R.}\ \bibnamefont
  {Choi}}\ and\ \bibinfo {author} {\bibfnamefont {K.~H.}\ \bibnamefont
  {Yeon}},\ }\bibfield  {{SU(1,1) Coherent States for the Generalized Two-Mode
  Time-Dependent Quadratic Hamiltonian System}} {\emph {\bibinfo {title}
  {{SU(1,1) Coherent States for the Generalized Two-Mode Time-Dependent
  Quadratic Hamiltonian System}},\ }}\href {\doibase 10.1007/s10773-007-9634-5}
  {\bibfield  {journal} {\bibinfo  {journal} {Int. J. Theor. Phys.}\ }\textbf
  {\bibinfo {volume} {47}},\ \bibinfo {pages} {1891} (\bibinfo {year}
  {2008})}\BibitemShut {NoStop}%
\bibitem [{\citenamefont {Wang}\ \emph {et~al.}(2016)\citenamefont {Wang},
  \citenamefont {Gao}, \citenamefont {Reinhold}, \citenamefont {Heeres},
  \citenamefont {Ofek}, \citenamefont {Chou}, \citenamefont {Axline},
  \citenamefont {Reagor}, \citenamefont {Blumoff}, \citenamefont {Sliwa},
  \citenamefont {Frunzio}, \citenamefont {Girvin}, \citenamefont {Jiang},
  \citenamefont {Mirrahimi}, \citenamefont {Devoret},\ and\ \citenamefont
  {Schoelkopf}}]{Wang2016}%
  \BibitemOpen
  \bibfield  {author} {\bibinfo {author} {\bibfnamefont {C.}~\bibnamefont
  {Wang}}, \bibinfo {author} {\bibfnamefont {Y.~Y.}\ \bibnamefont {Gao}},
  \bibinfo {author} {\bibfnamefont {P.}~\bibnamefont {Reinhold}}, \bibinfo
  {author} {\bibfnamefont {R.~W.}\ \bibnamefont {Heeres}}, \bibinfo {author}
  {\bibfnamefont {N.}~\bibnamefont {Ofek}}, \bibinfo {author} {\bibfnamefont
  {K.}~\bibnamefont {Chou}}, \bibinfo {author} {\bibfnamefont {C.}~\bibnamefont
  {Axline}}, \bibinfo {author} {\bibfnamefont {M.}~\bibnamefont {Reagor}},
  \bibinfo {author} {\bibfnamefont {J.}~\bibnamefont {Blumoff}}, \bibinfo
  {author} {\bibfnamefont {K.~M.}\ \bibnamefont {Sliwa}}, \bibinfo {author}
  {\bibfnamefont {L.}~\bibnamefont {Frunzio}}, \bibinfo {author} {\bibfnamefont
  {S.~M.}\ \bibnamefont {Girvin}}, \bibinfo {author} {\bibfnamefont
  {L.}~\bibnamefont {Jiang}}, \bibinfo {author} {\bibfnamefont
  {M.}~\bibnamefont {Mirrahimi}}, \bibinfo {author} {\bibfnamefont {M.~H.}\
  \bibnamefont {Devoret}}, \ and\ \bibinfo {author} {\bibfnamefont {R.~J.}\
  \bibnamefont {Schoelkopf}},\ }\bibfield  {{A Schrodinger cat living in two
  boxes}} {\emph {\bibinfo {title} {{A Schrodinger cat living in two boxes}},\
  }}\href {\doibase 10.1126/science.aaf2941} {\bibfield  {journal} {\bibinfo
  {journal} {Science (80-. ).}\ }\textbf {\bibinfo {volume} {352}},\ \bibinfo
  {pages} {1087} (\bibinfo {year} {2016})}\BibitemShut {NoStop}%
\bibitem [{\citenamefont {Hillery}\ \emph {et~al.}(1984)\citenamefont
  {Hillery}, \citenamefont {O'Connell}, \citenamefont {Scully},\ and\
  \citenamefont {Wigner}}]{Hillery1984}%
  \BibitemOpen
  \bibfield  {author} {\bibinfo {author} {\bibfnamefont {M.}~\bibnamefont
  {Hillery}}, \bibinfo {author} {\bibfnamefont {R.}~\bibnamefont {O'Connell}},
  \bibinfo {author} {\bibfnamefont {M.}~\bibnamefont {Scully}}, \ and\ \bibinfo
  {author} {\bibfnamefont {E.}~\bibnamefont {Wigner}},\ }\bibfield
  {{Distribution functions in physics: Fundamentals}} {\emph {\bibinfo {title}
  {{Distribution functions in physics: Fundamentals}},\ }}\href {\doibase
  10.1016/0370-1573(84)90160-1} {\bibfield  {journal} {\bibinfo  {journal}
  {Phys. Rep.}\ }\textbf {\bibinfo {volume} {106}},\ \bibinfo {pages} {121}
  (\bibinfo {year} {1984})}\BibitemShut {NoStop}%
\bibitem [{\citenamefont {Werner}(2016)}]{Werner2016}%
  \BibitemOpen
  \bibfield  {author} {\bibinfo {author} {\bibfnamefont {R.~F.}\ \bibnamefont
  {Werner}},\ }\bibfield  {{Uncertainty relations for general phase spaces}}
  {\emph {\bibinfo {title} {{Uncertainty relations for general phase spaces}},\
  }}\href {\doibase 10.1007/s11467-016-0558-5} {\bibfield  {journal} {\bibinfo
  {journal} {Front. Phys.}\ }\textbf {\bibinfo {volume} {11}},\ \bibinfo
  {pages} {110305} (\bibinfo {year} {2016})}\BibitemShut {NoStop}%
\bibitem [{\citenamefont {Brif}\ and\ \citenamefont
  {Ben-Aryeh}(1994)}]{Brif1994}%
  \BibitemOpen
  \bibfield  {author} {\bibinfo {author} {\bibfnamefont {C.}~\bibnamefont
  {Brif}}\ and\ \bibinfo {author} {\bibfnamefont {Y.}~\bibnamefont
  {Ben-Aryeh}},\ }\bibfield  {{Subcoherent p-representation for non-classical
  photon states}} {\emph {\bibinfo {title} {{Subcoherent p-representation for
  non-classical photon states}},\ }}\href {\doibase 10.1088/0954-8998/6/5/001}
  {\bibfield  {journal} {\bibinfo  {journal} {Quantum Opt.}\ }\textbf {\bibinfo
  {volume} {6}},\ \bibinfo {pages} {391} (\bibinfo {year} {1994})}\BibitemShut
  {NoStop}%
\bibitem [{\citenamefont {{D. Popov}}(2001)}]{DusanPopov2001}%
  \BibitemOpen
  \bibfield  {author} {\bibinfo {author} {\bibnamefont {{D. Popov}}},\
  }\bibfield  {{Barut-Girardello coherent states of the pseudoharmonic
  oscillator}} {\emph {\bibinfo {title} {{Barut-Girardello coherent states of
  the pseudoharmonic oscillator}},\ }}\href {\doibase
  http://dx.doi.org/10.1088/0305-4470/34/25/310} {\bibfield  {journal}
  {\bibinfo  {journal} {J. Phys. A Math. Gen.}\ }\textbf {\bibinfo {volume}
  {34}},\ \bibinfo {pages} {5283} (\bibinfo {year} {2001})}\BibitemShut
  {NoStop}%
\bibitem [{\citenamefont {D'Ariano}\ \emph {et~al.}(2001)\citenamefont
  {D'Ariano}, \citenamefont {{De Vito}},\ and\ \citenamefont
  {Maccone}}]{Maccone2001}%
  \BibitemOpen
  \bibfield  {author} {\bibinfo {author} {\bibfnamefont {G.~M.}\ \bibnamefont
  {D'Ariano}}, \bibinfo {author} {\bibfnamefont {E.}~\bibnamefont {{De Vito}}},
  \ and\ \bibinfo {author} {\bibfnamefont {L.}~\bibnamefont {Maccone}},\
  }\bibfield  {{SU(1,1) tomography}} {\emph {\bibinfo {title} {{SU(1,1)
  tomography}},\ }}\href {\doibase 10.1103/PhysRevA.64.033805} {\bibfield
  {journal} {\bibinfo  {journal} {Phys. Rev. A}\ }\textbf {\bibinfo {volume}
  {64}},\ \bibinfo {pages} {033805} (\bibinfo {year} {2001})}\BibitemShut
  {NoStop}%
\bibitem [{\citenamefont {Carmeli}\ \emph {et~al.}(2009)\citenamefont
  {Carmeli}, \citenamefont {Cassinelli},\ and\ \citenamefont
  {Zizzi}}]{Carmeli2009}%
  \BibitemOpen
  \bibfield  {author} {\bibinfo {author} {\bibfnamefont {C.}~\bibnamefont
  {Carmeli}}, \bibinfo {author} {\bibfnamefont {G.}~\bibnamefont {Cassinelli}},
  \ and\ \bibinfo {author} {\bibfnamefont {F.}~\bibnamefont {Zizzi}},\
  }\bibfield  {{Generalized Orthogonality Relations and SU(1,1)-Quantum
  Tomography}} {\emph {\bibinfo {title} {{Generalized Orthogonality Relations
  and SU(1,1)-Quantum Tomography}},\ }}\href {\doibase
  10.1007/s10701-009-9290-0} {\bibfield  {journal} {\bibinfo  {journal} {Found.
  Phys.}\ }\textbf {\bibinfo {volume} {39}},\ \bibinfo {pages} {521} (\bibinfo
  {year} {2009})}\BibitemShut {NoStop}%
\bibitem [{\citenamefont {Ferrie}(2011)}]{Ferrie2011}%
  \BibitemOpen
  \bibfield  {author} {\bibinfo {author} {\bibfnamefont {C.}~\bibnamefont
  {Ferrie}},\ }\bibfield  {{Quasi-probability representations of quantum theory
  with applications to quantum information science}} {\emph {\bibinfo {title}
  {{Quasi-probability representations of quantum theory with applications to
  quantum information science}},\ }}\href {\doibase
  10.1088/0034-4885/74/11/116001} {\bibfield  {journal} {\bibinfo  {journal}
  {Rep. Prog. Phys.}\ }\textbf {\bibinfo {volume} {74}},\ \bibinfo {pages}
  {116001} (\bibinfo {year} {2011})}\BibitemShut {NoStop}%
\bibitem [{\citenamefont {Haroche}\ and\ \citenamefont
  {Raimond}(2006)}]{catbook}%
  \BibitemOpen
  \bibfield  {author} {\bibinfo {author} {\bibfnamefont {S.}~\bibnamefont
  {Haroche}}\ and\ \bibinfo {author} {\bibfnamefont {J.-M.}\ \bibnamefont
  {Raimond}},\ }\href
  {http://www.oxfordscholarship.com/view/10.1093/acprof:oso/9780198509141.001.0001/acprof-9780198509141}
  {\emph {\bibinfo {title} {{Exploring the quantum: atoms, cavities, and
  photons}}}}\ (\bibinfo  {publisher} {Oxford University Press},\ \bibinfo
  {address} {Oxford},\ \bibinfo {year} {2006})\BibitemShut {NoStop}%
\bibitem [{\citenamefont {Gottesman}(1997)}]{gottesman_thesis}%
  \BibitemOpen
  \bibfield  {author} {\bibinfo {author} {\bibfnamefont {D.}~\bibnamefont
  {Gottesman}},\ }\emph {\bibinfo {title} {{Stabilizer codes and quantum error
  correction}}},\ \href {https://arxiv.org/abs/quant-ph/9705052} {Ph.D.
  thesis},\ \bibinfo  {school} {California Institute of Technology} (\bibinfo
  {year} {1997})\BibitemShut {NoStop}%
\bibitem [{\citenamefont {Knill}(1996)}]{Knill1996}%
  \BibitemOpen
  \bibfield  {author} {\bibinfo {author} {\bibfnamefont {E.}~\bibnamefont
  {Knill}},\ }\bibfield  {{Group Representations, Error Bases and Quantum
  Codes}} {\emph {\bibinfo {title} {{Group Representations, Error Bases and
  Quantum Codes}},\ }}\href {http://arxiv.org/abs/quant-ph/9608049} {\
  (\bibinfo {year} {1996})},\ \Eprint {http://arxiv.org/abs/9608049}
  {arXiv:9608049 [quant-ph]} \BibitemShut {NoStop}%
\bibitem [{\citenamefont {Pollatsek}\ and\ \citenamefont
  {Ruskai}(2004)}]{Pollatsek2004}%
  \BibitemOpen
  \bibfield  {author} {\bibinfo {author} {\bibfnamefont {H.}~\bibnamefont
  {Pollatsek}}\ and\ \bibinfo {author} {\bibfnamefont {M.~B.}\ \bibnamefont
  {Ruskai}},\ }\bibfield  {{Permutationally invariant codes for quantum error
  correction}} {\emph {\bibinfo {title} {{Permutationally invariant codes for
  quantum error correction}},\ }}\href {\doibase 10.1016/j.laa.2004.06.014}
  {\bibfield  {journal} {\bibinfo  {journal} {Linear Algebr. Appl.}\ }\textbf
  {\bibinfo {volume} {392}},\ \bibinfo {pages} {255} (\bibinfo {year}
  {2004})}\BibitemShut {NoStop}%
\bibitem [{\citenamefont {Looi}\ \emph {et~al.}(2008)\citenamefont {Looi},
  \citenamefont {Yu}, \citenamefont {Gheorghiu},\ and\ \citenamefont
  {Griffiths}}]{Looi2008}%
  \BibitemOpen
  \bibfield  {author} {\bibinfo {author} {\bibfnamefont {S.~Y.}\ \bibnamefont
  {Looi}}, \bibinfo {author} {\bibfnamefont {L.}~\bibnamefont {Yu}}, \bibinfo
  {author} {\bibfnamefont {V.}~\bibnamefont {Gheorghiu}}, \ and\ \bibinfo
  {author} {\bibfnamefont {R.~B.}\ \bibnamefont {Griffiths}},\ }\bibfield
  {{Quantum-error-correcting codes using qudit graph states}} {\emph {\bibinfo
  {title} {{Quantum-error-correcting codes using qudit graph states}},\ }}\href
  {\doibase 10.1103/PhysRevA.78.042303} {\bibfield  {journal} {\bibinfo
  {journal} {Phys. Rev. A}\ }\textbf {\bibinfo {volume} {78}},\ \bibinfo
  {pages} {042303} (\bibinfo {year} {2008})}\BibitemShut {NoStop}%
\bibitem [{\citenamefont {Cross}\ \emph {et~al.}(2009)\citenamefont {Cross},
  \citenamefont {Smith}, \citenamefont {Smolin},\ and\ \citenamefont
  {Zeng}}]{Cross2009}%
  \BibitemOpen
  \bibfield  {author} {\bibinfo {author} {\bibfnamefont {A.}~\bibnamefont
  {Cross}}, \bibinfo {author} {\bibfnamefont {G.}~\bibnamefont {Smith}},
  \bibinfo {author} {\bibfnamefont {J.~A.}\ \bibnamefont {Smolin}}, \ and\
  \bibinfo {author} {\bibfnamefont {B.}~\bibnamefont {Zeng}},\ }\bibfield
  {{Codeword Stabilized Quantum Codes}} {\emph {\bibinfo {title} {{Codeword
  Stabilized Quantum Codes}},\ }}\href {\doibase 10.1109/TIT.2008.2008136}
  {\bibfield  {journal} {\bibinfo  {journal} {IEEE Trans. Inf. Theory}\
  }\textbf {\bibinfo {volume} {55}},\ \bibinfo {pages} {433} (\bibinfo {year}
  {2009})}\BibitemShut {NoStop}%
\bibitem [{\citenamefont {Kruszynska}\ and\ \citenamefont
  {Kraus}(2009)}]{alKruszynska2009}%
  \BibitemOpen
  \bibfield  {author} {\bibinfo {author} {\bibfnamefont {C.}~\bibnamefont
  {Kruszynska}}\ and\ \bibinfo {author} {\bibfnamefont {B.}~\bibnamefont
  {Kraus}},\ }\bibfield  {{Local entanglability and multipartite entanglement}}
  {\emph {\bibinfo {title} {{Local entanglability and multipartite
  entanglement}},\ }}\href {\doibase 10.1103/PhysRevA.79.052304} {\bibfield
  {journal} {\bibinfo  {journal} {Phys. Rev. A}\ }\textbf {\bibinfo {volume}
  {79}},\ \bibinfo {pages} {052304} (\bibinfo {year} {2009})}\BibitemShut
  {NoStop}%
\bibitem [{\citenamefont {Rossi}\ \emph {et~al.}(2013)\citenamefont {Rossi},
  \citenamefont {Huber}, \citenamefont {Bru{\ss}},\ and\ \citenamefont
  {Macchiavello}}]{Rossi2013}%
  \BibitemOpen
  \bibfield  {author} {\bibinfo {author} {\bibfnamefont {M.}~\bibnamefont
  {Rossi}}, \bibinfo {author} {\bibfnamefont {M.}~\bibnamefont {Huber}},
  \bibinfo {author} {\bibfnamefont {D.}~\bibnamefont {Bru{\ss}}}, \ and\
  \bibinfo {author} {\bibfnamefont {C.}~\bibnamefont {Macchiavello}},\
  }\bibfield  {{Quantum hypergraph states}} {\emph {\bibinfo {title} {{Quantum
  hypergraph states}},\ }}\href {\doibase 10.1088/1367-2630/15/11/113022}
  {\bibfield  {journal} {\bibinfo  {journal} {New J. Phys.}\ }\textbf {\bibinfo
  {volume} {15}},\ \bibinfo {pages} {113022} (\bibinfo {year}
  {2013})}\BibitemShut {NoStop}%
\bibitem [{\citenamefont {Ni}\ \emph {et~al.}(2015)\citenamefont {Ni},
  \citenamefont {Buerschaper},\ and\ \citenamefont {{Van den Nest}}}]{Ni2015}%
  \BibitemOpen
  \bibfield  {author} {\bibinfo {author} {\bibfnamefont {X.}~\bibnamefont
  {Ni}}, \bibinfo {author} {\bibfnamefont {O.}~\bibnamefont {Buerschaper}}, \
  and\ \bibinfo {author} {\bibfnamefont {M.}~\bibnamefont {{Van den Nest}}},\
  }\bibfield  {{A non-commuting stabilizer formalism}} {\emph {\bibinfo {title}
  {{A non-commuting stabilizer formalism}},\ }}\href {\doibase
  10.1063/1.4920923} {\bibfield  {journal} {\bibinfo  {journal} {J. Math.
  Phys.}\ }\textbf {\bibinfo {volume} {56}},\ \bibinfo {pages} {052201}
  (\bibinfo {year} {2015})}\BibitemShut {NoStop}%
\bibitem [{\citenamefont {Leung}\ \emph {et~al.}(1997)\citenamefont {Leung},
  \citenamefont {Nielsen}, \citenamefont {Chuang},\ and\ \citenamefont
  {Yamamoto}}]{Leung1997}%
  \BibitemOpen
  \bibfield  {author} {\bibinfo {author} {\bibfnamefont {D.~W.}\ \bibnamefont
  {Leung}}, \bibinfo {author} {\bibfnamefont {M.~A.}\ \bibnamefont {Nielsen}},
  \bibinfo {author} {\bibfnamefont {I.~L.}\ \bibnamefont {Chuang}}, \ and\
  \bibinfo {author} {\bibfnamefont {Y.}~\bibnamefont {Yamamoto}},\ }\bibfield
  {{Approximate quantum error correction can lead to better codes}} {\emph
  {\bibinfo {title} {{Approximate quantum error correction can lead to better
  codes}},\ }}\href {\doibase 10.1103/PhysRevA.56.2567} {\bibfield  {journal}
  {\bibinfo  {journal} {Phys. Rev. A}\ }\textbf {\bibinfo {volume} {56}},\
  \bibinfo {pages} {2567} (\bibinfo {year} {1997})}\BibitemShut {NoStop}%
\bibitem [{\citenamefont {Cr{\'{e}}peau}\ \emph {et~al.}(2005)\citenamefont
  {Cr{\'{e}}peau}, \citenamefont {Gottesman},\ and\ \citenamefont
  {Smith}}]{Crepeau2005}%
  \BibitemOpen
  \bibfield  {author} {\bibinfo {author} {\bibfnamefont {C.}~\bibnamefont
  {Cr{\'{e}}peau}}, \bibinfo {author} {\bibfnamefont {D.}~\bibnamefont
  {Gottesman}}, \ and\ \bibinfo {author} {\bibfnamefont {A.}~\bibnamefont
  {Smith}},\ }in\ \href {\doibase 10.1007/11426639_17} {\emph {\bibinfo
  {booktitle} {Adv. Cryptol. - EUROCRYPT 2005. Lect. Notes Comput. Sci. vol.
  3494}}},\ \bibinfo {editor} {edited by\ \bibinfo {editor} {\bibfnamefont
  {R.}~\bibnamefont {Cramer}}}\ (\bibinfo  {publisher} {Springer},\ \bibinfo
  {address} {Berlin, Heidelberg},\ \bibinfo {year} {2005})\ pp.\ \bibinfo
  {pages} {285--301}\BibitemShut {NoStop}%
\bibitem [{\citenamefont {An}(2003)}]{an2003}%
  \BibitemOpen
  \bibfield  {author} {\bibinfo {author} {\bibfnamefont {N.~B.}\ \bibnamefont
  {An}},\ }\bibfield  {{Even and odd trio coherent states: number distribution,
  squeezing and realization scheme}} {\emph {\bibinfo {title} {{Even and odd
  trio coherent states: number distribution, squeezing and realization
  scheme}},\ }}\href {\doibase 10.1016/S0375-9601(03)00652-2} {\bibfield
  {journal} {\bibinfo  {journal} {Phys. Lett. A}\ }\textbf {\bibinfo {volume}
  {312}},\ \bibinfo {pages} {268} (\bibinfo {year} {2003})}\BibitemShut
  {NoStop}%
\bibitem [{\citenamefont {Niu}\ \emph {et~al.}(2018{\natexlab{b}})\citenamefont
  {Niu}, \citenamefont {Chuang},\ and\ \citenamefont {Shapiro}}]{niu1}%
  \BibitemOpen
  \bibfield  {author} {\bibinfo {author} {\bibfnamefont {M.~Y.}\ \bibnamefont
  {Niu}}, \bibinfo {author} {\bibfnamefont {I.~L.}\ \bibnamefont {Chuang}}, \
  and\ \bibinfo {author} {\bibfnamefont {J.~H.}\ \bibnamefont {Shapiro}},\
  }\bibfield  {{Qudit-Basis Universal Quantum Computation Using chi-squared
  Interactions}} {\emph {\bibinfo {title} {{Qudit-Basis Universal Quantum
  Computation Using chi-squared Interactions}},\ }}\href {\doibase
  10.1103/PhysRevLett.120.160502} {\bibfield  {journal} {\bibinfo  {journal}
  {Phys. Rev. Lett.}\ }\textbf {\bibinfo {volume} {120}},\ \bibinfo {pages}
  {160502} (\bibinfo {year} {2018}{\natexlab{b}})}\BibitemShut {NoStop}%
\bibitem [{\citenamefont {Ng}\ and\ \citenamefont {Mandayam}(2010)}]{Ng2010}%
  \BibitemOpen
  \bibfield  {author} {\bibinfo {author} {\bibfnamefont {H.~K.}\ \bibnamefont
  {Ng}}\ and\ \bibinfo {author} {\bibfnamefont {P.}~\bibnamefont {Mandayam}},\
  }\bibfield  {{Simple approach to approximate quantum error correction based
  on the transpose channel}} {\emph {\bibinfo {title} {{Simple approach to
  approximate quantum error correction based on the transpose channel}},\
  }}\href {\doibase 10.1103/PhysRevA.81.062342} {\bibfield  {journal} {\bibinfo
   {journal} {Phys. Rev. A}\ }\textbf {\bibinfo {volume} {81}},\ \bibinfo
  {pages} {062342} (\bibinfo {year} {2010})}\BibitemShut {NoStop}%
\bibitem [{\citenamefont {Mirrahimi}\ and\ \citenamefont
  {Rouchon}(2015)}]{Mirrahimi2015}%
  \BibitemOpen
  \bibfield  {author} {\bibinfo {author} {\bibfnamefont {M.}~\bibnamefont
  {Mirrahimi}}\ and\ \bibinfo {author} {\bibfnamefont {P.}~\bibnamefont
  {Rouchon}},\ }\href
  {http://cas.ensmp.fr/$\sim$rouchon/quantumsyst/controlquantsyst_28nov2010.pdf}
  {\bibinfo {title} {{Modeling and Control of Quantum Systems}},\ } (\bibinfo
  {year} {2015})\BibitemShut {NoStop}%
\bibitem [{\citenamefont {Carmichael}(2008)}]{carmichael2}%
  \BibitemOpen
  \bibfield  {author} {\bibinfo {author} {\bibfnamefont {H.~J.}\ \bibnamefont
  {Carmichael}},\ }\href {https://www.springer.com/us/book/9783540713197}
  {\emph {\bibinfo {title} {{Statistical Methods in Quantum Optics 2:
  Non-classical fields}}}}\ (\bibinfo  {publisher} {Springer-Verlag},\ \bibinfo
  {address} {Berlin/Heidelberg},\ \bibinfo {year} {2008})\BibitemShut {NoStop}%
\bibitem [{\citenamefont {Verstraete}\ \emph {et~al.}(2009)\citenamefont
  {Verstraete}, \citenamefont {Wolf},\ and\ \citenamefont
  {Cirac}}]{Verstraete2009}%
  \BibitemOpen
  \bibfield  {author} {\bibinfo {author} {\bibfnamefont {F.}~\bibnamefont
  {Verstraete}}, \bibinfo {author} {\bibfnamefont {M.~M.}\ \bibnamefont
  {Wolf}}, \ and\ \bibinfo {author} {\bibfnamefont {J.~I.}\ \bibnamefont
  {Cirac}},\ }\bibfield  {{Quantum computation and quantum-state engineering
  driven by dissipation}} {\emph {\bibinfo {title} {{Quantum computation and
  quantum-state engineering driven by dissipation}},\ }}\href {\doibase
  10.1038/nphys1342} {\bibfield  {journal} {\bibinfo  {journal} {Nat. Phys.}\
  }\textbf {\bibinfo {volume} {5}},\ \bibinfo {pages} {633} (\bibinfo {year}
  {2009})}\BibitemShut {NoStop}%
\bibitem [{\citenamefont {Hatridge}\ \emph {et~al.}(2013)\citenamefont
  {Hatridge}, \citenamefont {Shankar}, \citenamefont {Mirrahimi}, \citenamefont
  {Schackert}, \citenamefont {Geerlings}, \citenamefont {Brecht}, \citenamefont
  {Sliwa}, \citenamefont {Abdo}, \citenamefont {Frunzio}, \citenamefont
  {Girvin}, \citenamefont {Schoelkopf},\ and\ \citenamefont
  {Devoret}}]{Hatridge2013}%
  \BibitemOpen
  \bibfield  {author} {\bibinfo {author} {\bibfnamefont {M.}~\bibnamefont
  {Hatridge}}, \bibinfo {author} {\bibfnamefont {S.}~\bibnamefont {Shankar}},
  \bibinfo {author} {\bibfnamefont {M.}~\bibnamefont {Mirrahimi}}, \bibinfo
  {author} {\bibfnamefont {F.}~\bibnamefont {Schackert}}, \bibinfo {author}
  {\bibfnamefont {K.}~\bibnamefont {Geerlings}}, \bibinfo {author}
  {\bibfnamefont {T.}~\bibnamefont {Brecht}}, \bibinfo {author} {\bibfnamefont
  {K.~M.}\ \bibnamefont {Sliwa}}, \bibinfo {author} {\bibfnamefont
  {B.}~\bibnamefont {Abdo}}, \bibinfo {author} {\bibfnamefont {L.}~\bibnamefont
  {Frunzio}}, \bibinfo {author} {\bibfnamefont {S.~M.}\ \bibnamefont {Girvin}},
  \bibinfo {author} {\bibfnamefont {R.~J.}\ \bibnamefont {Schoelkopf}}, \ and\
  \bibinfo {author} {\bibfnamefont {M.~H.}\ \bibnamefont {Devoret}},\
  }\bibfield  {{Quantum Back-Action of an Individual Variable-Strength
  Measurement}} {\emph {\bibinfo {title} {{Quantum Back-Action of an Individual
  Variable-Strength Measurement}},\ }}\href {\doibase 10.1126/science.1226897}
  {\bibfield  {journal} {\bibinfo  {journal} {Science (80-. ).}\ }\textbf
  {\bibinfo {volume} {339}},\ \bibinfo {pages} {178} (\bibinfo {year}
  {2013})}\BibitemShut {NoStop}%
\bibitem [{\citenamefont {Scully}\ and\ \citenamefont
  {Zubairy}(1997)}]{scully_book}%
  \BibitemOpen
  \bibfield  {author} {\bibinfo {author} {\bibfnamefont {M.~O.}\ \bibnamefont
  {Scully}}\ and\ \bibinfo {author} {\bibfnamefont {M.~S.}\ \bibnamefont
  {Zubairy}},\ }\href
  {https://www.amazon.com/Quantum-Optics-Marlan-Scully/dp/0521435951} {\emph
  {\bibinfo {title} {{Quantum Optics}}}}\ (\bibinfo  {publisher} {Cambridge
  University Press},\ \bibinfo {address} {Cambridge},\ \bibinfo {year}
  {1997})\BibitemShut {NoStop}%
\end{thebibliography}%

\end{document}